\documentclass[12pt]{article}

\usepackage{lmodern}
\usepackage{graphicx}             
\DeclareGraphicsExtensions{.pdf}  
\usepackage{tikz}                 
\usetikzlibrary{shapes,arrows}
\usepackage{subfigure}            
\usepackage{amsmath}              
\usepackage{amssymb}              
\usepackage{amsthm}               
\usepackage[squaren,mediumqspace]{SIunits}     
\usepackage{multirow}             
\usepackage{tabularx}             
\usepackage{slashed}              
\usepackage{booktabs}             
\usepackage{import}               
\usepackage{xcolor}    
\usepackage{bigdelim}             
\usepackage{bbm}                  
\usepackage{feynmp}               
\usepackage{wasysym}              
\usepackage{cite}                 
\usepackage{hyphenat}             
\usepackage{datetime}
\usepackage[utf8]{inputenc}
\usepackage{ulem}

\newcommand{\MHexp}{125 \, \giga\electronvolt}

\newcommand{\order}[1]{\ensuremath{{\cal O}(#1)}}
\newcommand{\cp}{\ensuremath{{\cal CP}}}
\newcommand{\al}{\alpha}
\newcommand{\als}{\al_s}
\newcommand{\alt}{\al_t}
\newcommand{\alb}{\al_b}
\newcommand{\be}{\beta}
\newcommand{\tb}{\tan\be}
\newcommand{\la}{\lambda}
\newcommand{\ka}{\kappa}

\newcommand{\gev}{\giga\electronvolt}
\newcommand{\mev}{\mega\electronvolt}

\newcommand\sw{s_\mathrm{w}}

\newcommand\cw{c_\mathrm{w}}
\newcommand\MW{M_W}
\newcommand\MZ{M_Z}

\newcommand\MA{M_A}
\newcommand\MHp{M_{H^\pm}}

\newcommand{\mt}{m_t}
\newcommand{\tst}{\ensuremath{\theta_{\tilde t}}}
\newcommand{\At}{A_t}
\newcommand{\yt}{Y_t}

\newcommand{\Hi}{h_i}

\newcommand{\lsim}
{\;\raisebox{-.3em}{$\stackrel{\displaystyle <}{\sim}$}\;}

\newcommand{\DRbar}{\ensuremath{\overline{\mathrm{DR}}}}
\newcommand{\MSbar}{\ensuremath{\overline{\mathrm{MS}}}}

\newcommand{\fh}{{\tt Feyn\-Higgs}}
\newcommand{\FH}{{\tt Feyn\-Higgs}}

\newcommand{\NC}{\texttt{NMSSM\-Calc}}
\newcommand{\FA}{\texttt{Feyn\-Arts}}
\newcommand{\FC}{\texttt{Form\-Calc}}
\newcommand{\NFH}{\texttt{NMSSM\hyp{}Feyn\-Higgs}}

\def\refeq#1{\mbox{eq.~(\ref{#1})}}

\def\reffi#1{\mbox{fig.~\ref{#1}}}

\def\refta#1{\mbox{tab.~\ref{#1}}}

\def\refse#1{\mbox{sect.~\ref{#1}}}

\newcommand{\mathi}{i}

\newcommand{\UY}{\ensuremath{U(1)_\text{Y}}}
\newcommand{\SUL}{\ensuremath{SU(2)_\text{L}}}
\newcommand{\SUc}{\ensuremath{SU(3)_\text{c}}}

\newcommand{\mueff}{\ensuremath{\mu_{\text{eff}}}}

\newcommand{\dvSqZ}{\ensuremath{\frac{\delta{v^2}}{v^2}}}

\newcommand{\CB}[1]{\ensuremath{\cos^{#1}{\beta}}}

\newcommand{\SB}[1]{\ensuremath{\sin^{#1}{\beta}}}

\newcommand{\dSWSq}{\ensuremath{\delta{\sw^2}}}
\newcommand{\dSWSqZ}{\ensuremath{\frac{\delta{\sw^2}}{\sw^2}}}
\newcommand{\dMWSq}{\ensuremath{\delta{\MW^2}}}
\newcommand{\dMWSqZ}{\ensuremath{\frac{\delta{\MW^2}}{\MW^2}}}

\newcommand{\fmfvcenter}[1]{\vcenter{\hbox{\fmfreuse{#1}}}}

\usepackage[pdftitle={Letter}, pdfauthor={Peter Drechsel}, pdfkeywords={higgs mass, precision calculation, NMSSM}, bookmarks=true, pdfborder={0 0 0}]{hyperref}
\usepackage[all]{hypcap}

\evensidemargin \oddsidemargin
\begin{document}

\thispagestyle{empty}

\def\thefootnote{\fnsymbol{footnote}}

\begin{flushright}
DESY 15-069 
\end{flushright}

\vspace{0.5cm}

\begin{center}
{\large\sc  {\bf  Precise  Predictions for  \\[.5em]  the  Higgs-Boson
    Masses in the NMSSM}}
\vspace{0.5cm}

\vspace{1cm}

{\sc
P.~Drechsel$^{1}$%
\footnote{email: peter.drechsel@desy.de}%
, L.~Galeta$^{2}$%
\footnote{email: leo@ifca.unican.es}%
, S.~Heinemeyer$^{2,3}$%
\footnote{email: Sven.Heinemeyer@cern.ch}%
,~and G.~Weiglein$^{1}$%
\footnote{email: Georg.Weiglein@desy.de}
}

\vspace*{.7cm}

{\sl
$^1$DESY, Notkestra\ss e 85, D--22607 Hamburg, Germany

\vspace*{0.1cm}

$^2$Instituto de F\'isica de Cantabria (CSIC-UC), Santander,  Spain

\vspace*{0.1cm}

$^3$Instituto de F\'isica Te\'orica, (UAM/CSIC), Universidad
   Aut\'onoma de Madrid,\\ Cantoblanco, E-28049 Madrid, Spain
}

\end{center}

\vspace*{0.1cm}

\begin{abstract}
\noindent
The particle discovered in the Higgs  boson searches at the LHC with a
mass of about $\MHexp$ can be identified with one of the neutral Higgs
bosons of  the Next-to-Minimal Supersymmetric Standard  Model (NMSSM).
We calculate predictions for the Higgs-boson masses in the NMSSM using
the Feynman-diagrammatic  approach.  The predictions are  based on the
full  NMSSM one-loop  corrections supplemented  with the  dominant and
sub-dominant  two-loop corrections  within the  Minimal Supersymmetric
Standard    Model   (MSSM).     These    include   contributions    at
$\mathcal{O}{\left(\alpha_t      \alpha_s,     \alpha_b      \alpha_s,
  \alpha_t^2,\alpha_t\alpha_b\right)}$,  as well  as a  resummation of
leading  and subleading  logarithms  from the  top/scalar top  sector.
Taking these corrections  into account in the prediction  for the mass
of the Higgs  boson in the NMSSM that is  identified with the observed
signal is crucial in order to reach  a precision at a similar level as
in the  MSSM.  The quality of  the approximation made at  the two-loop
level is  analysed on the  basis of the  full one-loop result,  with a
particular focus on  the prediction for the  Standard Model-like Higgs
boson  that is  associated  with the  observed  signal.  The  obtained
results will be used as a basis  for the extension of the code \fh\ to
the NMSSM.
\end{abstract}

\def\thefootnote{\arabic{footnote}}
\setcounter{page}{0}
\setcounter{footnote}{0}

\newpage

\newsavebox{\SE}
\sbox{\SE}{
\begin{fmffile}{Feynman/SE}

\begin{fmfgraph*}(45,30)
  \fmfkeep{ScalarSESq}
  \fmfleft{i1}
  \fmfright{o1}
  \fmf{dashes}{i1,v1}
  \fmf{dashes}{v2,o1}
  \fmf{dashes,left,tension=.4}{v1,v2}
  \fmf{dashes,left,tension=.4}{v2,v1}
\end{fmfgraph*}

\begin{fmfgraph*}(45,30)
  \fmfkeep{ScalarSEA0Sq}
  \fmfleft{i1}
  \fmfright{o1}
  \fmf{dashes}{i1,v1,o1}
  \fmf{dashes}{v1,v1}
\end{fmfgraph*}

\begin{fmfgraph*}(45,30)
  \fmfkeep{ScalarSEQ}
  \fmfleft{i1}
  \fmfright{o1}
  \fmf{dashes}{i1,v1}
  \fmf{dashes}{v2,o1}
  \fmf{plain,left,tension=.4}{v1,v2}
  \fmf{plain,left,tension=.4}{v2,v1}
\end{fmfgraph*}

\end{fmffile}
}

\section{Introduction}
\label{sec:intro}
 
The spectacular  discovery of a boson  with a mass around  $\MHexp$ by
the ATLAS and CMS experiments~\cite{Aad:2012tfa,Chatrchyan:2012ufa} at
CERN  constitutes  a milestone  in  the  quest for  understanding  the
physics  of  electroweak  symmetry  breaking.   Any  model  describing
electroweak physics  needs to provide  a state that can  be identified
with  the  observed signal.   While  within  the present  experimental
uncertainties the properties of the observed state are compatible with
the        predictions       of        the       Standard        Model
(SM)~\cite{LHCP-ATLAS1,LHCP-CMS1},  many   other  interpretations  are
possible as well, in particular as  a Higgs boson of an extended Higgs
sector.

One of the prime candidates for physics beyond the SM is supersymmetry
(SUSY), which  doubles the particle  degrees of freedom  by predicting
two scalar partners for all SM fermions, as well as fermionic partners
to all bosons.  The most widely  studied SUSY framework is the Minimal
Supersymmetric                      Standard                     Model
(MSSM)~\cite{Nilles:1983ge,Haber:1984rc},  which keeps  the number  of
new fields  and couplings  to a  minimum.  In  contrast to  the single
Higgs  doublet of  the  (minimal) SM,  the Higgs  sector  of the  MSSM
contains two Higgs doublets, which  in the $\cp$ conserving case leads
to a physical spectrum consisting of two $\cp$-even, one $\cp$-odd and
two charged Higgs  bosons.  The light $\cp$-even MSSM  Higgs boson can
be    interpreted    as    the     signal    discovered    at    about
$125~\giga\electronvolt$,       see      e.g.~\cite{Heinemeyer:2011aa,
  Jakobs:2015hja}.
   
Going beyond  the MSSM, this  model has a well-motivated  extension in
the  Next-to-Minimal   Supersymmetric  Standard  Model   (NMSSM),  see
e.g.~\cite{Ellwanger:2009dp,Maniatis:2009re}  for reviews.   The NMSSM
provides  in  particular  a  solution  for  naturally  associating  an
adequate  scale  to   the  $\mu$  parameter  appearing   in  the  MSSM
superpotential~\cite{Ellis:1988er,Miller:2003ay}.   In the  NMSSM, the
introduction of  a new singlet  superfield, which only couples  to the
Higgs- and  sfermion-sectors, gives  rise to an  effective $\mu$-term,
generated  in a  similar  way as  the Yukawa  mass  terms of  fermions
through  its vacuum  expectation value.   In the  case where  $\cp$ is
conserved, which  we assume  throughout the paper,  the states  in the
NMSSM Higgs sector can be classified as three $\cp$-even Higgs bosons,
$\Hi$ ($i  = 1,2,3$), two $\cp$-odd  Higgs bosons, $A_j$ ($j  = 1,2$),
and  the charged  Higgs boson  pair  $H^\pm$.  In  addition, the  SUSY
partner  of  the  singlet  Higgs (called  the  singlino)  extends  the
neutralino sector  to a total of  five neutralinos.  In the  NMSSM the
lightest but also  the second lightest $\cp$-even  neutral Higgs boson
can be  interpreted as  the signal observed  at about  $125~\gev$, see,
e.g., \cite{King:2012is, Domingo:2015eea}.

The measured mass value of the observed signal has already reached the
level  of a  precision observable,  with an  experimental accuracy  of
better than  $300~\mev$~\cite{Aad:2015zhl}, and by itself  provides an
important test for  the predictions of models  of electroweak symmetry
breaking. In the MSSM the masses of the $\cp$-even Higgs bosons can be
predicted  at   lowest  order   in  terms   of  two   SUSY  parameters
characterising the MSSM Higgs sector,  e.g.\ $\tan\beta$, the ratio of
the vacuum expectation values of the two doublets, and the mass of the
$\cp$-odd Higgs boson, $\MA$, or  the charged Higgs boson, $\MHp$. These
relations, which in particular 
give rise to an upper  bound on the mass of the light
$\cp$-even  Higgs boson  given by  the $Z$-boson  mass, receive  large
corrections  from  higher-order  contributions.    In  the  NMSSM  the
corresponding predictions are modified both  at the tree-level and the
loop-level.   In  order   to  fully  exploit  the   precision  of  the
experimental mass value for constraining the available parameter space
of the considered  models, the theoretical predictions  should have an
accuracy that ideally is at the  same level of accuracy or even better
than the one of the  experimental value.  The theo\-retical uncertainty,
on the other hand, is composed  of two sources, the parametric and the
intrinsic uncertainty.   The theoretical uncertainties induced  by the
parametric  errors  of  the  input parameters  are  dominated  by  the
experimental error  of the top-quark  mass (where the latter  needs to
include  the systematic  uncertainty from  relating the  measured mass
parameter    to   a    theoretically   well-defined    quantity,   see
e.g.~\cite{Skands:2007zg,Hoang:2008xm,Hoang:2014oea}).   However,  the
largest  theoretical  uncertainty  at   present  arises  from  unknown
higher-order corrections, as will be discussed below.

In the MSSM%
\footnote{As mentioned  above, we focus in  this paper on the  case of
  real  parameters,  i.e.\  the $\cp$-conserving  case.}   beyond  the
one-loop    level,    the    dominant    two-loop    corrections    of
$\order{\alt\als}$~\cite{Heinemeyer:1998jw,         Heinemeyer:1998kz,
  Heinemeyer:1998np,  Zhang:1998bm, Espinosa:1999zm,  Degrassi:2001yf}
and \order{\alt^2}~\cite{Espinosa:2000df, Brignole:2001jy}  as well as
the              corresponding             corrections              of
$\order{\alb\als}$~\cite{Brignole:2002bz,    Heinemeyer:2004xw}    and
\order{\alt\alb}~\cite{Brignole:2002bz}  are known  since more  than a
decade.  (Here we use $\al_f  = Y_f^2/(4\pi)$, with $Y_f$ denoting the
fermion  Yukawa   coupling.)   These  corrections,  together   with  a
resummation of  leading and subleading logarithms  from the top/scalar
top    sector~\cite{Hahn:2013ria}   (see    also~\cite{Draper:2013oza,
  Lee:2015uza}  for  more  details  on   this  type  of  approach),  a
resummation  of leading  contributions from  the bottom/scalar  bottom
sector~\cite{Brignole:2002bz,   Heinemeyer:2004xw,   Hempfling:1993kv,
  Hall:1993gn,        Carena:1994bv,        Carena:1999py}        (see
also~\cite{Noth:2008tw, Noth:2010jy})  and momentum-dependent two-loop
contributions~\cite{Borowka:2014wla,       Borowka:2015ura}       (see
also~\cite{Degrassi:2014pfa})    are    included   in    the    public
code~\fh~\cite{Heinemeyer:1998yj,    Hahn:2009zz,   Heinemeyer:1998np,
  Degrassi:2002fi,  Frank:2006yh,   Hahn:2013ria,  feynhiggs-www}.   A
(nearly)  full two-loop  EP  calculation, including  even the  leading
three-loop         corrections,          has         also         been
published~\cite{Martin:2005eg,Martin:2007pg},  which is,  however, not
publicly available  as a computer code.   Furthermore, another leading
three-loop calculation of \order{\alt\als^2}, depending on the various
SUSY           mass           hierarchies,          has           been
performed~\cite{Harlander:2008ju,Kant:2010tf},  resulting in  the code
{\tt H3m} (which adds the three-loop corrections to the \fh\ result up
to the two-loop  level).  The theoretical uncertainty  on the lightest
$\cp$-even Higgs-boson mass within  the MSSM from unknown higher-order
contributions is  still at the level  of about $3~\gev$ for  scalar top
masses at the $\tera\electronvolt$-scale, where the actual uncertainty
depends  on  the  considered  parameter  region~\cite{Degrassi:2002fi,
  Heinemeyer:2004gx, Hahn:2013ria, Buchmueller:2013psa}.

Within   the    NMSSM   beyond   the   well    known   full   one-loop
results~\cite{Degrassi:2009yq,       Staub:2010ty,       Ender:2011qh,
  Graf:2012hh} several codes exist that  calculate the Higgs masses in
the pure \DRbar\  scheme with different contributions  at the two-loop
level.    Amongst   these  codes   \texttt{SPheno}~\cite{Porod:2003um,
  Porod:2011nf} incorporates the most complete results at the two-loop
level, including SUSY-QCD contributions  from the fermion/sfermions of
$\order{\alt  \als,  \alb\als}$,  as  well  as  pure  fermion/sfermion
contributions  of $\order{\alt^2,  \alb^2,  \alt \alb,  \alpha_\tau^2,
  \alpha_\tau \alb}$, and contributions from the Higgs/higgsino sector
in  the   gauge-less  limit  of   $\order{\alpha_\la^2,  \alpha_\ka^2,
  \alpha_\la  \alpha_\ka}$~\cite{Goodsell:2014pla}  as well  as  mixed
contributions from the latter  two sectors of $\order{\alpha_\la \alt,
  \alpha_\la  \alb}$.  The  included Higgs/higgsino  contributions are
genuine to  the NMSSM, they  are proportional to the  NMSSM parameters
$\lambda^2  =  4\pi\cdot  \alpha_\lambda$ and  $\kappa^2  =  4\pi\cdot
\alpha_\kappa$.                        The                       tools
\texttt{FlexibleSUSY}~\cite{Athron:2014yba},
\texttt{NMSSMTools}~\cite{Ellwanger:2006rn,    Domingo:2015qaa}    and
\texttt{SOFTSUSY}~\cite{Allanach:2001kg,             Allanach:2013kza,
  Allanach:2014nba}  include NMSSM  corrections of  $\order{\alt\als}$
and  $\order{\alb\als}$  supplemented  by  certain  MSSM  corrections.
\texttt{NMSSMCalc}~\cite{Ender:2011qh,   Graf:2012hh,  Baglio:2013iia,
  Muhlleitner:2014vsa} provides the option  to perform the NMSSM Higgs
mass    calculation     up    to    $\order{\alt\als}$     with    the
\DRbar\ renormalisation scheme applied  to the top-/stop-sector, while
in the  electroweak sector at  one-loop order on-shell  conditions are
used.  It has  been noticed in a comparison of  spectrum generators in
the NMSSM  that are  currently publicly  available that  the numerical
differences between the  various codes can be  very significant, often
exceeding  $3~\giga\electronvolt$ in  the  prediction  of the  SM-like
Higgs even for  the set-up where all predictions  were obtained within
the  \DRbar\ renormalisation  scheme~\cite{Staub:2015aea}.  While  the
sources  of  discrepancies  between   the  different  codes  could  be
identified~\cite{Staub:2015aea}, a reliable  estimate of the remaining
theoretical uncertainties should of course also address issues related
to  the use  of different  renormalisation schemes.   Beyond the  pure
\DRbar\       scheme,      so       far       only      the       code
\texttt{NMSSMCalc}~\cite{Ender:2011qh,   Graf:2012hh,  Baglio:2013iia,
  Muhlleitner:2014vsa}    provides   a    prediction   in    a   mixed
OS/\DRbar\ scheme,  where genuine two-loop contributions  in the NMSSM
up to $\order{\alt \als}$ have  been incorporated.  The resummation of
logarithmic contributions beyond the two-loop level is not included so
far  in any  of the  public codes  for Higgs-mass  predictions in  the
NMSSM.   Accordingly, at  present the  theoretical uncertainties  from
unknown higher-order corrections in the NMSSM are expected to be still
larger than for the MSSM.

Concerning the phenomenology of the NMSSM it is of particular interest
whether this model  can be distinguished from the  MSSM by confronting
Higgs sector  measurements with  the corresponding predictions  of the
two models. In order to facilitate the identification of genuine NMSSM
contributions in this context it is important to treat the predictions
for the  MSSM and the NMSSM  within a coherent framework  where in the
MSSM limit of  the NMSSM the state-of-the-art prediction  for the MSSM
is recovered.

With this goal in mind, we seek  to extend the public tool \fh\ to the
case of  the NMSSM. As  a first step in  this direction we  present in
this paper  a full one-loop  calculation of the Higgs-boson  masses in
the NMSSM,  where the  renormalisation scheme  and all  parameters and
conventions are chosen such that the well-known MSSM result of \fh\ is
obtained  for the  MSSM limit  of the  NMSSM. We  supplement the  full
one-loop result in the NMSSM with all higher-order corrections of MSSM
type that are implemented in \fh, as described above. In our numerical
evaluation we use our full one-loop  result in the NMSSM to assess the
quality of  the approximation that we  make at the two-loop  level. We
find  that for  a  SM-like Higgs  boson that  is  compatible with  the
detected signal  at about $125~\giga\electronvolt$  this approximation
works indeed very well. We analyse in this context which genuine NMSSM
contributions are  most relevant  when going beyond  the approximation
based on MSSM-type  higher-order corrections.  We then  apply our most
accurate prediction  including all higher-order contributions  to four
phenomenologically interesting  scenarios.  We compare  our prediction
both  with   the  result  in  the   MSSM  limit  and  with   the  code
\texttt{NMSSMCalc}~\cite{Baglio:2013iia}.  We discuss  in this context
the  impact   of  higher-order   contributions  beyond  the   ones  of
$\order{\alt \als}$, that are not implemented in \texttt{NMSSMCalc}.

The paper is organised as follows. In \refse{sec:calc} we describe our
full one-loop  calculation in  the NMSSM, specify  the renormalisation
scheme  that we  have  used  and discuss  the  contributions that  are
expected  to  be numerically  dominant  at  the one-loop  level.   The
incorporation of higher-order contributions  of MSSM-type is addressed
in \refse{sec:calcho}.   Our numerical analysis for  the prediction at
the  one-loop level,  including a  discussion  of the  quality of  the
approximation in  terms of MSSM-type  contributions, and for  our most
accurate prediction including higher-order corrections is presented in
\refse{sec:1L}.  The conclusions can be found in \refse{sec:concl}.

\section{One-loop result in the NMSSM}
\label{sec:calc}

For the sectors that are identical for the calculation within the MSSM
the  conventions  as  implemented  in  \fh\  are  used,  as  described
in~\cite{Frank:2006yh}.  Therefore  the present section  is restricted
to the quantities genuine to the NMSSM. For a more detailed discussion
of the NMSSM, see e.g.~\cite{Ellwanger:2009dp}.

\subsection{The relevant NMSSM sectors}

The  superpotential   of  the  NMSSM  for  the   third  generation  of
fermions/sfermions reads
\begin{align}
\label{eq:superpot}
  W = 
  \yt \left(\hat{H}_2\cdot \hat{Q}_3 \right)\hat{u}_3 
  - Y_d \left(\hat{H}_1\cdot \hat{Q}_3 \right)\hat{d}_3 
  - Y_\tau \left(\hat{H}_1\cdot \hat{L}_3 \right)\hat{e}_3
  + \lambda \hat{S} \left(\hat{H}_2 \cdot \hat{H}_1 \right) + \frac{1}{3} \kappa \hat{S}^3,
\end{align}
\noindent
with  the  quark  and  lepton  superfields  $\hat{Q}_3$,  $\hat{u}_3$,
$\hat{d}_3$,  $\hat{L}_3$,  $\hat{e}_3$   and  the  Higgs  superfields
$\hat{H}_1$, $\hat{H}_s$,  $\hat{S}$. The $\SUL$-invariant  product is
denoted by a  dot. The Higgs singlet and doublets  are decomposed into
\cp-even  and  \cp-odd  neutral  scalars $\phi_i$  and  $\chi_i$,  and
charged states $\phi^\pm_i$,
\begin{align}
  H_1 = \begin{pmatrix} v_1 + \frac{1}{\sqrt{2}}\left( \phi_1 - i\chi_1\right)  \\ 
    -\phi_1^- \end{pmatrix} , \quad
  H_2 = \begin{pmatrix} \phi_2^+ \\ v_2 + \frac{1}{\sqrt{2}}\left( \phi_2+i\chi_2\right) \end{pmatrix} , \quad
  S = v_s + \frac{1}{\sqrt{2}}\left( \phi_s+i\chi_s\right) ,
\end{align}
\noindent
with  the real  vacuum expectation  values  for the  doublet- and  the
singlet-fields, $v_{\{1,2\}}$  and $v_s$.  Since  $\hat{S}$ transforms
as a  singlet, the  $D$-terms remain  identical to  the ones  from the
MSSM.  Compared to  the \cp-conserving MSSM the  superpotential of the
\cp-conserving  NMSSM  contains  additional  dimensionless  parameters
$\lambda$ and $\kappa$, while the  $\mu$-term is absent.  This term is
effectively generated via the  vacuum expectation-value of the singlet
field,
\begin{align}
  \mueff = \lambda v_s .
\end{align}
\noindent
As in the MSSM it is convenient to define the ratio
\begin{align}
  \tb = \frac{v_2}{v_1}.
\end{align}
\noindent
Soft  SUSY-breaking in  the NMSSM  gives  rise to  the real  trilinear
soft-breaking parameters $A_\lambda$ and $A_\kappa$, as well as to the
soft-breaking mass term $m_S^2$ of the scalar singlet-field,
\begin{align}
  \mathcal{L}_{\text{soft}} = 
  - m_1^2 H_{1i}^\dagger H_{1i}  - m_2^2 H_{2i}^\dagger H_{2i}  - m_S^2 |S|^2  
  - \left[
    \lambda A_\lambda S \left(H_2 \cdot H_1\right) + \frac{1}{3} \kappa A_\kappa S^3.
    + \text{h.c.}
    \right]
\end{align}
\noindent
The Higgs potential $V_H$ can be written in powers of the fields,
\begin{align}
  \label{eq:HiggsPot}
  &V_{\text{H}} =
  \ldots - T_{\phi_1} \phi_1 - T_{\phi_2} \phi_2 - T_{\phi_S} \phi_s \\
  &\quad + \frac{1}{2} \begin{pmatrix} \phi_1,\phi_2, \phi_s \end{pmatrix}
  \mathbf{M}_{\phi\phi} \begin{pmatrix} \phi_1 \\ \phi_2 \\ \phi_s \end{pmatrix} +
  \frac{1}{2} \begin{pmatrix} \chi_1, \chi_2, \chi_s \end{pmatrix}
  \mathbf{M}_{\chi\chi}
  \begin{pmatrix} \chi_1 \\ \chi_2 \\ \chi_s \end{pmatrix} +
  \begin{pmatrix} \phi^-_1,\phi^-_2  \end{pmatrix}
  \mathbf{M}_{\phi^\pm\phi^\pm}
  \begin{pmatrix} \phi^+_1 \\ \phi^+_2  \end{pmatrix} + \cdots,  \nonumber
\end{align}
\noindent
where the  coefficients bilinear in  the fields are the  mass matrices
$\text{\textbf{M}}_{\phi\phi}$,   $\text{\textbf{M}}_{\chi\chi}$   and
$\text{\textbf{M}}_{\phi^\pm\phi^\pm}$.  For  the \cp-even  fields the
(symmetric) mass matrix reads
\begin{align}
  \label{eq:MaMaPhi}
  \textbf{M}_{\phi\phi}  &=
  \begin{pmatrix}
    \hat{m}_A^2 s_\beta^2 - \MZ^2 c_\beta^2
    &
    \left(\hat{m}_A^2 + \MZ^2 \right) s_\beta c_\beta
    &
    \mueff \left(2 \lambda v c_\beta - \kappa v s_\beta \right) +\frac{\lambda v}{\mueff} c_\beta
    \\  
    .
    &
    \hat{m}_A^2 c_\beta^2 - \MZ^2 s_\beta^2
    &
    \mueff \left(2 \lambda v s_\beta - \kappa v c_\beta \right) +\frac{\lambda v}{\mueff} s_\beta
    \\
    .
    &
    .
    & 
    \lambda \kappa v^2 c_\beta s_\beta + \frac{\lambda^2 v^2}{\mueff^2} \hat{m}_A^2 + 
    \frac{\kappa}{\lambda} \mueff \left(4 \frac{\kappa}{\lambda} \mueff +  A_\kappa\right)
  \end{pmatrix},
\end{align}
\noindent
where $s_\beta$ and $c_\beta$ denote the  sine and cosine of the angle
$\beta$, and
\begin{align}
  \label{eq:maHat}
  \hat{m}_A^2 = \MHp^2 - \MW^2 + \lambda^2 v^2.
\end{align}
\noindent
The lower triangle in \refeq{eq:MaMaPhi} is filled with the transposed
matrix  element.  For  the  \cp-conserving case  the  mixing into  the
eigenstates of mass  and \cp\ can be described at  lowest order by the
following unitary transformations
\begin{align}
  \label{eq:MixMatrices}
  \begin{pmatrix} h_1 \\ h_2 \\ h_3 \end{pmatrix}
  = \mathbf{U}_{e(0)}
  \begin{pmatrix} \phi_1 \\ \phi_2 \\ \phi_s \end{pmatrix}, \quad
  \begin{pmatrix}  A_1 \\ A_2 \\ G_0 \end{pmatrix}
  = \mathbf{U}_{o(0)}
  \begin{pmatrix} \chi_1 \\ \chi_2 \\ \chi_s \end{pmatrix}, \quad
  \begin{pmatrix} H^\pm \\ G^\pm  \end{pmatrix}
  = \mathbf{U}_{c(0)}
  \begin{pmatrix} \phi^\pm_1 \\ \phi^\pm_2 \end{pmatrix} .
\end{align}
\noindent
The  matrices $\mathbf{U}_{\{e,o,c\}(0)}$  transform the  Higgs fields
such that the  mass matrices are diagonalised at tree  level.  The new
fields correspond  to the five  neutral Higgs bosons $h_i$  and $A_j$,
the charged pair $H^\pm$, and the Goldstone bosons $G^0$ and $G^\pm$.

In \refeq{eq:MaMaPhi}  the third row  and column depend  explicitly on
$\mueff$.  The numerical value of  $\mueff$ has an important impact on
the  singlet admixture  after performing  the rotation  into the  mass
eigenstate basis.  For instance, for  values of $\mueff$  large enough
that
\begin{align}
  \left(\textbf{M}_{\phi\phi}\right)_{33} \gg \left(\textbf{M}_{\phi\phi}\right)_{i3},
  \quad
  i \in \left\{1,2\right\},
\end{align}
\noindent
the mass of the singlet becomes decoupled from the doublet masses.

The superpartner of the scalar  singlet appears as a fifth neutralino.
The corresponding $5\times5$ mass-matrix reads
\begin{align}
  \label{eq:MaMaNeu}
  \textbf{Y} =
  \begin{pmatrix}
    M_1  & 0  & -\MZ \sw \cos{\beta} & \MZ \sw \sin{\beta} & 0\\
    0  & M_2  & \MZ \cw \cos{\beta} & -\MZ \cw \sin{\beta}  & 0\\
    -\MZ \sw \cos{\beta} & \MZ \cw \cos{\beta} & 0 & -\mueff  & \lambda v \sin{\beta}\\
    \MZ \sw \sin{\beta} & -\MZ \cw \sin{\beta} &  -\mueff & 0 & \lambda v \cos{\beta} \\
    0 & 0 & \lambda v \sin{\beta} & \lambda v \cos{\beta} & -2\frac{\kappa}{\lambda} \mueff
  \end{pmatrix} \ .
\end{align}
\noindent
It is diagonalised by a unitary matrix
\begin{align}
  \textbf{D}_{\textbf{Y}}
  = \textbf{N}^* \textbf{Y} \textbf{N}^\dagger
  = \text{diag}{
    \left\{m_{\tilde{\chi}^0_1},m_{\tilde{\chi}^0_2},m_{\tilde{\chi}^0_3},m_{\tilde{\chi}^0_4},m_{\tilde{\chi}^0_5}\right\}} \ .
\end{align}
\noindent
Also in  \refeq{eq:MaMaNeu} $\mueff$ can have  a significant influence
on the  mixing between the  singlino and the doublet  higgsino fields.
For instance, for sufficiently large values of $\mueff$ such that
\begin{align}
  \left(\textbf{Y}\right)_{55} \gg \left(\textbf{Y}\right)_{i5},
  \quad
  i \in \left\{3,4\right\},
\end{align}
\noindent
the singlino mass decouples from the masses of higgsinos and gauginos.

\subsection{Renormalisation Scheme}
\label{sec:RenScheme}

In order to derive the counterterms entering the 1-loop corrections to
the  Higgs-boson masses  the independent  parameters appearing  in the
linear   and    bilinear   terms    of   the   Higgs    potential   in
\refeq{eq:HiggsPot} have  to be renormalised.  The  set of independent
parameters from the Higgs-sector used for the presented calculation is
formed by
\begin{alignat}{4}
  \begin{matrix}
    &\text{MSSM-like:}&
    &T_{h_{1,2}}  \ , \ \mueff \  \ , \ \MHp^2 \ , \ \tb \ , \ \MW^2 \ , \ \MZ^2\\
    &\text{genuine NMSSM:}&
    &T_{h_3} \ , \ \kappa \ , \ \lambda \ , \ A_\kappa \ , \ v = \sqrt{v_1^2+v_2^2}\ .
  \end{matrix}
\end{alignat}
\noindent
Here $T_{h_i}$ denotes the tadpole coefficient for the field $h_i$ (as
indicated  by  the  subscript)  in the  mass  eigenstate  basis.   The
relation to  the tadpoles in  the interaction basis,  $T_{\phi_i}$, is
given by
\begin{align}
  \begin{pmatrix}
    T_{h_1} \\ T_{h_2} \\ T_{h_3}
  \end{pmatrix}
  =
  \mathbf{U}_{e(0)}
  \begin{pmatrix}
    T_{\phi_1} \\ T_{\phi_2} \\ T_{\phi_s}
  \end{pmatrix}.
\end{align}
\noindent
The soft-breaking mass terms are related to the tadpole coefficients
by
\begin{subequations}
  \label{eq:TadpoleExpressionsMB}
  \begin{align}
    m_1 &= 
    -\frac{T_{\phi_1}}{\sqrt{2} v \sin{\beta}} - \mueff^2
    + \mueff B \tan{\beta} - \lambda^2 v^2 \sin^2{\beta}
    + \frac{1}{4} M_Z^2 \left(\sin^2{\beta} - \cos^2{\beta}\right)\\
    m_2 &= 
    -\frac{T_{\phi_2}}{\sqrt{2} v \cos{\beta}} - \mueff^2
    + \mueff B \cot{\beta} - \lambda^2 v^2 \cos^2{\beta}
    + \frac{1}{4} M_Z^2 \left(\cos^2{\beta} - \sin^2{\beta}\right)\\
    m_s^2 &= 
    -\frac{\lambda T_{\phi_s}}{\sqrt{2} \mueff} + 
    \left(
    \mueff B \frac{\lambda^2 v^2}{\mueff^2} +
    \lambda \kappa v^2\right)\sin{\beta} \cos{\beta}
    - \lambda^2 v^2  - 
    2 \frac{\kappa^2}{\lambda^2} \mueff^2 +  \frac{\kappa}{\lambda} \mueff A_\kappa,
  \end{align}
\end{subequations}
\noindent
where
\begin{align}
  \label{eq:mueffBMassBasis}
  \mueff B ={}&
   \frac{1}{\sqrt{2}v}\left(\SB{3} \ T_{\phi_1} + \CB{3} \ T_{\phi_2} \right)
   + \left( M_{H^\pm}^2 - M_W^2 + \lambda^2 v^2 \right) \SB{} \CB{} ,
 \end{align}
 \noindent
 and
 \begin{align}
   \mueff B = \mueff\left(\frac{\kappa}{\lambda}\mueff + A_\lambda  \right).
 \end{align}
\noindent
Using  these  equations  the original  soft-breaking  mass  parameters
$m_1$, $m_2$ and $m_s^2$ are replaced  by $T_{h_1, h_2, h_3}$, and the
soft-breaking trilinear parameter $A_\lambda$ is replaced by $\MHp$.

Parameters that  do not enter  the MSSM calculation are  considered as
genuine  of  the  calculation  in  the  NMSSM.   Although  the  vacuum
expectation value  $v$ is not  a parameter  genuine to the  NMSSM, its
appearance as  an independent parameter  is a specific feature  of the
NMSSM Higgs-mass calculation, see below.

For all  parameters appearing  only in the  NMSSM-calculation, besides
the additional tadpole coefficient, a $\overline{\text{DR}}$-scheme is
applied.   This   is  a  difference  to   the  calculations  performed
in~\cite{Ender:2011qh,           Graf:2012hh,          Baglio:2013iia,
  Muhlleitner:2014vsa}, where the electric  charge $e$ is renormalised
instead of  the parameter  $v$.  These two  parameters are  related to
each other by
\begin{align}
\label{eq:vevDef}
  v = \frac{\sqrt{2} \sw \MW}{e}
  \rightarrow
  v \left(1 + \frac{\delta{v}}{v}\right) 
  = v\left[1 + \frac{1}{2}\left(\dMWSqZ + \dSWSqZ - 2 \delta{Z_e}\right)\right],
\end{align}
\noindent
with the  renormalisation constants for the  $W$-boson mass, $\dMWSq$,
the sine of  the weak mixing angle, $\dSWSq$, where  $\sw^2 \equiv 1 -
\MW^2/\MZ^2, \sw^2 + \cw^2 =  1$, and the electric charge renormalised
as
\begin{align}
  e \rightarrow e\left(1 + \delta{Z_e}\right).
\end{align}
\noindent
Considering $\dMWSq$ and $\dSWSq$ already fixed by on-shell conditions
for the  gauge-boson masses~\cite{Frank:2006yh},  either $\delta{Z_e}$
or $\delta{v}$  in \refeq{eq:vevDef}  can be  fixed by  an independent
renormalisation  condition  (and  the  other  counterterm  is  then  a
dependent           quantity).           The           renormalisation
prescription~\cite{Ender:2011qh}  where  $\delta{Z_e}$   is  fixed  by
renormalising    $e$   in    the   static    limit   results    in   a
non-$\overline{\text{DR}}$ renormalisation  for $\delta{v}$.   For the
self-energies in the Higgs  sector $\delta{v}$ enters the counterterms
for the renormalised Higgs potential,
\begin{align}
  V_{\rm H} \rightarrow V_{\rm H} + \delta{V_{\rm H}},
\end{align}
\noindent
with coefficients involving $\lambda$ and $\kappa$, like
\begin{align}
  \label{eq:renConsdV}
  \left. \frac{\delta^{(2)}}{\delta{\phi_s}\delta{\phi_i}} \delta{V_{\rm H}}
  \right|_{\phi_l, \chi_m, \phi_n^\pm = 0}
  \supset
  - \kappa \mueff \left\{\sin{\beta},\cos{\beta}\right\} 
  \left(\delta{v} + \ldots\right), 
\end{align}
\noindent
for the self-energies with each an external doublet and singlet field.
The  ellipsis  in  \refeq{eq:renConsdV} denote  other  renormalisation
constants that are fixed in the $\overline{\text{DR}}$-scheme and thus
do not contribute  with a finite part. However,  a finite contribution
from  $\delta{v}$ would  lead  to a  $\kappa$-dependence  of all  loop
contributions  entering via  $\delta  v$, in  particular  also of  the
corrections from  the fermions  and sfermions  (while the  fermion and
sfermion   contributions  to   the   unrenormalised  self-energy   are
$\kappa$-independent).  A  finite contribution from  $\delta{v}$ would
furthermore  imply the  rather artificial  feature that  a self-energy
involving an external gauge singlet  field would receive a counterterm
contribution involving the  renormalisation constant $\delta{Z_e}$ for
the   electric    charge.    We   therefore   prefer    to   use   the
$\overline{\text{DR}}$-scheme  for the  renormalisation of  $v$, which
means  that  we  use  a  scheme where  $\delta{Z_e}$  is  a  dependent
counterterm.  This leads to the relation
\begin{align}
  \label{eq:dZEdep}
  \delta{Z_e^{\text{dep}}} &= 
  \frac{1}{2}\left[\dSWSqZ + \dMWSqZ - \dvSqZ \right] \ ,
\end{align}
\noindent
which implies
\begin{align}
  \label{eq:dZEII}
  \left. \delta{Z_e^{\text{dep}}} \right|^{\text{fin}} &= 
  \frac{1}{2}\left[\dSWSqZ + \dMWSqZ  \right]^{\text{fin}}
\end{align}
\noindent
for the finite part of $\delta{Z_e^{\text{dep}}}$.  In this scheme the
numerical value for  the electric charge $e$ (and  accordingly for the
electromagnetic coupling  constant $\alpha$) is  determined indirectly
via  \refeq{eq:dZEII}.  In  order  to avoid  a non-standard  numerical
value  for $\alpha$  in our  numerical  results, we  apply a  two-step
procedure:  in  the  first  step  we  apply  a  $\overline{\text{DR}}$
renormalisation for $v$  as outlined above.  As a  second step we then
reparametrise this result in terms of a suitably chosen expression for
$\alpha$.   By default  we use  the same  convention as  for  the MSSM
result   that  is  implemented   in  \texttt{FeynHiggs},   namely  the
expression  for the  electric charge  in terms  of the  Fermi constant
$G_F$,   in   order  to   facilitate   the   comparison  between   the
\texttt{FeynHiggs}  result in  the  MSSM  and our  new  result in  the
NMSSM. Taking the MSSM limit of  our new NMSSM result, the MSSM result
as  implemented  in  FeynHiggs   is  recovered,  since  the  described
calculational  differences are  genuine NMSSM  effects that  vanish in
this limit.  For the  numerical comparison with  \texttt{NMSSMCalc} we
will    use   instead   $\alpha(M_Z)$.     The   procedure    of   the
reparametrisation is outlined in the following section.

\subsection{Reparametrisation of the electromagnetic coupling}
\label{sec:reparametrisation}

The  couplings $g^I$  and  $g^{II}$ in  two different  renormalisation
schemes are in general related to each other by
\begin{align}
  \label{eq:bareRenG}
  g^{I} \left(1 + \delta{Z^{I}_g}\right) = g^{II} \left(1 + \delta{Z^{II}_g}
  \right) \ ,
\end{align}
  \noindent
because of the equality of the bare couplings. The corresponding shift
in the numerical  values of the coupling definitions  is obtained from
the finite difference  of the two counterterms,  $\Delta \equiv g^{II}
\delta{Z^{II}_g}   -    g^{I}   \delta{Z^{I}_g}$.     Accordingly,   a
reparametrisation from  the numerical  value of  the coupling  used in
scheme I to the one of scheme II can be performed via
\begin{align}
  \label{eq:reparam}
  g^{I} = g^{II} + \Delta \ .
\end{align}
\noindent
Since $\Delta$ is  of one-loop order, its insertion  into a tree-level
expression generates a term of one-loop order, etc.

In  our  calculation  the  reparametrisation  of  the  electromagnetic
coupling  is  only necessary  up  to  the  one-loop level,  since  all
corrections  of  two-loop  and  higher  order that  we  are  going  to
incorporate have been  obtained in the gauge-less limit  (some care is
necessary regarding  the incorporation of the  MSSM-type contributions
of  \order{\alt^2},   see~\cite{Brignole:2001jy,Hollik:2014wea}).   At
this order  the shift $\Delta$  can simply  be expressed as  $\Delta =
g^{II}      \left(\delta{Z^{II}_g}     -      \delta{Z^{I}_g}\right)$.
Specifically,  for   the  reparametrisation  of   the  electromagnetic
coupling 
constant $G_F$ the parameter shift $\Delta_{G_F}$ reads
\begin{align}
  \label{eq:reparamGF}
  \Delta_{G_F} = e \left(\delta Z_e - \delta Z_e^{\rm dep} - \frac{1}{2}
  \Delta r^{\rm NMSSM} \right) \ . 
\end{align}
\noindent
Here $\delta  Z_e$ is  the counterterm  of the  charge renormalisation
within the NMSSM according to the static (Thomson) limit,
\begin{align}
  \label{eq:ZeThomson}
  \delta Z_e = \frac{1}{2} \Pi^{\gamma\gamma}(0) + \frac{\sw}{\cw}
  \frac{\Sigma_T^{\gamma Z}(0)}{\MZ^2} \ ,
\end{align}
\noindent
and   $\Pi^{\gamma\gamma}(0)$,   $\Sigma_T^{\gamma  Z}(0)$   are   the
derivative of  the transverse part  of the photon self-energy  and the
transverse  part  of  the  photon--$Z$ self-energy  at  zero  momentum
transfer,  respectively. The  counterterm $\delta  Z_e^{\rm dep}$  has
been defined in \refeq{eq:dZEdep}, and for the quantity $\Delta r^{\rm
  NMSSM}$  we   use  the  result  of   \cite{Stal:2015zca}  (see  also
\cite{Domingo:2011uf}).%
\footnote{For      the      sample       scenario      defined      in
  tab.~\ref{tab:SampleScenario} below  the numerical value  of $\Delta
  r^{\rm NMSSM}$  from \cite{Stal:2015zca}  turns out  to be  close to
  $3.8 \%$, with only a weak  dependence on $\lambda$ for the range of
  $\lambda$ values discussed in this  paper.}  The numerical value for
the electromagnetic  coupling $e$ in this  parametrisation is obtained
from the Fermi constant  in the usual way as $e =  2 \MW \sw \sqrt{G_F
  \sqrt{2}}$.

Similarly, for  the reparametrisation of the  electromagnetic coupling
defined  in  the  previous  section  in  terms  of  $\alpha(\MZ)$  the
parameter shift $\Delta_{\alpha(\MZ)}$ reads
\begin{align}
  \label{eq:reparamalphaMZ}
  \Delta_{\alpha(\MZ)} = e \left(\delta Z_e - \delta Z_e^{\rm dep} - \frac{1}{2}
  \Delta \alpha \right) \ . 
\end{align}
\noindent
The numerical  value of $e$  in this parametrisation is  obtained from
$\alpha(\MZ) = \alpha(0)/(1 - \Delta  \alpha)$, and $\alpha(0)$ is the
value of the fine-structure constant in the Thomson limit.

\subsection{Dominant Contributions at One-Loop Order}
\label{sec:leadingContributions}

As explained above,  we will supplement our full  one-loop result with
all available higher-order contributions of  MSSM type.  This means in
particular  that the  two-loop contributions  are approximated  by the
two-loop  corrections  in  the  MSSM  (i.e.\  omitting  genuine  NMSSM
corrections) as  included in \fh,  and further corrections  beyond the
two-loop level are included.  In  order to validate this approximation
we analyse at the one-loop level the size of genuine NMSSM corrections
w.r.t.\ the MSSM-like contributions.

\begin{table}
  \centering
  \begin{tabular}{c|cccc}
    \toprule
    (i,j) & $(1|2,1|2)$ & $(1,s)$ & $(2,s)$ & $(s,s)$\\\midrule
    order &
    $\mathcal{O}{\left(\yt^2\right)}$ & $\mathcal{O}{\left(\lambda \yt\right)}$
    & $\mathcal{O}{\left(\lambda \yt\right)}$  & $\mathcal{O}{\left(\lambda^2\right)}$ \\
    fields & top/stop & stop & stop & stop \\
    \multirow{6}{*}{topologies} 
    & $\fmfvcenter{ScalarSESq}$ & $\fmfvcenter{ScalarSESq}$ & $\fmfvcenter{ScalarSESq}$ & $\fmfvcenter{ScalarSESq}$\\
    & $\fmfvcenter{ScalarSEA0Sq}$ & $\fmfvcenter{ScalarSEA0Sq}$ & & \\
    & $\fmfvcenter{ScalarSEQ}$ &  & & \\\bottomrule
  \end{tabular}
  \caption{Topologies and their order in terms of the couplings in the
    top/stop  sector  that  contribute  to the  self-energies  of  the
    \cp-even  fields  $\phi_i$ at  one-loop  order  in the  gauge-less
    limit. The numbers  1 and 2 denote the  doublet-states as external
    fields, while $s$ denotes an  external singlet. The internal lines
    depict either a top (solid) or a scalar top (dashed).}
  \label{tab:topologies1L}
\end{table}

Since the corrections from the top/stop  sector are usually the by far
dominant  ones,  we  start  with a  qualitative  discussion  of  those
contributions before we perform a  numerical analysis in the following
section. In the MSSM the  leading corrections from the top/stop sector
are commonly  denoted as $\order{\alt}$, indicating  the occurrence of
two Yukawa couplings $\yt$. In the limit where all other masses of the
SM particles and  the external momentum are neglected  compared to the
top-quark mass, for dimensional reasons  the correction to the squared
Higgs-boson mass  furthermore receives a contribution  proportional to
$\mt^2$.  This gives rise to the well-known coefficient $G_F \mt^4$ of
the leading one-loop contributions. In  the NMSSM the formally leading
contributions  either are  of  $\order{\yt^2}$  (involving two  Yukawa
couplings), of  $\order{\lambda\yt}$ (involving one  Yukawa coupling),
or of $\order{\lambda^2}$ (involving no Yukawa coupling).  The various
contributions   from   the   top/stop   sector   are   summarised   in
\refta{tab:topologies1L}.  The contributions in  the second column are
the ones  of MSSM-type, while the  entries in the third  through fifth
column represent the genuine  NMSSM corrections, involving only scalar
tops.\footnote{We    discuss    here     only    the    Higgs    boson
  self-energies. However,  the same line  of argument can be  made for
  the tadpole contributions.}

For  the  doublet  fields,  the   couplings  between  the  Higgs-  and
stop-fields in the gauge-less limit  are proportional to the top-quark
Yukawa-coupling,
\begin{subequations}
  \label{eq:HSfCoupling}
  \begin{align}
    \label{eq:HSfCouplingdoub}
    \mathi \Gamma_{\phi_2 \tilde{t}_i \tilde{t}_j} = 
    \mathi \sqrt{2} \yt \left[ \At \cdot c^{\phi_2}_{ij}{(\theta_{\tilde{t}})}
      +
      m_t \cdot \left(-1\right)^{1-i} \delta_{ij} \right]
    , \quad
    \mathi \Gamma_{\phi_1 \tilde{t}_i \tilde{t}_j} = 
    \mathi \sqrt{2} \yt \mueff \cdot c^{\phi_1}_{ij}{(\theta_{\tilde{t}})},
  \end{align}
  \noindent
  while the corresponding coupling for the singlet field reads
  \begin{align}
    \mathi \Gamma_{\phi_s \tilde{t}_i \tilde{t}_j} = 
    \mathi \sqrt{2} \lambda \cot{\beta}\ m_t \cdot c^{\phi_s}_{ij}{(\theta_{\tilde{t}})}.
    \label{eq:HSfCouplingsing}
  \end{align}
\end{subequations}
\noindent
The non-vanishing quartic Higgs--stop couplings read
\begin{align}
  \mathi \Gamma_{\phi_2 \phi_2 \tilde{t}_i \tilde{t}_j} = 
  -\mathi \yt^2 \cdot \delta_{ij}
  ,\quad
  \mathi \Gamma_{\phi_1 \phi_s \tilde{t}_i \tilde{t}_j} = 
  -\mathi \lambda \yt \cdot c^{\phi_1 \phi_s}_{ij}{(\theta_{\tilde{t}})}.
\end{align}
\noindent
Here  functions of  the mixing  angle  of the  stop-sector, \tst,  are
denoted by $c$  with the appropriate indices and  superscripts for the
involved fields.   These functions $c$  can never be larger  than $1$.
In the singlet--stop coupling we  have explicitly spelled out a factor
$\lambda  v_1  \yt  =  \lambda   m_t  \cot{\beta}$  to  highlight  the
appearance of  the factor $m_t$ in  \refeq{eq:HSfCouplingsing} instead
of the usual factor $m_t/M_W \sim \yt$ in \refeq{eq:HSfCouplingdoub}.

The genuine NMSSM  couplings of a singlet to stops  are seen to follow
the pattern mentioned above, i.e.\  they give rise to contributions of
\order{\la\yt} (third  and fourth column  in \refta{tab:topologies1L})
or \order{\la^2}  (fifth column), whereas the  MSSM-like contributions
are of \order{\yt^2} (second column).  Those different patterns do not
only  indicate a  distinction between  the MSSM-like  and the  genuine
NMSSM contributions,  but also  give rise  to a  significant numerical
suppression%
\footnote{For the  trilinear couplings  in eq.~\eqref{eq:HSfCoupling},
  comparing  the  Higgs  singlet   with  the  doublet,  an  additional
  potential    suppression    factor   of    \order{\mt/\At}    and/or
  \order{\mt/\mueff}  appears.}  of  the  genuine NMSSM  contributions
w.r.t.\  the  MSSM-like ones  for  $\lambda  <  \yt$. If  one  demands
validity of perturbation theory up to  the GUT scale, this relation is
always fulfilled,  since then  $\lambda$ and  $\kappa$ are  bound from
above~\cite{Miller:2003ay} by
\begin{align}
  \lambda^2 + \kappa^2 \lesssim 0.5,
\end{align}
\noindent 
so  that $\lambda  \lesssim 0.7$,  where the  largest values  are only
allowed  for  vanishing  $\kappa$.   The size  of  the  genuine  NMSSM
contributions will be discussed numerically in the following sections.

\section{Incorporation of higher-order contributions}
\label{sec:calcho}

The masses of the \cp-even Higgs  bosons are obtained from the complex
poles of  the full propagator  matrix.  The inverse  propagator matrix
for    the     three    \cp-even     Higgs    bosons     $h_i$    from
eq.~\eqref{eq:MixMatrices} is a $3 \times 3$ matrix that reads
\begin{align}
  \Delta^{-1}{\left(k^2\right)} = \mathi
  \left[k^2\mathbbm{1} - 
    \mathcal{M}_{hh}
    + \hat{\Sigma}_{hh}{\left(k^2\right)}
    \right].
\end{align}
\noindent
Here $\mathcal{M}_{hh}$  denotes the  diagonalised mass matrix  of the
\cp-even Higgs  fields at tree level,  and $\hat{\Sigma}_{hh}$ denotes
their  renormalised  self-energy  corrections\footnote{Details on  the
  calculation  of the renormalised  self-energy contributions  will be
  presented in a future publication.}.  The three complex poles of the
propagator in the \cp-even Higgs sector are given by the values of the
external  momentum  $k$  for  which  the determinant  of  the  inverse
propagator-matrix vanishes,
\begin{align}
  \det\left[
    \Delta^{-1}{\left(k^2\right)}
    \right]_{k^2 = m_{h_i}^2 - \mathi \Gamma_{h_i} m_{h_i}} \overset{!}{=} 0
  , \ i \in \{1,2,3\}.
\end{align}
\noindent
The real parts  of the three poles are identified  with the the square
of the  Higgs-boson masses in  the \cp-even sector.   The renormalised
self-energy  matrix $\hat{\Sigma}_{hh}$  is evaluated  by taking  into
account the full  contributions from the NMSSM at  one-loop order and,
as an approximation, the MSSM-like  contributions at two-loop order of
$\mathcal{O}{\left(\alpha_t      \alpha_s,     \alpha_b      \alpha_s,
  \alpha_t^2,\alpha_t\alpha_b\right)}$ at  vanishing external momentum
taken     over    from     \FH~\cite{Heinemeyer:1998yj,
  Hahn:2009zz,   Heinemeyer:1998np,   Degrassi:2002fi,   Frank:2006yh,
  Hahn:2013ria, feynhiggs-www}, where also  the resummation of leading
and  subleading   logarithms  from   the  top/scalar  top   sector  is
incorporated~\cite{Hahn:2013ria},
\footnote{In the public version of \FH\  for the NMSSM also the recent
  results for momentum-dependent two-loop contributions in the MSSM of
  \cite{Borowka:2014wla, Borowka:2015ura} will be implemented.}
\begin{align}
  \label{eq:SEapprox}
  \hat{\Sigma}_{hh}{\left(k^2\right)}
  \approx
  \left.
  \hat{\Sigma}^{(\text{1L})}_{hh}{\left(k^2\right)}
  \right|^{\text{NMSSM}} +
  \left.
  \hat{\Sigma}^{(\text{2L + beyond})}_{hh}{\left(k^2\right)}
  \right|_{k^2=0}^{\text{MSSM}}.
\end{align}
\noindent
In order  to facilitate  the incorporation  of the  MSSM-like two-loop
contributions  from \FH,  the  renormalisation scheme  chosen for  the
NMSSM contributions closely follows  the \FH\ conventions as described
in \cite{Frank:2006yh}.  Accordingly, the stop masses are renormalised
on-shell.   For our  numerical  evaluation below  we  employ the  MSSM
contributions    obtained   from    the   version    \texttt{FeynHiggs
  2.10.2}\footnote{More  recent  updates  of \FH\  contain  additional
  contributions that  however do not significantly  modify the results
  of our present investigation.}  The  poles of the inverse propagator
matrix are  determined numerically.  The algorithm  for this procedure
is  the same  as the  one described  in~\cite{Ender:2011qh}.  For  the
generation and  calculation of the  self energies the tools  \FA\ {\tt
  3.9}~\cite{Kublbeck:1990xc,     Hahn:2000kx}    and     \FC\    {\tt
  7.4}~\cite{Hahn:1998yk,   Agrawal:2012cv}  have   been  used.    The
implementation  of the  NMSSM  with  real parameters  was  based on  a
\FA\   model  file   generated  by   {\tt  SARAH}~\cite{Staub:2013tta,
  Staub:2012pb, Staub:2010jh, Staub:2009bi}.

\section{Numerical Results}
\label{sec:1L}

A particular  goal of our  numerical analysis is  to test the  kind of
approximation in terms of MSSM-type contributions that we have used at
the two-loop level.  For this purpose a genuine NMSSM scenario will be
studied, which gives rise to a  SM-like Higgs with a predicted mass at
the   two-loop  level   of  around   $125~\giga\electronvolt$   and  a
singlet-like Higgs  field with a mass  that can be above  or below the
one of  the SM-like state.  In  order to investigate  the influence of
the extended  Higgs and higgsino sector  of the NMSSM  compared to the
MSSM the  parameter $\lambda$ will  be varied.  In the  limit $\lambda
\rightarrow 0$ and constant  $\mueff$ all singlet fields decouple from
the  remaining  field spectrum.   Increasing  the  value of  $\lambda$
directly   translates   to  increasing   the   influence  of   genuine
NMSSM-effects.   A  detailed study  of  the  one-loop  result and  the
quality  of  approximations based  on  partial  contributions will  be
presented here.  In order to study the approximation of restricting to
MSSM-like      contributions     beyond     one-loop      order     at
$\mathcal{O}{(\alpha_t  \alpha_s)}$, we will  compare our  result with
the   public   tool  \texttt{NMSSMCalc}~\cite{Baglio:2013iia},   which
incorporates     the    genuine     NMSSM-type     contributions    of
$\mathcal{O}{\left(\alpha_t    \alpha_s\right)}$   using    a   hybrid
$\DRbar$/on-shell renormalisation  scheme. While for  the MSSM various
other   higher-order   corrections  are   implemented   in  \FH,   the
corresponding  contributions  have  not  been taken  into  account  in
\texttt{NMSSMCalc}.   We will  compare in  this context  the numerical
effect of the  NMSSM-type contributions of $\mathcal{O}{\left(\alpha_t
  \alpha_s\right)}$  as  implemented  in \texttt{NMSSMCalc}  with  the
MSSM-type  contributions of this  order, and  we will  investigate the
numerical    impact    of    the    MSSM-type    corrections    beyond
$\mathcal{O}{\left(\alpha_t \alpha_s\right)}$.

In our numerical discussion below we  will just focus on the masses of
the two lighter \cp-even states.  The effects discussed below turn out
to be  very small for the  heaviest \cp-even state, amounting  to less
than 1\permil\ for the considered scenarios.

\subsection{Numerical Scenarios and Treatment of Input Parameters}
\label{subsec:numScenario}

In  our study  we will  discuss four  different scenarios.   The first
``sample scenario'',  S, for  our study is  defined by  the parameters
given  in  tab.~\ref{tab:SampleScenario}.   It   has  been  chosen  to
exemplify typical features  of NMSSM phenomenology and  is well suited
for  studying  the  magnitude   of  the  NMSSM-contributions  and  the
behaviour  of  the  employed  approximation.   The  second  and  third
scenario   are   the   benchmark   scenarios   P1   and   P9   defined
in~\cite{Domingo:2016unq},  where the  parameter $\lambda$  is varied.
While the  original motivation  for these  scenarios arising  from the
diphoton  excess  that  was   observed  by  ATLAS~\cite{ATLAS750}  and
CMS~\cite{CMS:2015dxe} in the 2015 Run~2 data has not received support
from the latest data, we use those scenarios here to serve as examples
of possible  NMSSM phenomenology in order  to test to what  extent the
features  visible  for  the  ``sample   scenario''  S  also  apply  to
completely different  scenarios.  The fourth  scenario A1 is  based on
P1, but  permits much  larger values of  $\lambda$.  The  Higgs sector
parameters  of  P1,  P9  and  A1  are  given  in  tab.~\ref{tab:P1P9}.
Throughout  our analysis  the  parameter $\lambda$  is  varied if  not
stated otherwise.  We will show in our numerical discussion below that
the  qualitative features  of  the  scenarios P1,  P9  and  A1 can  be
understood from the discussion of the ``sample scenario''.

The choice  for the top-quark mass  in the loop contributions  will be
the   pole   mass   $m_t^{\text{OS}}$    for   the   comparison   with
\texttt{NMSSMCalc}   and  $m^{\overline{\text{MS}}}_t(m_t)$   for  the
remaining  studies. Using  the  $\overline{\text{MS}}$ top-quark  mass
allows us  to include the  resummation of leading  and next-to-leading
logarithms  implemented  in \texttt{FeynHiggs}.   The  renormalisation
scale for the studies in this chapter  will be fixed at the used value
of the top-quark mass.
\begin{table}[htb]
  \centering
  \begin{tabular}{cc}
    Higgs sector parameters: & heavy fermion masses:\\ 
    \begin{tabular}{ccccc}
      \\\toprule
      $\MHp$ & $\tan{\beta}$ & $\mueff$ & $A_\kappa$ & $\kappa$\\\midrule
      $1000$ & $8$ & $125$ & $-300$ & $0.2$\\
      \bottomrule\\\\
    \end{tabular}
    &
    \begin{tabular}{cccc}
      \\\toprule
      $m_t^{\text{OS}}$ & $m_t^{\overline{\text{MS}}}{\left(m_t\right)}$ & $m_b^{\overline{\text{MS}}}{\left(m_b\right)}$ & $m_\tau$\\\midrule
      $173.2$ & $167.48$ & $4.2$ & $1.78$\\
      \bottomrule\\\\
    \end{tabular}
  \end{tabular}\\
  sfermion- and gaugino-parameters:
  \\
  \begin{tabular}{cccccccc}
    \\\toprule
    $M_{\tilde{q}}$ & $M_{\tilde{l}}$ & $A_t$ & $A_{\tau}$, $A_{b}$, $A_{q}$ 
    & $A_{l}$ & $M^{(\text{GUT})}_1$ & $M_2$ & $M_3$\\\midrule
    $1500$  & $200$ & $-2000$ & $-1500$ & $-100$ & $\approx 143$  & $300$ & $1500$\\
    \bottomrule\\\\
  \end{tabular}
  \begin{tabular}{ll}
    \toprule
    $M_{\tilde{q}}$ & universal squark mass breaking parameter\\
    $M_{\tilde{l}}$ & universal slepton mass breaking parameter\\
    $A_{t/b/q}$ & trilinear breaking term for stop-/sbottom/the lighter squark-generations\\
    $A_{\tau/l}$ & trilinear breaking term for stau/the two lighter slepton-generations\\
    $M_{\{1,2,3\}}$ & Gaugino mass breaking parameters for \UY, \SUL, \SUc .\\\bottomrule\\
  \end{tabular}
  \caption{Definition  of the  sample scenario,  S.  All  dimensionful
    parameters    are    given     in    $\giga\electronvolt$.     All
    $\overline{\text{DR}}$-parameters       are       defined       at
    $m_t^{\overline{\text{MS}}}{\left(m_t\right)}$.                All
    stop-parameters  are on-shell  parameters.   As  indicated by  the
    superscript ``(GUT)'', $M_1$ is related  to $M_2$ by the usual GUT
    relation,  $M_1^{(\text{GUT})} =  5  s_{\rm  w}^2/(3 c_{\rm  w}^2)
    M_2$.}
  \label{tab:SampleScenario}
\end{table}
\begin{table}[htb]
  \centering
  \begin{tabular}{cc}
    Higgs sector parameters: & sfermion- and gaugino-parameters:\\
    \begin{tabular}{cccccc}
      \\\toprule
      & $\hat{m}_A$ & $\tan{\beta}$ & $\mueff$ & $A_\kappa$ & $\kappa$\\\midrule
      P1 & \ $760$ & $10$ & $150$ & $0$ & $0.25$\\
      P9 & \ $765$ & $14$ & $110$ & $0$ & $0.17$\\
      \bottomrule
    \end{tabular}
    &
    \begin{tabular}{cccccc}
      \\\toprule
      $M_{\tilde{q}}$ & $M_{\tilde{l}}$ & $A_t$ & $M_1$ & $M_2$ & $M_3$\\\midrule
      $1750$ & $300$ & $-4000$ & $500$ & $1000$ & $3000$\\
      $2050$ & $400$ & $-4000$ & $500$ & $1000$ & $3000$\\
      \bottomrule
    \end{tabular}
    \\
    \begin{tabular}{cccccc}
      \\\toprule
      & $\MHp$ & $\tan{\beta}$ & $\mueff$ & $A_\kappa$ & $\kappa$\\\midrule
      A1 & $1500$ & $10$ & $150$ & $0$ & $0.25$\\
      \bottomrule\\
    \end{tabular}
    &
    \begin{tabular}{cccccc}
      \\\toprule
      $M_{\tilde{q}}$ & $M_{\tilde{l}}$ & $A_t$ & $M_1$ & $M_2$ & $M_3$\\\midrule
      $1750$ & $300$ & $-4000$ & $500$ & $1000$ & $3000$\\
      \bottomrule\\
    \end{tabular}
  \end{tabular}
  \caption{Definition of  the analysed scenarios  P1, P9 and  A1.  All
    dimensionful  parameters are  given in  $\giga\electronvolt$.  All
    $\overline{\text{DR}}$-parameters are considered  to be defined at
    $m_t^{\overline{\text{MS}}}{\left(m_t\right)}$,       and      all
    stop-parameters  are considered  to be  on-shell parameters.   The
    remaining  trilinear  breaking parameters  are  chosen  as $A_f  =
    1500~\giga\electronvolt$. The parameter  $\hat{m}_A$ is related to
    the charged Higgs $M_{H^\pm}$ mass by eq.~\eqref{eq:maHat}.}
  \label{tab:P1P9}
\end{table}

\subsubsection{Sample Scenario S}

The sample scenario S for our study is defined by the parameters given
in tab.~\ref{tab:SampleScenario}.   For values $\lambda  \gtrsim 0.32$
the mass  of the  lightest state becomes  tachyonic at  tree-level for
this scenario, and  therefore the analyses will be  performed only for
values of $\lambda$ up to $0.32$.

The viability of the discussed scenario is tested with the full set of
experimental   data  implemented   in  the   tool  \texttt{HiggsBounds
  4.1.3}~\cite{Bechtle:2008jh,     Bechtle:2011sb,    Bechtle:2013wla,
  Bechtle:2013gu, Bechtle:2015pma}.  In order  to obtain the necessary
input  for  \texttt{HiggsBounds}  we made  use  of  \texttt{NMSSMTools
  4.4.0}~\cite{Ellwanger:2009dp}      and      linked     it      with
\texttt{HiggsBounds}.   While  our  calculation  assumes  an  on-shell
renormalised  stop-sector as  in~\cite{Frank:2006yh},  the SLHA  input
file   for  \texttt{NMSSMTools}   needs   \DRbar-parameters  for   the
stop-sector.    Thus  a   conversion  from   the  on-shell   into  the
\DRbar\ scheme is necessary for  the parameters of the sample scenario
given  in tab.~\ref{tab:SampleScenario}.   We only  accounted for  the
dominant effect  of these  conversions that  occurs for  $X_t =  A_t -
\mueff  \cot{\beta}$ by  applying the  on-shell to  \DRbar\ conversion
outlined in~\cite{Williams:2011bu}.   We find that the  scenario is in
agreement    with    the    experimental   limits    implemented    in
\texttt{HiggsBounds 4.1.3}.

\subsubsection{Scenarios  with \boldmath{$A_\kappa = 0$} and very
    large \boldmath{$|A_t|$}}

The  scenarios P1  and  P9  are defined  by  the  parameters given  in
tab.~\ref{tab:P1P9}.   They are  characterised  in  particular by  the
choice of $A_\kappa = 0$ and very large (negative) $A_t$. While in the
original definition of~\cite{Domingo:2016unq} the values $\lambda=0.1$
and  $\lambda=0.05$  were   chosen  for  the  scenarios   P1  and  P9,
respectively, we  vary the parameter $\lambda$  here.  We nevertheless
refer to the scenarios as P1 and P9 also for other values of $\lambda$
for simplicity.

In  the scenario  P1  for all  values of  $\lambda  \gtrsim 0.43$  the
lightest Higgs  state becomes tachyonic,  for scenario P9 this  is the
case  for $\lambda  \gtrsim  0.35$.  The  analyses  will therefore  be
restricted to  values of $\lambda  \lesssim 0.43$ for the  scenario P1
and $\lambda  \lesssim 0.35$  for the  scenario P9,  respectively.  The
parameters  entering   at  higher  order   are  chosen  as   given  in
tab.~\ref{tab:P1P9} in the same fashion as above.

\subsubsection{Example of a scenario with large values of $\lambda$}

The scenario A1 is based on  P1, but with a substantially larger value
of  $\MHp$, which  prevents tachyonic  Higgs-masses at  the tree-level
even for large  values of $\lambda$.  The parameters are  given in the
lower part of  \refta{tab:P1P9}. Although we found  that this scenario
is in  disagreement with experimental  data from the Tevatron  and the
LHC Run~1 for  $\lambda \gtrsim 0.75$, it permits the  analysis of the
MSSM-approximation also for very large values of $\lambda$.

\subsection{Full Results at Two-Loop Order}
\label{subsec:2loopOverall}

\begin{figure}[htbp]
  \centering
  \includegraphics[width=.49\textwidth]{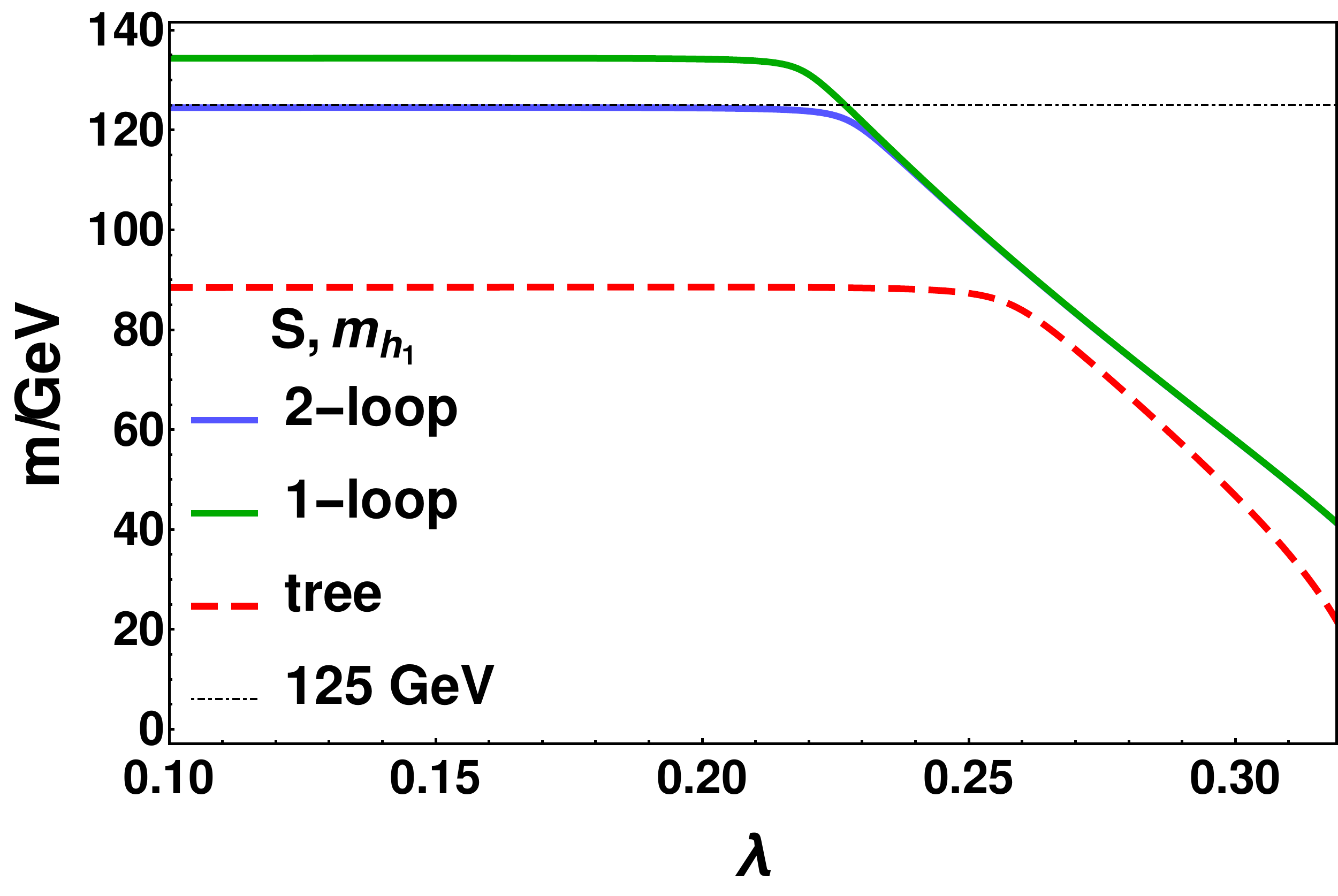}
  \includegraphics[width=.49\textwidth]{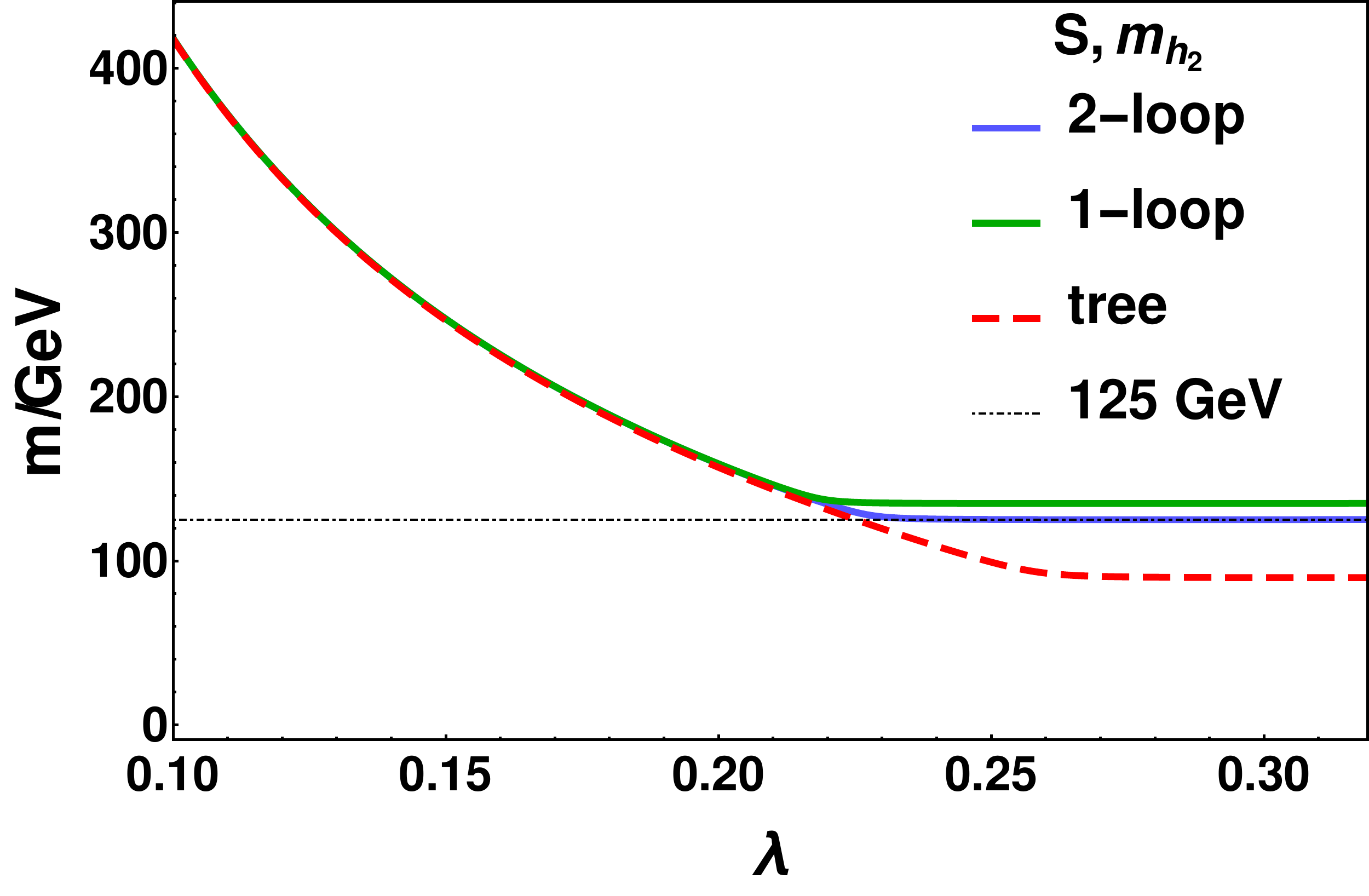}
  \\
  \includegraphics[width=.49\textwidth]{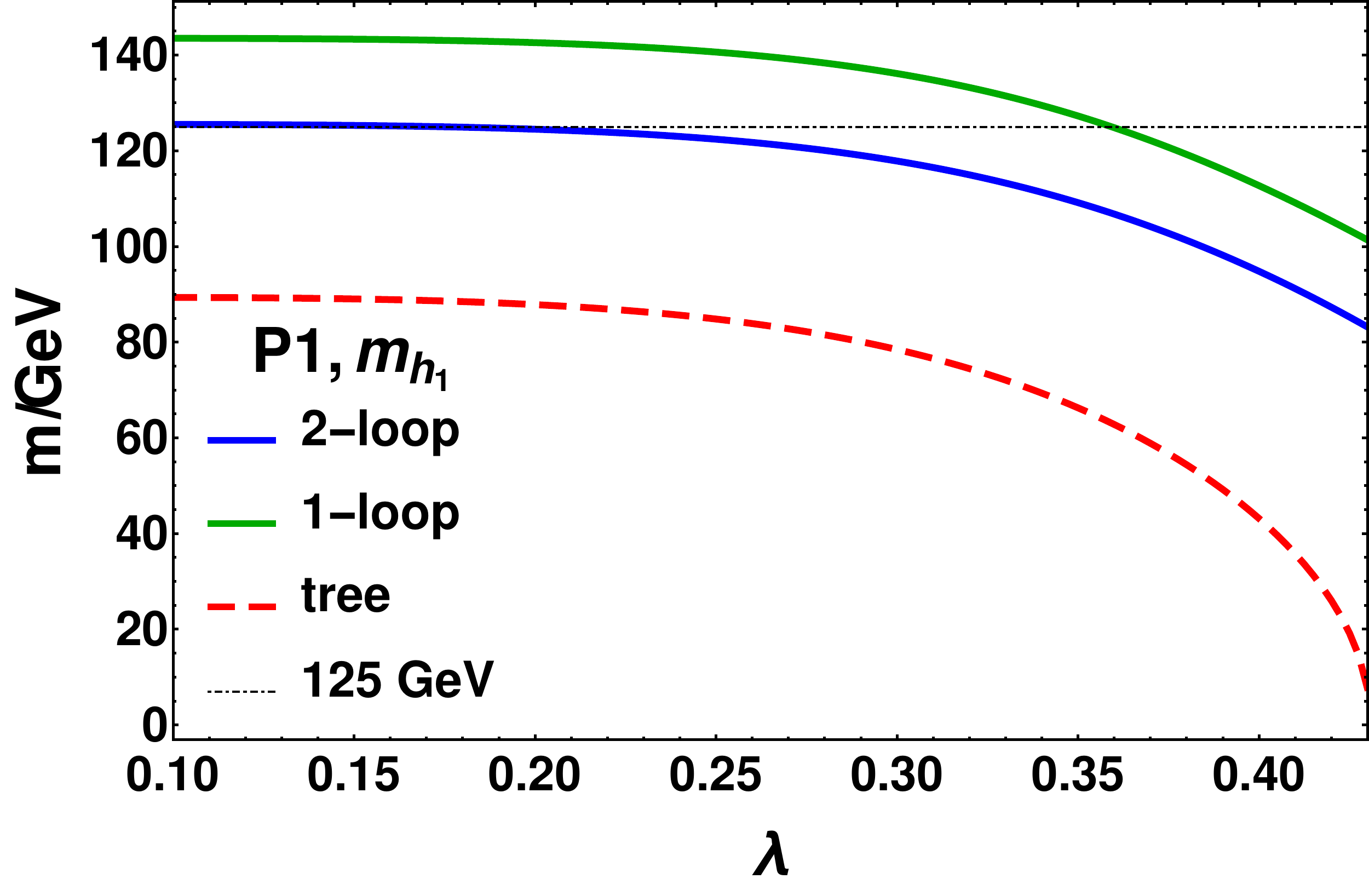}
  \includegraphics[width=.49\textwidth]{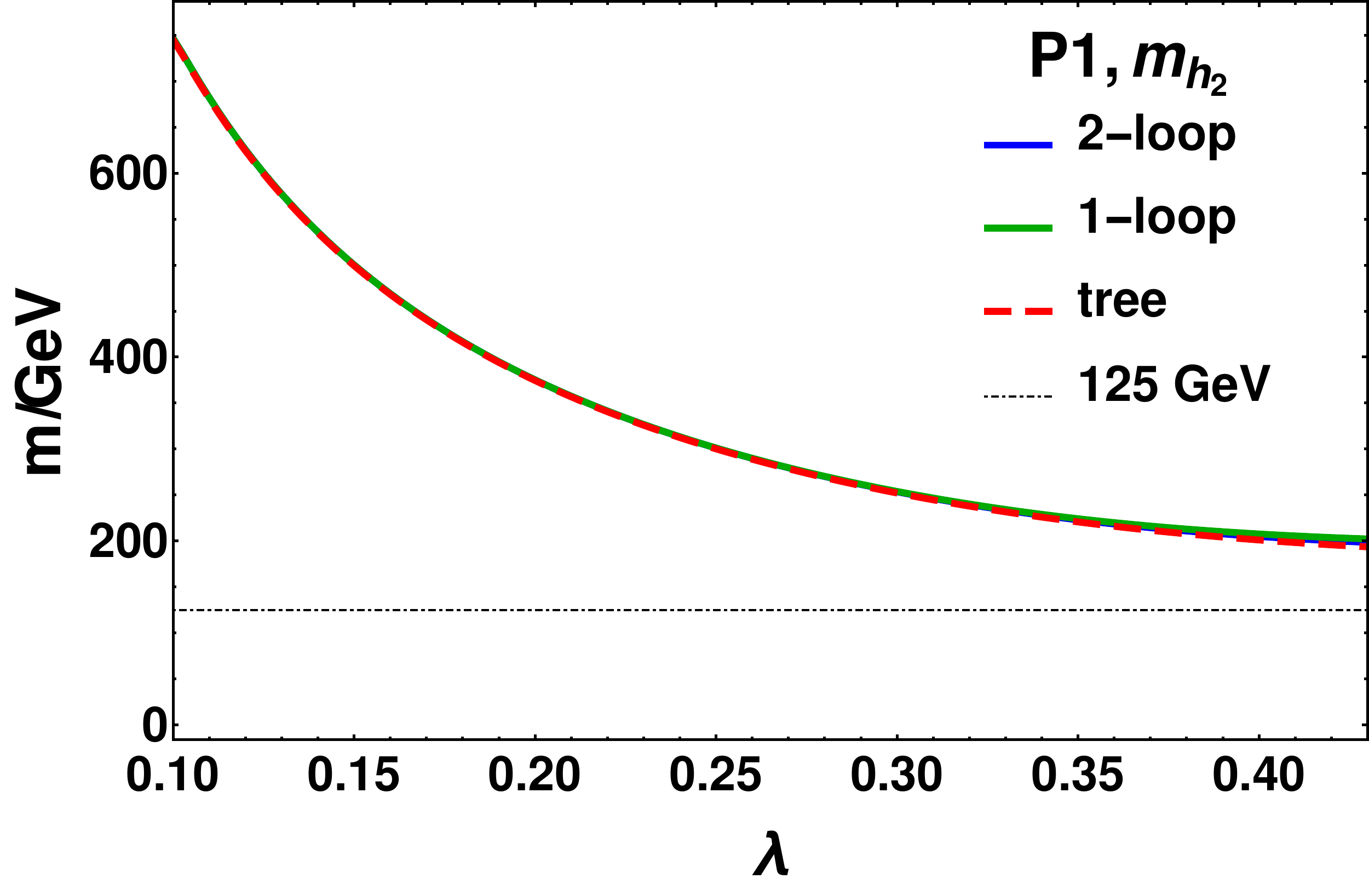}
  \\
  \includegraphics[width=.49\textwidth]{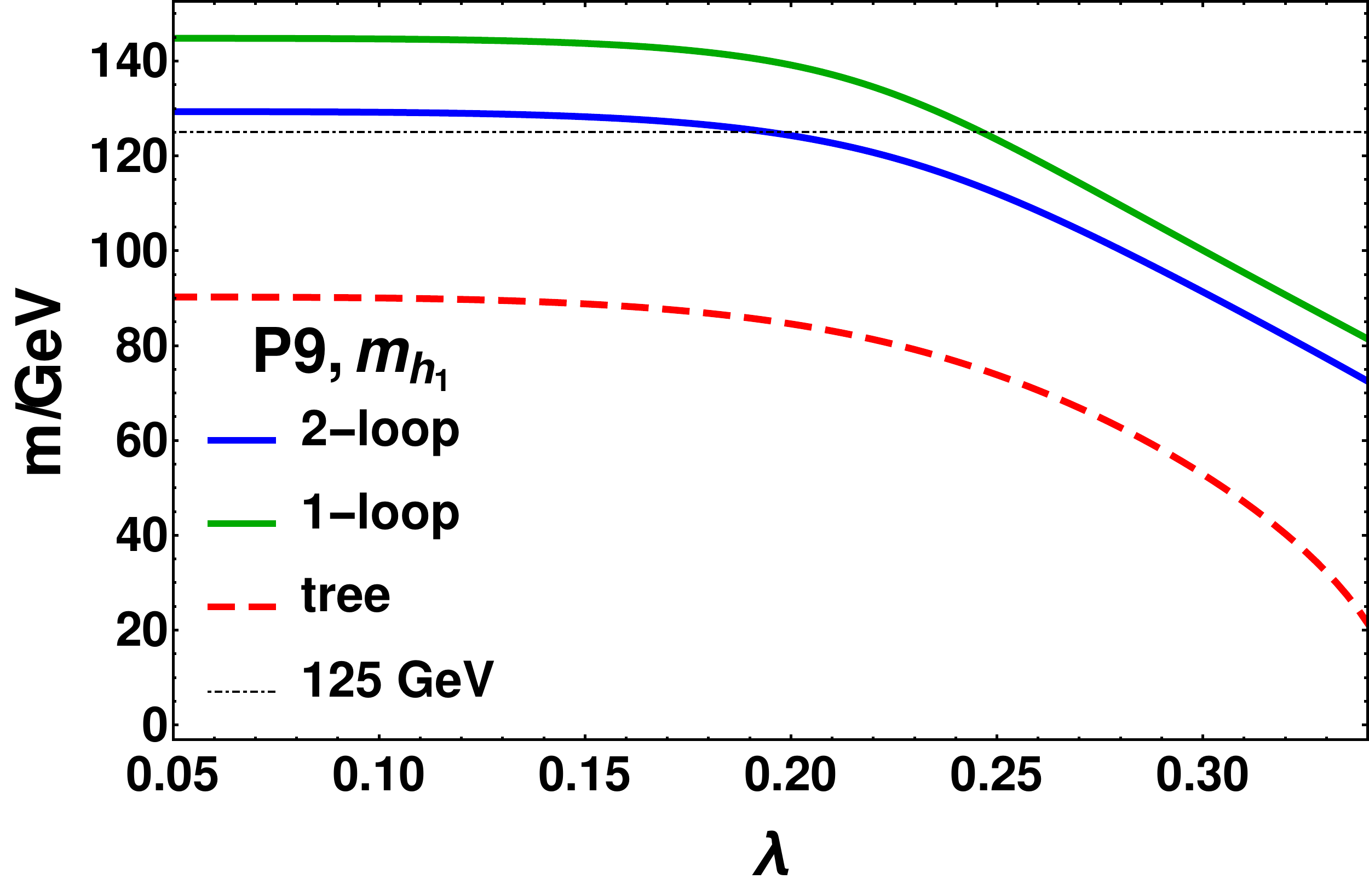}
  \includegraphics[width=.49\textwidth]{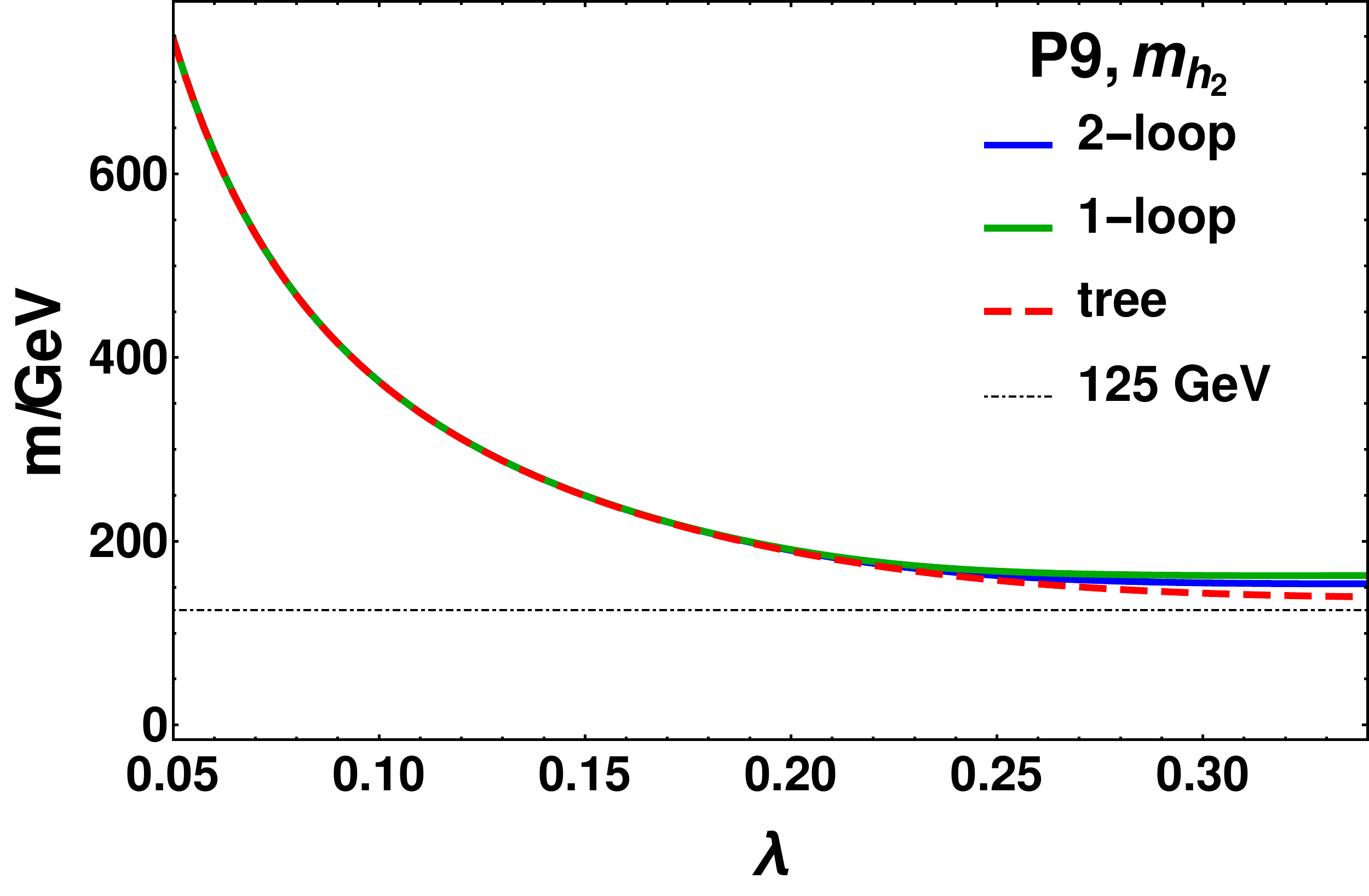}
  \caption{Mass  of   the  lightest  and  next-to   lightest  \cp-even
    Higgs-states,   $m_{h_1}$  (left)   and   $m_{h_2}$  (right),   at
    tree-level, one-loop  and two-loop  order for the  sample scenario
    (first row), the scenarios P1 (second row) and P9 (third row).  At
    one-loop  order all  corrections of  the NMSSM  are included  with
    their   momentum-dependence.    The   two-loop   corrections   are
    approximated     by     the     MSSM-type     contributions     of
    $\mathcal{O}{\left(\alpha_t    \alpha_s,     \alpha_b    \alpha_s,
      \alpha_t^2,\alpha_t\alpha_b\right)}$  including the  resummation
    of  the leading  and next-to-leading  logarithms (see  text).  The
    dotted line  represents $125\ \giga\electronvolt$.   The $\lambda$
    values for which a cross-over  behaviour between the masses occurs
    in  the  sample  scenario  are  at  the  tree-level  $\lambda_{\rm
      c}^{(0)} \approx 0.26$, at one-loop order $\lambda_{\rm c}^{(1)}
    \approx 0.22$ and at two-loop order $\lambda_{\rm c}^{(2)} \approx
    0.23$.   In  the scenario  P9  a  cross-over behaviour  occurs  at
    $\lambda_{\rm  c}^{(0)} >  0.34$ at  tree-level, at  $\lambda_{\rm
      c}^{(1)} \approx  0.25$ at one-loop order,  and at $\lambda_{\rm
      c}^{(2)} \approx  0.26$ at two-loop  order.  In the  scenario P1
    the cross-over behaviour occurs outside of the plotted interval.}
  \label{fig:h12Total}
\end{figure}

The  full results  for the  tree-level, one-  and two-loop  Higgs-mass
predictions     in    the     discussed    scenarios     defined    in
tabs.~\ref{tab:SampleScenario}  and~\ref{tab:P1P9}  are   shown  as  a
function of  $\lambda$ in fig.~\ref{fig:h12Total} for  the two lighter
\cp-even  fields.  The  term ``full  result'' refers  to all  one-loop
corrections in the  NMSSM (including the full  momentum dependence and
also the reparametrisation of the electromagnetic coupling in terms of
the Fermi constant), supplemented  with all available contributions of
$\mathcal{O}{\left(\alpha_t      \alpha_s,     \alpha_b      \alpha_s,
  \alpha_t^2,\alpha_t\alpha_b\right)}$  from the  MSSM, and  including
the resummation of large logarithms.

\subsubsection{Sample Scenario S}

The   results   for   the   sample   scenario   S,   as   defined   in
tab.~\ref{tab:SampleScenario},  are   shown  in   the  first   row  of
fig.~\ref{fig:h12Total}.  For  this study  the parameter  $\lambda$ is
varied between  $0.1$ and  $0.32$.  The lower  limit on  the parameter
$\lambda$ has been chosen such that in the considered parameter region
a cross-over  type behaviour occurs  only for the two  smaller masses,
$m_{h_1}$ and $m_{h_2}$  (for values $\lambda < 0.1$  there is another
point with cross-over behaviour of  the two larger Higgs-boson masses;
however, because of the small values  of $\lambda$ this region is less
suitable for studying the  behaviour of the genuine NMSSM-corrections,
which scale with $\lambda$).

The variation  of the two  masses with $\lambda$  in the first  row of
fig.~\ref{fig:h12Total}  clearly  shows  a cross-over  type  behaviour
between  the masses,  which is  correlated to  their mixing  character
w.r.t.\ the singlet field and the doublet fields.  For small values of
$\lambda$ the field $h_1$ is  doublet-like in this scenario and, based
on   the   prediction   incorporating   all   available   higher-order
corrections, can  be identified with  the signal that was  detected at
the  LHC  at  about  $125\  \giga\electronvolt$.  The  prediction  for
$m_{h_1}$ varies only very little  with $\lambda$ in this region.  The
field $h_2$, on the other  hand, is predominantly singlet-like in this
parameter  region,   and  its  mass  prediction   falls  steeply  with
increasing $\lambda$.  The cross-over occurs at $\lambda_{\rm c}^{(0)}
\approx 0.26$  at tree-level, at $\lambda_{\rm  c}^{(1)} \approx 0.22$
at  one-loop order,  and at  $\lambda_{\rm c}^{(2)}  \approx 0.23$  at
two-loop order.  Above  the cross-over point the behaviour  of the two
masses and  the admixture of  the fields $h_1$  and $h_2$ in  terms of
singlet and  doublet fields are  reversed.  The two fields  are evenly
mixed between  singlet- (i.e.,  genuine NMSSM-type)  and doublet-field
(i.e., MSSM-type) components for $\lambda_{\rm c}^{(n)}$, with $n = 0,
1, 2$.  The heaviest $\cp$-even Higgs field, $h_3$, is doublet-like in
the depicted interval of $\lambda$.  As in the MSSM, the larger masses
(of doublet-like fields) are affected by higher-order corrections to a
lesser  extent  than  the  lighter  states.   Since  at  $\lambda_{\rm
  c}^{(n)}$ the  MSSM-type and genuine NMSSM-type  contributions enter
at  equal footing,  the SM-like  state  is most  sensitive to  genuine
NMSSM-type contributions in the region of the cross-over behaviour.

\subsubsection{Scenario P1}

The  results for  the  scenario P1  are  shown in  the  second row  of
fig.~\ref{fig:h12Total}.    The    lightest   field    is   dominantly
doublet-like, and  the second-lightest  state is singlet-like  for the
depicted  values  of $\lambda$.   The  cross-over  region between  the
doublet- and singlet-like state is rather wide in this case and starts
at $\lambda  \approx 0.2$.  The  cross-over would occur for  values of
$\lambda$ above 0.43,
\noindent
where the  lightest field  becomes tachyonic  at the  tree-level (this
parameter region  is therefore  not shown here).   Thus, even  for the
largest value of $\lambda \approx 0.43$ shown in the plot the lightest
field is  still dominantly  doublet-like at  all depicted  orders.  We
therefore find that the qualitative behaviour in this scenario is very
similar to the sample scenario, but  the allowed range of $\lambda$ is
restricted to the region below the cross-over point in this case.  For
small values  of $\lambda$ the  lightest field can be  identified with
the   signal    that   was    detected   at    the   LHC    at   about
$125\  \giga\electronvolt$.  The  heaviest \cp-even  Higgs field  (not
shown in the figure) remains  doublet-like with a nearly constant mass
of  $\approx  760~\giga\electronvolt$  for   the  depicted  values  of
$\lambda$.

\subsubsection{Scenario P9}

The  results  for the  scenario  P9  are shown  in  the  third row  of
fig.~\ref{fig:h12Total}.  Similarly  to scenario  P1 the  variation of
the  two masses  with $\lambda$  follows the  behaviour of  the sample
scenario.  The  interval in which  the cross-over behaviour  occurs is
larger than in  the sample scenario, but smaller than  in scenario P1.
The cross-over occurs at $\lambda_{\rm c}^{(0)} > 0.34$ at tree-level,
at  $\lambda_{\rm c}^{(1)}  \approx 0.25$  at one-loop  order, and  at
$\lambda_{\rm c}^{(2)} \approx 0.26$ at  two-loop order.  It thus lies
within the  displayed $\lambda$  range if  loop corrections  are taken
into account.   While as  before the character  of the  lightest field
$h_1$ changes from dominantly  doublet-like to dominantly singlet-like
when $\lambda$  is increased through  the cross-over region  (and vice
versa for $h_2$), $h_1$ retains  a doublet-admixture of more than 40\%
even  for  $\lambda$  values  above  the  cross-over  region  in  this
scenario.  Because of the sizable  admixture in this region, $m_{h_1}$
and $m_{h_2}$ each receive  significant self-energy contributions from
both  the singlet  and doublet  fields.  The  heaviest \cp-even  Higgs
field (not  shown in  the figure) remains  doublet-like with  a nearly
constant  mass of  $\approx 750~\giga\electronvolt$  for the  depicted
values of $\lambda$.

\subsubsection{Scenario A1}

\begin{figure}[htbp]
  \centering
  \includegraphics[width=.49\textwidth]{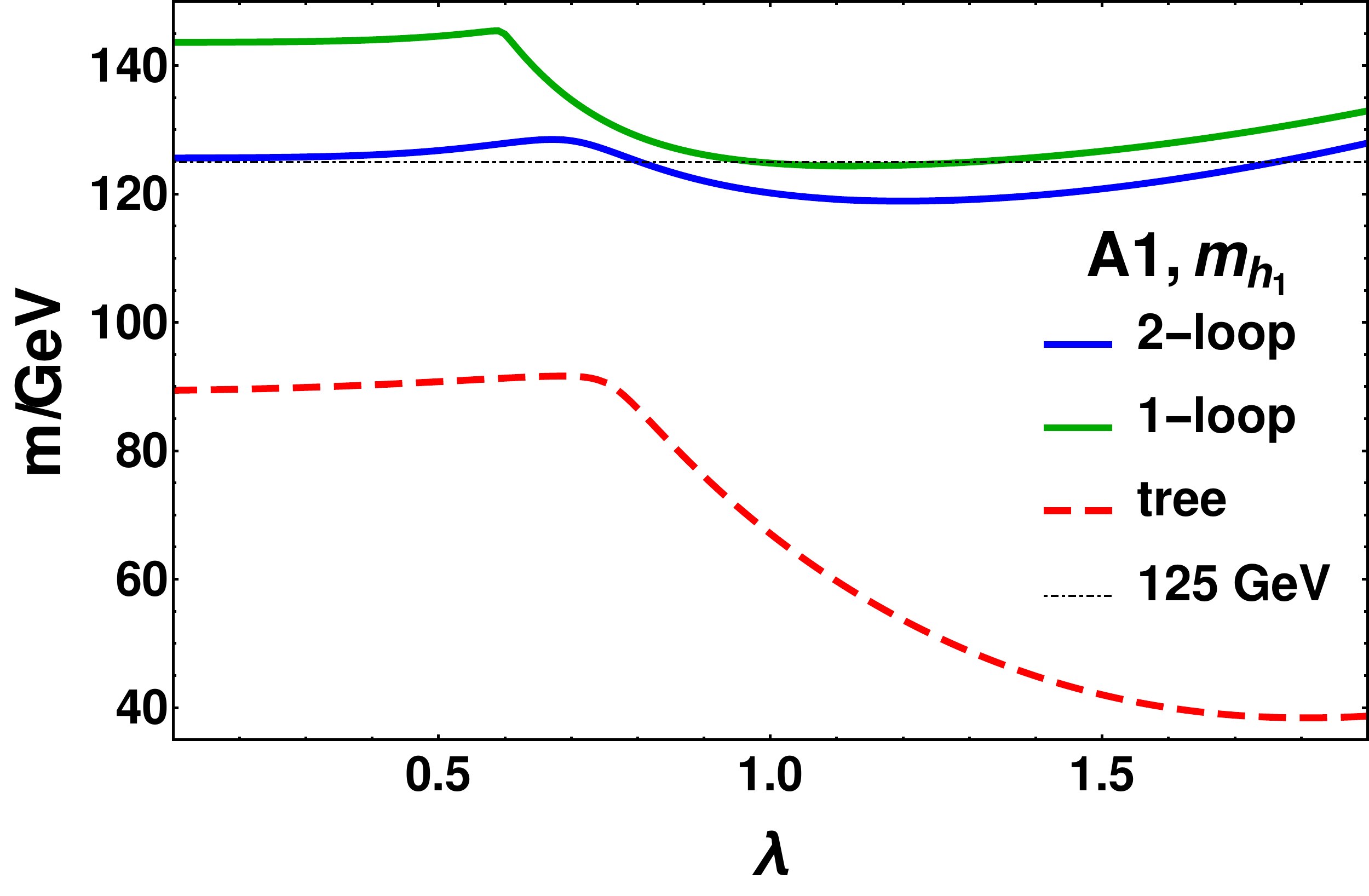}
  \includegraphics[width=.49\textwidth]{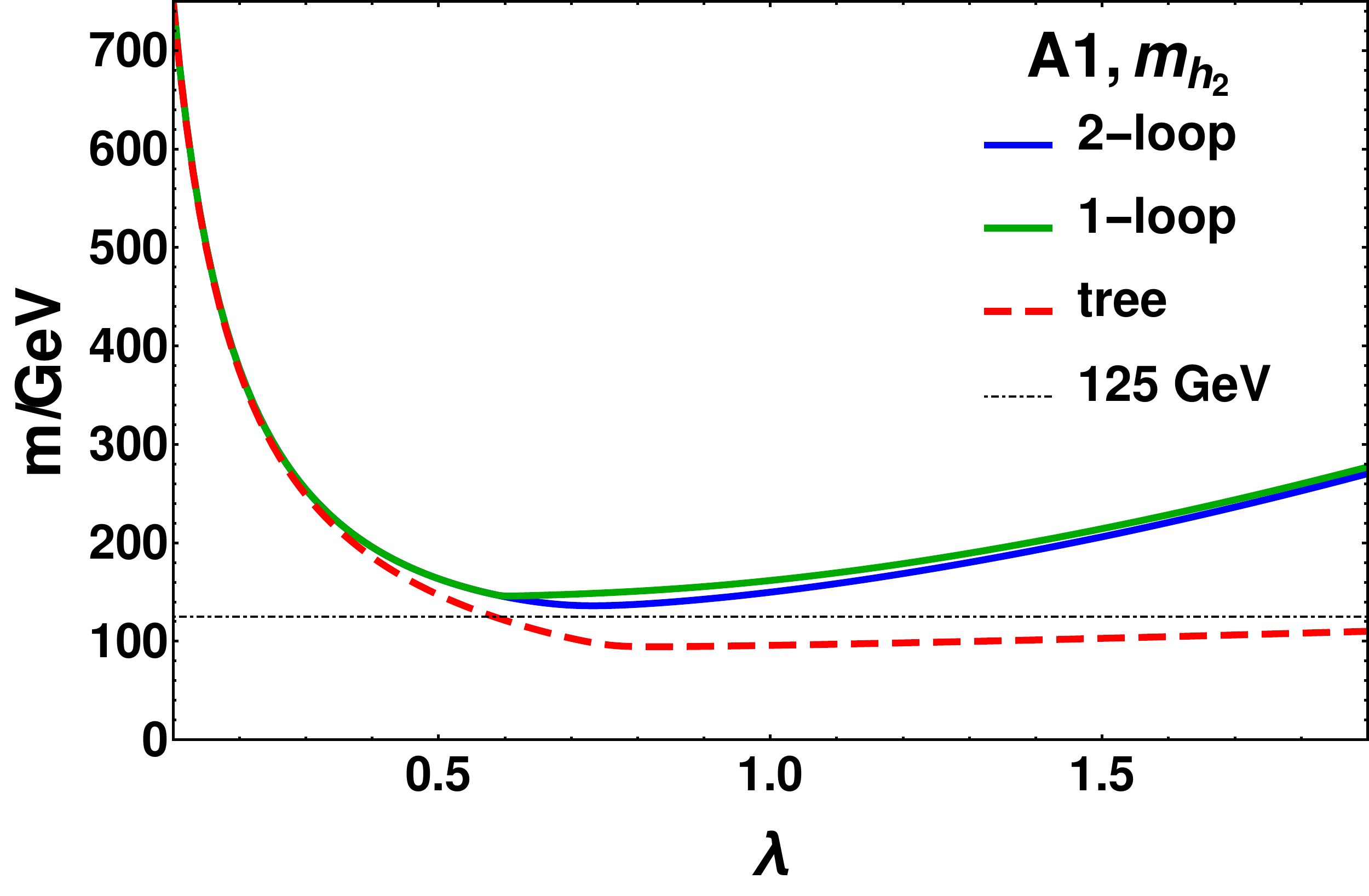}
  \caption{Mass of  the lightest  and next-to  lightest \cp-even
      Higgs-states,  $m_{h_1}$ (left), $m_{h_2}$  (right),
      two-loop order  for the  scenario A1.  The  included corrections
      are  identical  to  the ones  in  fig.~\ref{fig:h12Total}.   The
      $\lambda$ values  for which  a cross-over behaviour  between the
      masses  occurs  are  at the  tree-level  $\lambda_{\rm  c}^{(0)}
      \approx 0.75$, at one-loop  order $\lambda_{\rm c}^{(1)} \approx
      0.70$  and  at  two-loop order  $\lambda_{\rm  c}^{(2)}  \approx
      0.62$.
  }
  \label{fig:h12TotalPD}
\end{figure}

The    results     for    the    scenario    A1     are    shown    in
fig.~\ref{fig:h12TotalPD}.   For values  $\lambda  \lesssim 0.75$  the
variation of  the masses with  $\lambda$ follows the behaviour  of the
sample scenario.   In this region  the lightest field  is doublet-like
and the  second-lightest field  is singlet-like. The  $\lambda$ values
for  which  a  cross-over  behaviour  occurs  are  at  the  tree-level
$\lambda_{\rm c}^{(0)} \approx 0.75$,  at one-loop order $\lambda_{\rm
  c}^{(1)} \approx 0.70$ and  at two-loop order $\lambda_{\rm c}^{(2)}
\approx  0.62$.  For  larger values  of $\lambda$  the lightest  field
$h_1$ obtains  a singlet-admixture  of roughly  70\%, and  the next-to
lightest field $h_2$  obtains a doublet-admixture of the  same size. A
doublet-like Higgs field with a mass close to $125~\giga\electronvolt$
can  be   realised  only   for  values   of  $\lambda$   smaller  than
$\lambda_{\rm c}$ in this scenario.  The heaviest \cp-even Higgs field
remains  doublet-like with  a mass  increasing from  nearly $1500$  to
$1580~\giga\electronvolt$ for  the displayed values of  $\lambda$.  In
the following  we will  omit the discussion  of the  heaviest \cp-even
Higgs field, since it receives only very small two-loop contributions.

\subsection{Numerically leading Contributions at the one-loop Level}
\label{subsec:val1l}

For the prediction  in the MSSM the top/stop  sector contributions are
numerically   leading.    In   the   studied   scenarios,   given   in
tab.~\ref{tab:SampleScenario}  and  tab.~\ref{tab:P1P9},  the  genuine
NMSSM-corrections are  suppressed w.r.t.\ the  corresponding MSSM-like
stop-corrections  since $\lambda  \lesssim \lambda_{\rm  max} <  \yt$,
where $\lambda_{\rm max}  = 0.32, 0.43, 0.35$ in  the three scenarios,
see  the discussion  in  \refse{sec:leadingContributions}.  Thus,  the
genuine  NMSSM  corrections  from  this  sector  are  expected  to  be
sub-leading.

In order  to study the  impact of  the genuine NMSSM  contributions we
compare  the approximation  based  on the  leading MSSM-type  one-loop
corrections       in        the       gauge-less        limit       of
$\mathcal{O}{\left(\yt^2\right)}$,             labelled             as
``$t/\tilde{t}\text{-MSSM}$''  in \reffi{fig:1LoopMT4},  with the  one
where the genuine NMSSM corrections of $\mathcal{O}{\left(\lambda \yt,
  \lambda^2\right)}$  are incorporated.

\subsubsection{Sample Scenario}

\begin{figure}[htbp]
  \centering
  \includegraphics[width=.49\textwidth]{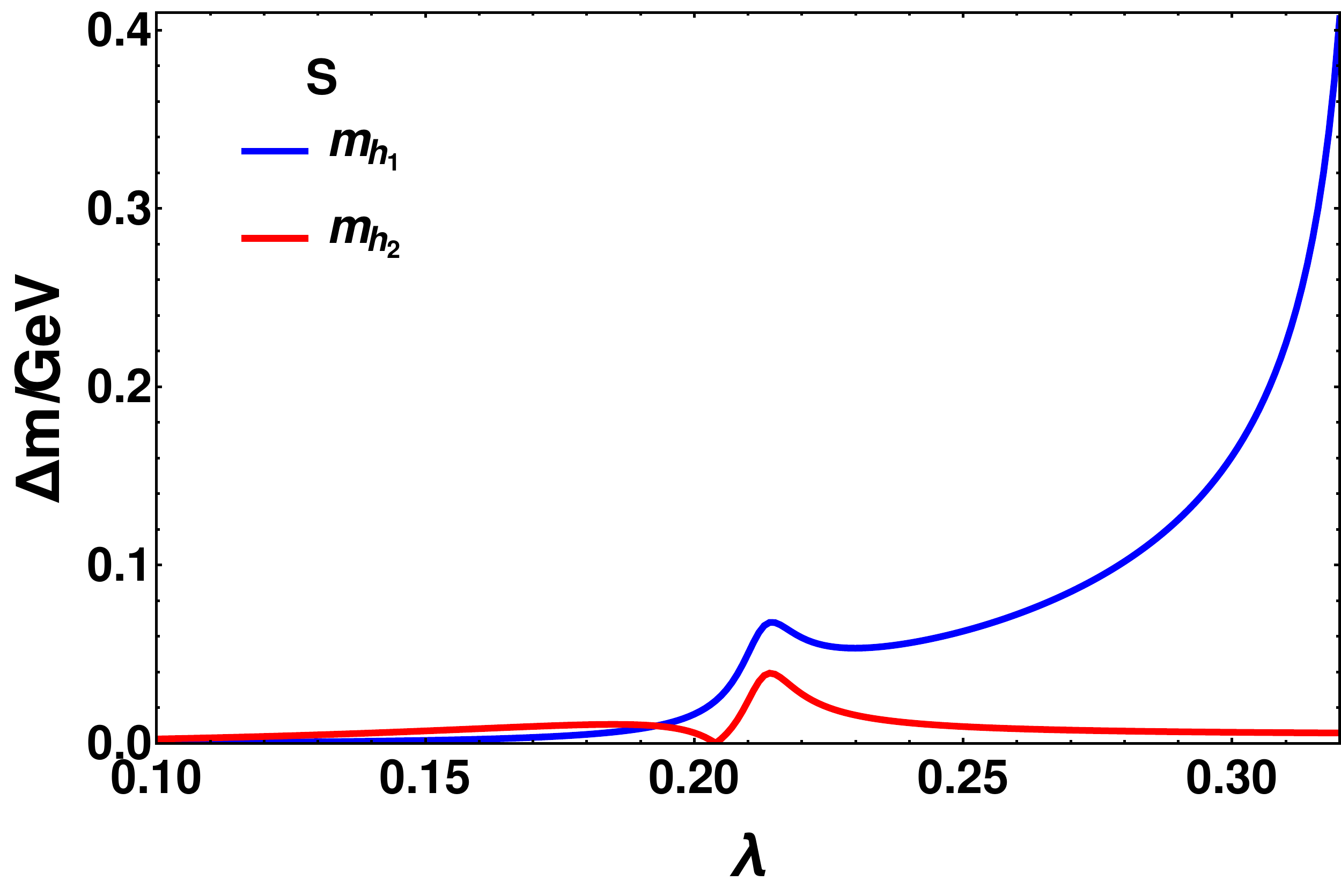}
  \includegraphics[width=.49\textwidth]{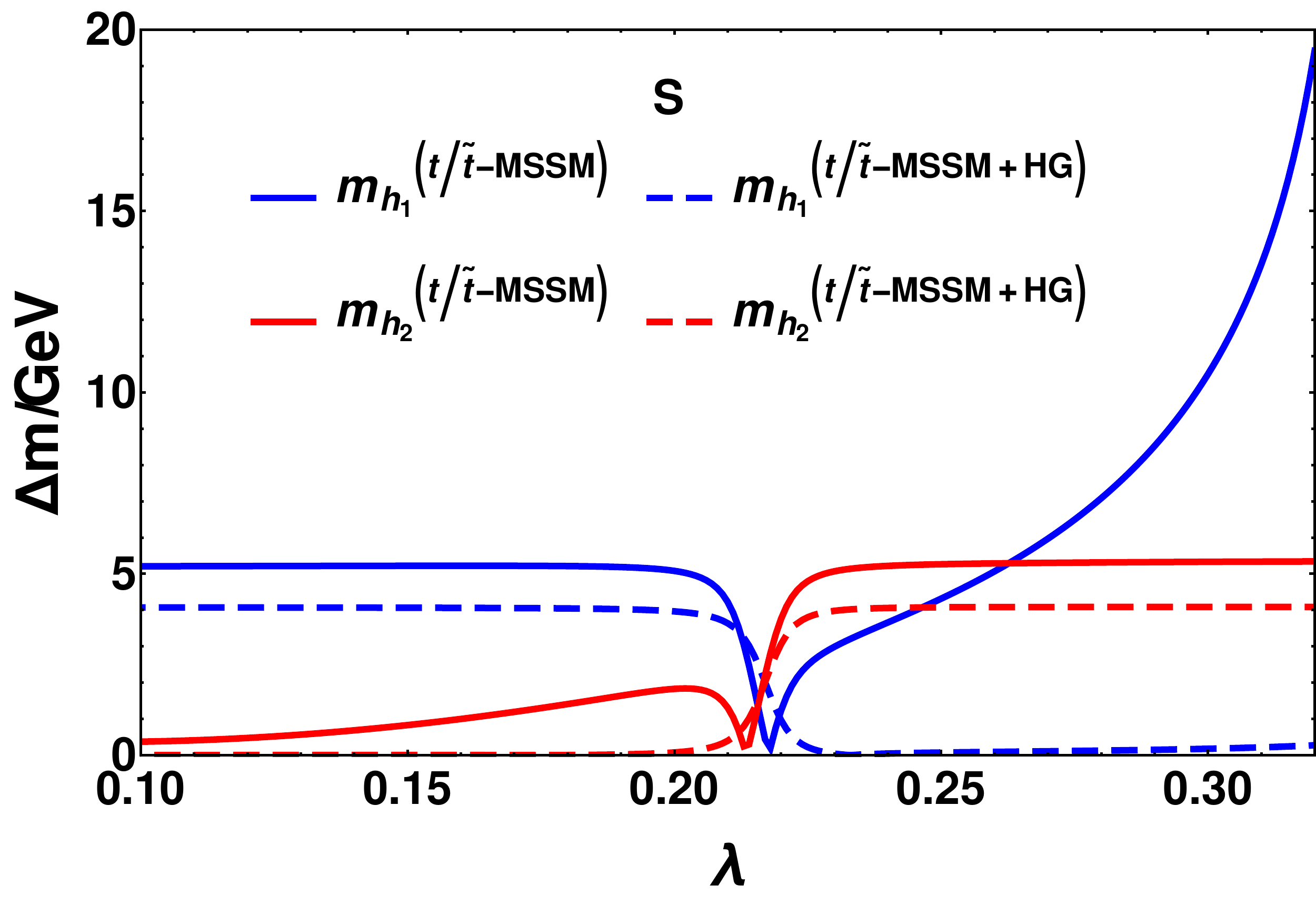}
  \includegraphics[width=.49\textwidth]{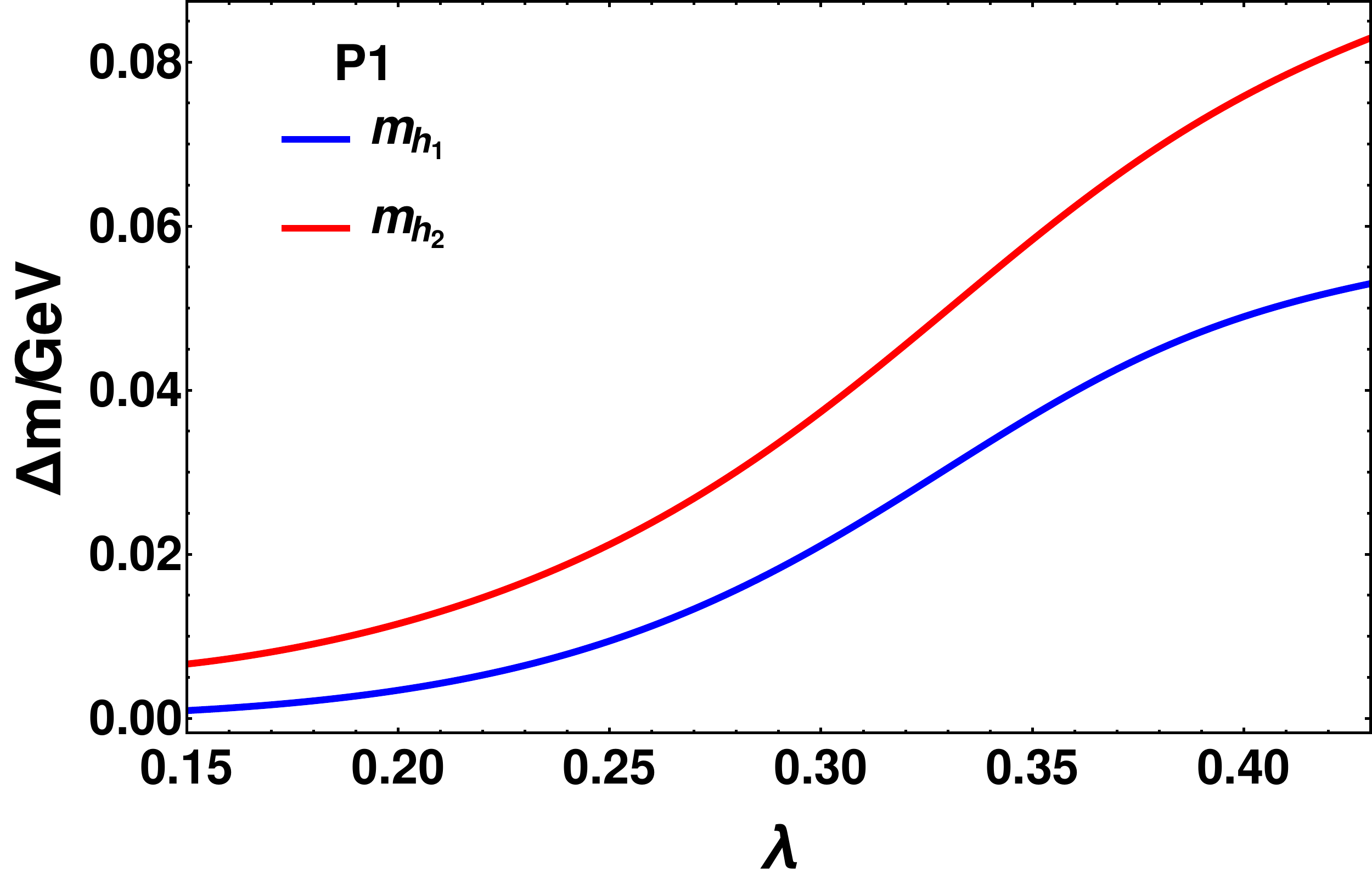}
  \includegraphics[width=.49\textwidth]{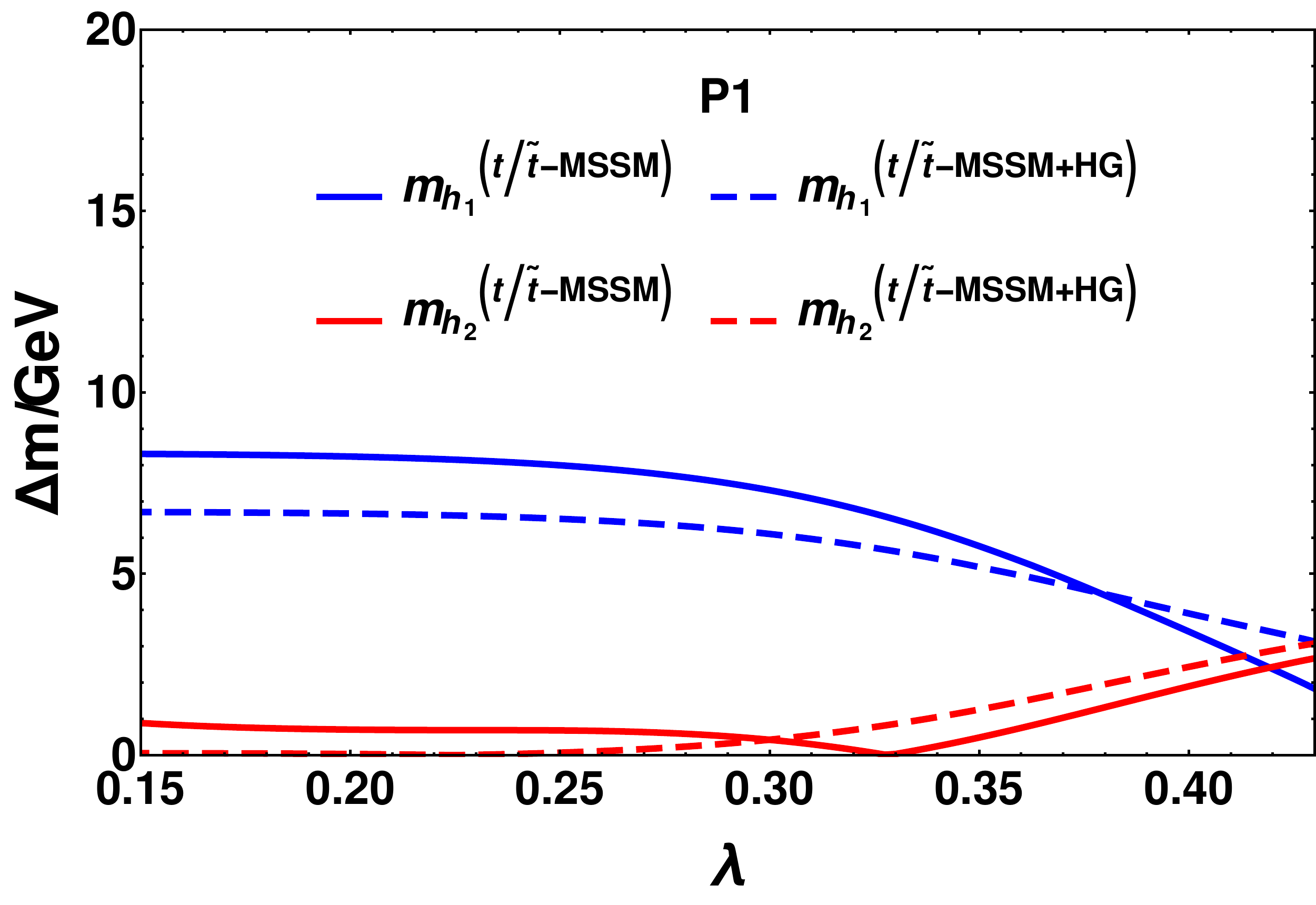}
  \includegraphics[width=.49\textwidth]{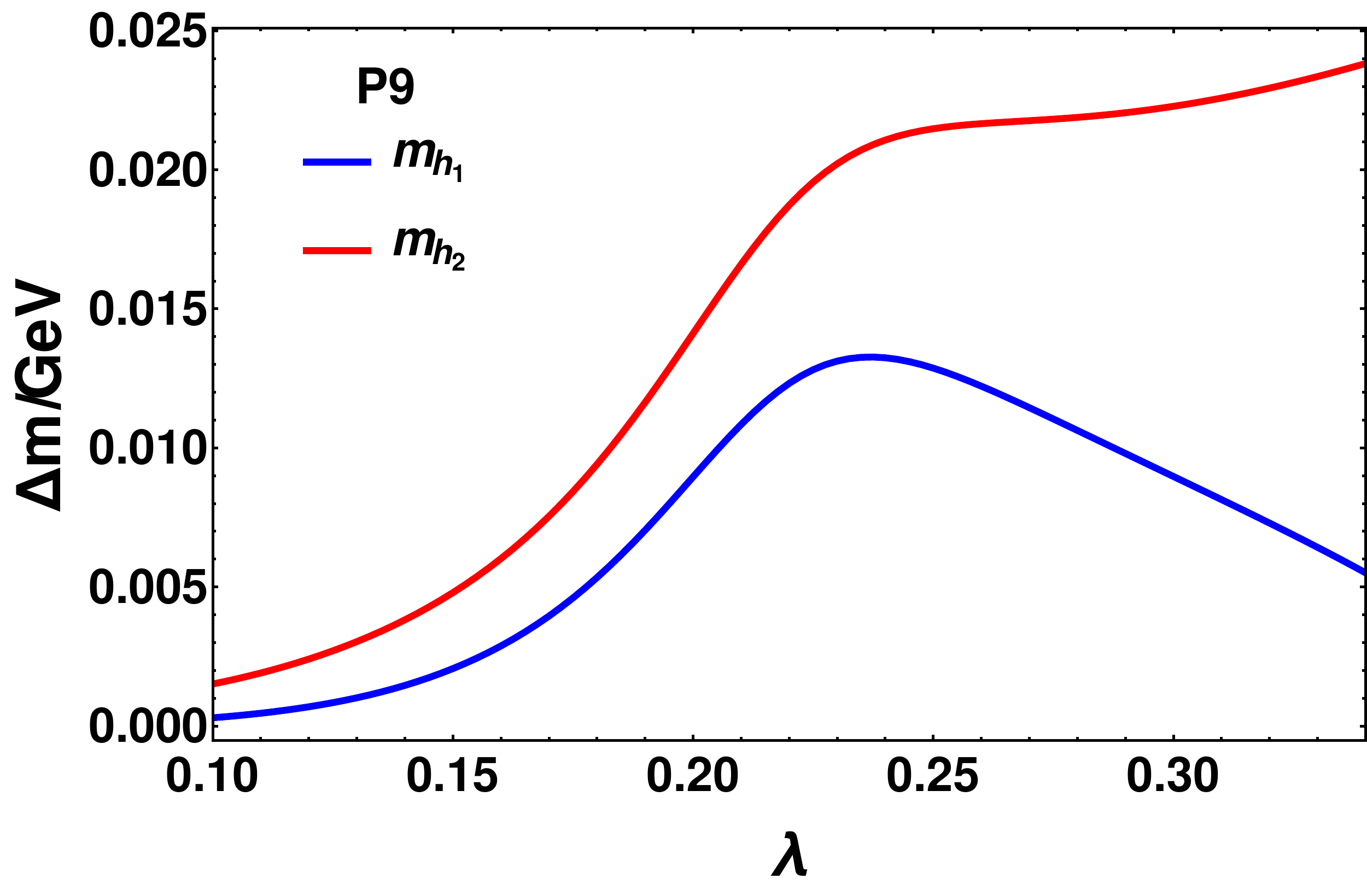}
  \includegraphics[width=.49\textwidth]{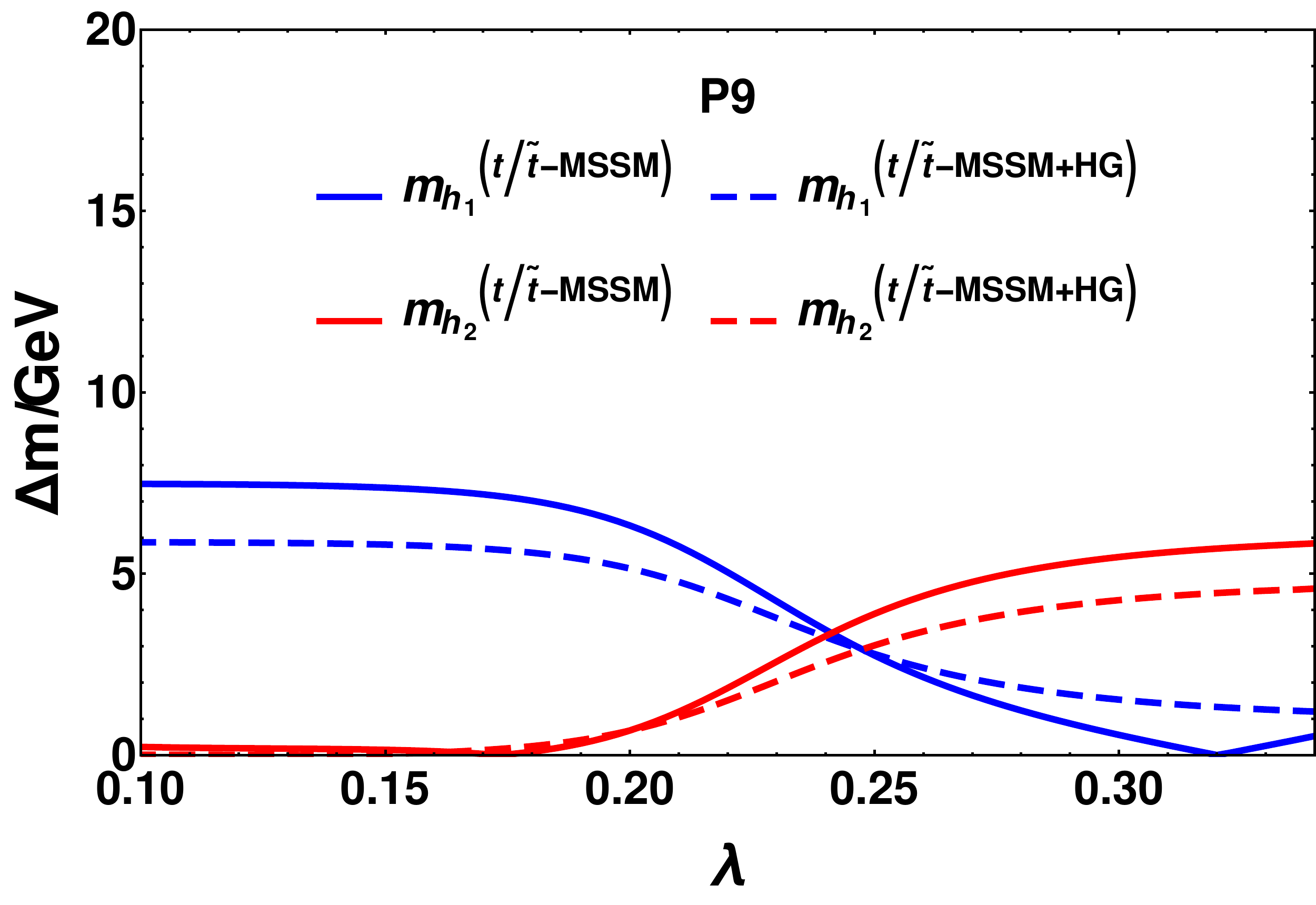}
  \caption{Absolute difference  between partial  and full  results for
    the  one-loop masses  of the  two lighter  \cp-even fields  in the
    sample scenario (first row) and  the scenarios P1 (second row) and
    P9 (third row).  \textit{Left column:} Absolute difference between
    the  mass predictions  including and  excluding the  genuine NMSSM
    contributions  from  the stops  of  $\mathcal{O}{\left(\lambda\yt,
      \lambda^2\right)}$ for  $m_{h_1}$ and  $m_{h_2}$.  \textit{Right
      column:} Absolute differences between the mass predictions based
    on  two different  one-loop approximations  and the  full one-loop
    result.        The       solid      lines,       labelled       as
    ``$t/\tilde{t}\text{-MSSM}$'', depict  the difference  between the
    full result and  the approximation based on  the leading MSSM-type
    contributions from  the top/stop sector,  \mbox{$\Delta{m_{h_i}} =
      \left| m_{h_i}^{(\text{1L})} - m_{h_i}^{t/\tilde{t}\text{-MSSM}}
      \right|$}.       The     dashed      lines,     labelled      as
    ``$t/\tilde{t}\text{-MSSM} + \text{HG}$'',  show the corresponding
    result where the leading MSSM-type contributions from the top/stop
    sector   are   supplemented   by  the   contributions   from   the
    Higgs-/higgsino-            and            gauge-/gaugino-sectors,
    \mbox{$\Delta{m_{h_i}}    =    \left|   m_{h_i}^{(\text{1L})}    -
      m_{h_i}^{t/\tilde{t}\text{-MSSM} + {\rm HG}} \right|$}.}
  \label{fig:1LoopMT4}
\end{figure}

For the sample scenario the difference between the mass predictions in
the  two approximations  is plotted  as  a function  of $\lambda$  for
$m_{h_1}$  and  $m_{h_2}$  in  the  left plot  of  the  first  row  in
\reffi{fig:1LoopMT4}.  We find  that for the whole  range of $\lambda$
in  the  plot   the  impact  of  the  genuine   NMSSM  corrections  of
$\mathcal{O}{\left(\lambda  \yt,\lambda^2\right)}$  remains less  than
0.5~\giga\electronvolt.   The  largest   difference  between  the  two
approximations occurs for the light  singlet-like state $h_1$ at large
values of $\lambda$ close to the upper limit of $\lambda \approx 0.32$
shown in the plot.  In fact,  for $m_{h_1}$ the difference between the
two approximations  is seen to  rise sharply for increasing  values of
$\lambda$.   On the  other  hand,  at the  $\lambda$  value where  the
cross-over behaviour  occurs, $\lambda_{\rm c}^{(1)}$,  the difference
between the  two approximations is  seen to  have a local  maximum but
remains small,  below $0.1~\giga\electronvolt$.  For  the doublet-like
state, which has a one-loop mass of more than $130~\giga\electronvolt$
(see    \reffi{fig:h12Total}),    the   corrections    from    genuine
NMSSM-contributions remain below the level of 1\permil\ over the whole
range of  $\lambda$.  Thus, the  approximation based on  the MSSM-type
contributions is  seen to provide  a very accurate prediction  for the
top/stop sector  contributions to  the mass  of a  doublet-like state.
For the singlet-like state, where  the deviation grows with $\lambda$,
the deviation  reaches $\approx  1 \%$  for the  one-loop mass  of the
singlet-like  state of  $\approx  40~\giga\electronvolt$ for  $\lambda
\approx 0.32$.

The sharp  increase of  the corrections of  $\mathcal{O}{(\lambda \yt,
  \lambda^2)}$ for the highest values of $\lambda$ that is visible for
the   light   singlet-like  field   in   the   upper  left   plot   of
\reffi{fig:1LoopMT4}  indicates that  the approximation  for the  stop
sector  of   restricting  to   the  MSSM-type   contributions  becomes
questionable for the  singlet-like state in this  region.  However, as
shown  in  the  upper  right plot  of  \reffi{fig:1LoopMT4},  in  this
parameter  region the  stop  sector as  a whole  ceases  to provide  a
reliable  approximation of  the full  one-loop contributions.   In the
right  plot   the  difference   between  the   full  result   and  the
approximation based  on the  leading MSSM-type contributions  from the
top/stop sector, \mbox{$\Delta{m_{h_i}} = \left| m_{h_i}^{(\text{1L})}
  - m_{h_i}^{t/\tilde{t}\text{-MSSM}}  \right|$},  is  shown  together
with   \mbox{$\Delta{m_{h_i}}   =   \left|   m_{h_i}^{(\text{1L})}   -
  m_{h_i}^{t/\tilde{t}\text{-MSSM} + {\rm HG}} \right|$}, where in the
latter  case the  leading  MSSM-type contributions  from the  top/stop
sector are  supplemented by  the contribution from  the Higgs-higgsino
and  gauge-/gaugino-sectors.  While  for  the  singlet-like state  the
deviation between  the leading contributions from  the top/stop sector
and the  full one-loop result becomes  huge for the largest  values of
$\lambda$,   reaching  the   level  of   $20~\giga\electronvolt$,  the
deviations stay small, far  below the level of $1~\giga\electronvolt$,
if the leading contributions from the top/stop sector are supplemented
by    the     contributions    from    the     Higgs-/higgsino-    and
gauge-gaugino-sectors.  This result for  the singlet-like state can be
understood from the fact that  the gauge couplings of the singlet-like
state  are   heavily  suppressed   and  that  therefore   the  leading
contributions for  large $\lambda$ arise  from the Higgs  and higgsino
sector.    Thus,  improving   on   the   approximation  of   MSSM-type
contributions in  the stop  sector requires  the incorporation  of the
contributions from  the Higgs and  higgsino sector, while  the genuine
NMSSM contributions  in the stop  sector are of minor  significance in
this context.

For the doublet-like state,  namely $h_1$ for values $\lambda \lesssim
\lambda_{\rm c}$ and $h_2$  for $\lambda \gtrsim \lambda_{\rm c}$, the
difference between  the full one-loop  result and the result  based on
the leading contributions from the  top/stop sector amounts to a shift
of  about $5~\giga\electronvolt$  that is  essentially  independent of
$\lambda$  except  for  the  region  where  the  cross-over  behaviour
occurs.  This nearly  constant  shift arises  mainly from  sub-leading
contributions  in the  top/stop sector.   As indicated  by  the dashed
lines, the inclusion of the contributions from the Higgs and the gauge
sector reduces  the difference  to the full  one-loop result  by about
$1~\giga\electronvolt$.

\subsubsection{Scenario P1}

The difference between the mass  predictions in the two approximations
in  the top/stop  sector is  plotted as  a function  of $\lambda$  for
$m_{h_1}$  and  $m_{h_2}$ in  the  left  plot  of  the second  row  in
\reffi{fig:1LoopMT4}.  As in  fig.~\ref{fig:h12Total}, the qualitative
behaviour is  similar to  the one  in the  sample scenario,  while the
allowed range in $\lambda$ in P1 is restricted to the region below the
cross-over point.  The impact of the genuine NMSSM corrections is even
smaller in  this case than in  the sample scenario, amounting  to less
than $100~\mega\electronvolt$ for both  lighter \cp-even Higgs fields.
In the  right plot of the  second row the difference  between the full
result  and   the  approximation   based  on  the   leading  MSSM-type
contributions  from  the top/stop  sector  and  the leading  MSSM-type
contributions   from  the   top/stop   sector   supplemented  by   the
contribution  from the  Higgs-higgsino and  gauge-/gaugino-sectors are
shown.   By  supplementing  the  partial  one-loop  results  with  the
Higgs/higgsino/gauge-boson/gaugino  contributions the  mass prediction
for    the    doublet-like    state   is    improved    by    $\approx
1.5\ \giga\electronvolt$.  As explained above for the sample scenario,
the difference  between the approximate  mass prediction and  the full
one-loop result for the doublet-like  state is mainly due to MSSM-type
sub-leading  contributions  in  the top/stop  sector.   The  different
variation with $\lambda$  in the right plot as compared  to the sample
scenario is related  to the much wider cross-over region  in this case
(starting at about $\lambda =  0.2$).  The large deviation encountered
in the sample scenario for the singlet-like state above the cross-over
region is obviously  not present in scenario P1, as  the latter one is
confined to $\lambda$ values below the cross-over region.

\subsubsection{Scenario P9}

For the scenario P9  the results are given in the  same fashion as for
the other two scenarios in  the third row of \reffi{fig:1LoopMT4}.  As
can  be  seen in  the  left  plot, the  impact  of  the genuine  NMSSM
corrections is  even still smaller  than in scenario P1,  amounting to
less  than $25~\mega\electronvolt$  for  both  lighter \cp-even  Higgs
fields.   By  supplementing  the  partial one-loop  results  with  the
Higgs/higgsino/gauge-boson/gaugino contributions (right plot) the mass
prediction  for  the  doublet-like   state  is  improved  by  $\approx
1.5~\giga\electronvolt$.    The  overall   variation  with   $\lambda$
resembles the one of the sample  scenario (upper right plot) if in the
latter  case one  focuses on  $\lambda$ values  up to  just above  the
cross-over  region. The  fact  that  in the  scenario  P9  there is  a
sizeable admixture  of singlet-  and doublet-components in  the states
$h_2$  and  $h_1$   above  the  cross-over  region   leads  to  slight
modifications. While  for the sample  scenario $\Delta m$ is  the same
above and below the cross-over  region for the doublet-like state, for
scenario  P9 we  find  that $\Delta  m$ is  somewhat  reduced for  the
doublet-like state above  the cross-over region. Thus,  in this region
the  singlet-admixture  of   more  than  40\%  to   $h_2$  shifts  the
approximate one-loop mass prediction closer  to the mass obtained with
the full  one-loop calculation.  It should  be noted that even  in the
case of  a sizeable  admixture of singlet-  and doublet-contributions,
which  is realised  in scenario  P9 above  the cross-over  region, the
genuine NMSSM-type contributions  have a minor impact  compared to the
Higgs/higgsino/gauge-boson/gaugino contributions.

\subsubsection{Scenario A1}

\begin{figure}[htbp]
  \centering
  \includegraphics[width=.49\textwidth]{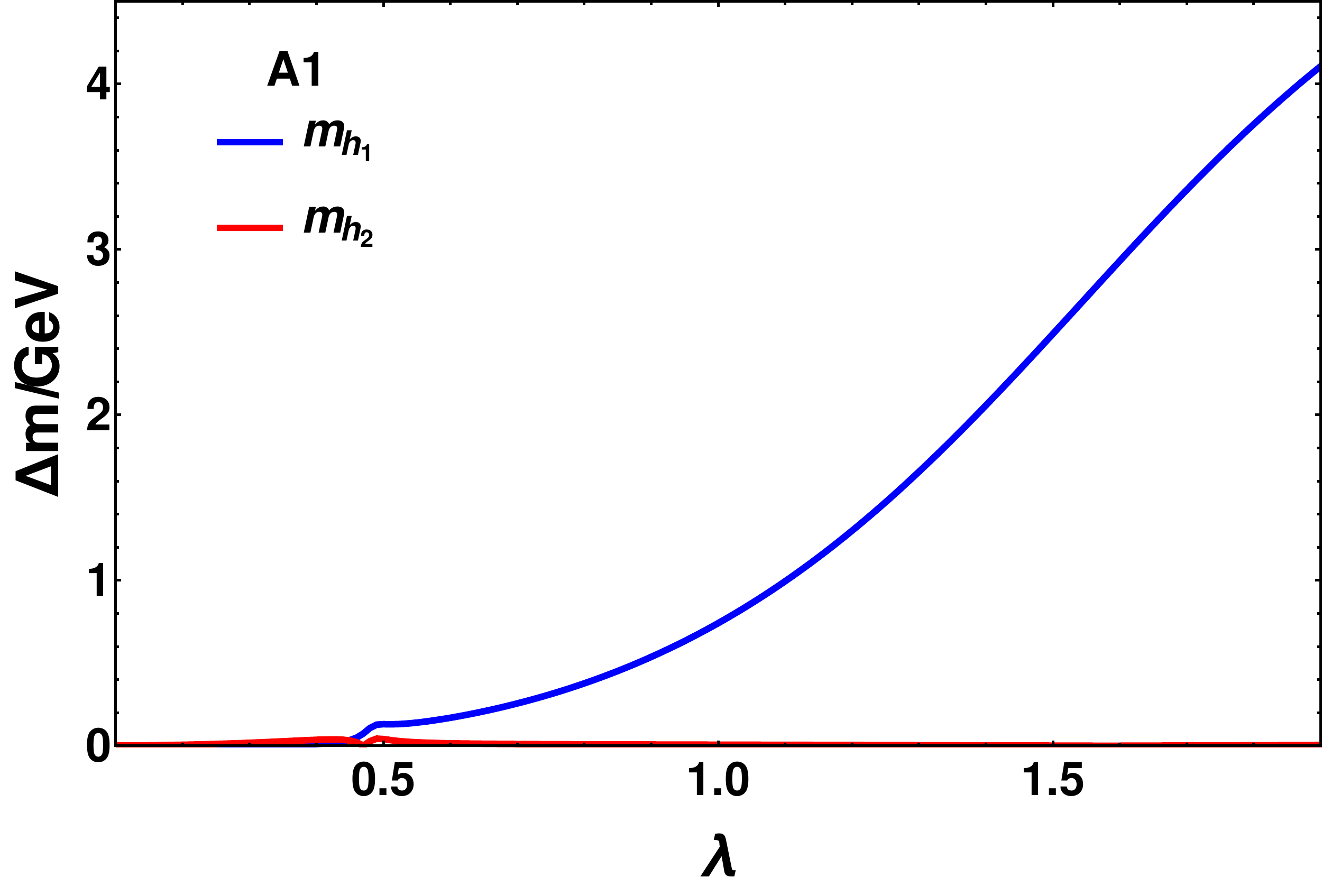}
  \includegraphics[width=.49\textwidth]{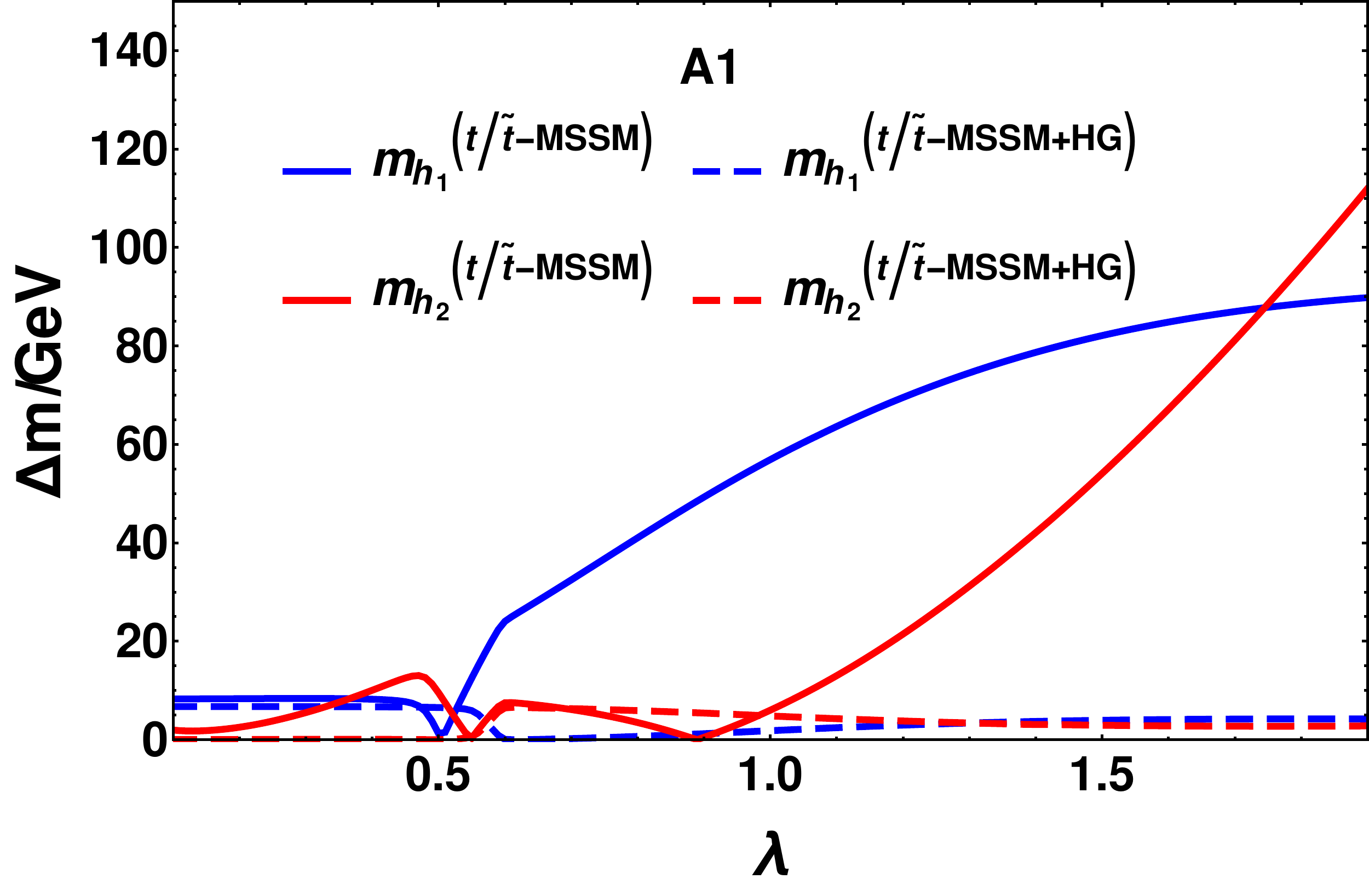}
  \caption{Absolute difference  between partial  and full  results for
    the  one-loop masses  of the  two lighter  \cp-even fields  in the
    scenario A1.  The  meaning of the displayed curves is  the same as
    in fig.~\ref{fig:1LoopMT4}.}
  \label{fig:1LoopMT4PD}
\end{figure}

For  the   scenario  A1  the   corresponding  analysis  is   shown  in
fig.~\ref{fig:1LoopMT4PD}.      As     in     figs.~\ref{fig:h12Total}
and~\ref{fig:h12TotalPD}, the  qualitative behaviour for  values close
to and below the cross-over region  is similar to the sample scenario.
Up to values $\lambda \approx  0.7$ the genuine NMSSM-type corrections
from  the top/stop  sector  are  of similar  size  as  for the  sample
scenario,  amounting up  to $\approx  100~\mega\electronvolt$ for  the
doublet-like   field    $h_1$   field    with   a   mass    close   to
$125~\giga\electronvolt$.  For  the singlet-like field  the NMSSM-type
contributions  from   the  top/stop  sector  amount   up  to  $\approx
500~\mega\electronvolt$ in  this region. The  NMSSM-type contributions
from the top/stop  sector increase sharply for  the singlet-like field
$h_1$   at   values  above   $\lambda_{\rm   c}$,   amounting  up   to
$4~\giga\electronvolt$  for  the  largest vaules  for  $\lambda$,  and
become  tiny for  the  doublet-like field  $h_2$,  staying well  below
$20~\mega\electronvolt$.

As before we observe also for  this scenario with very large values of
$\lambda$  that  other  contributions  beyond  the  leading  MSSM-type
contributions  from  the top/stop  sector  are  numerically much  more
important than  the leading genuine NMSSM-type  contributions from the
top/stop   sector.   As  can   be   seen   in   the  right   plot   of
fig.~\ref{fig:1LoopMT4PD},   the   difference  between   the   leading
MSSM-type contributions from the top/stop sector and the full one-loop
result amounts  to about  $8~\giga\electronvolt$ for  the doublet-like
field $h_1$ in the region where $\lambda \lesssim 0.5$.  Supplementing
the leading MSSM-type contributions from  the top/stop sector with the
Higgs /  higgsino / gauge-boson  / gaugino contributions  improves the
prediction by about 1--2~\giga\electronvolt.  As before, the remaining
difference in  this parameter region  is mainly caused  by sub-leading
contributions from  the top/stop  sector.  For the  singlet-like state
$h_2$ the discrepancy between the full one-loop result and the leading
MSSM-type  contributions   from  the  top/stop  sector   becomes  very
significant     for    increasing     $\lambda$,    reaching     about
$12~\giga\electronvolt$ for $\lambda \approx 0.45$.  This large effect
is caused by the contributions of the Higgs / higgsino / gauge-boson /
gaugino  sectors.   Incorporating   those  contributions  reduces  the
discrepancy below the level of $100~\mega\electronvolt$.  For $\lambda
\gtrsim 0.5$ the discrepancy between  the full one-loop result and the
leading MSSM-type contributions from  the top/stop sector becomes huge
for  $h_1$. The  same  is true  for  $h_2$ for  very  large values  of
$\lambda$  above  $1$.   This  huge  effect is  again  caused  by  the
contributions of the Higgs / higgsino / gauge-boson / gaugino sectors.
Incorporating those contributions reduces the discrepancy to the level
of  3--5~\giga\electronvolt.   Accordingly,   even  for  this  extreme
scenario the top/stop  sector is well described by  just the MSSM-type
contributions  in  those regions  of  the  parameter space  where  the
top/stop sector itself provides an  adequate approximation of the full
one-loop result. For the highest  values of $\lambda$ in this scenario
the contributions  beyond the top/stop sector  are huge, demonstrating
the necessity to use in this case a complete result incorporating also
the contributions from the Higgs  / higgsino and gauge-boson / gaugino
sectors.

\subsubsection{Conclusion}

As a result of the comparison  performed in this section the MSSM-type
top/stop sector  contributions of \order{\yt^2} have  been verified as
the leading  one-loop contributions to MSSM-like  fields.  The genuine
NMSSM top/stop sector contributions  of \order{\lambda \yt, \lambda^2}
have the  largest impact  on singlet-like fields  for large  values of
$\lambda$, where however an  approximation based only on contributions
from the  fermion/sfermion sector  is in  any case  insufficient.  Our
analysis at the one-loop level  therefore shows that approximating the
result for the top/stop sector  by the leading MSSM-type contributions
turns out  to work well  in the  parameter regions where  the top/stop
sector itself  yields a reasonable  approximation of the  full result.
These findings provide a strong  motivation for applying the same kind
of approximation also  at the two-loop level.  For  the description of
singlet-like fields in the region of large values of $\lambda$ we have
demonstrated the  importance of  incorporating also  the contributions
from the Higgs / higgsino and gauge-boson / gaugino sectors.

\subsection{Comparison  with \texttt{NMSSMCalc}}
\label{subsec:comparison}

For  the comparison  of  our  results with  available  tools the  code
\texttt{NMSSMCalc}~\cite{Baglio:2013iia}  is   particularly  suitable,
since it is the only public  tool using also a mixed $\DRbar$/on-shell
renormalisation  scheme.  In  this section  the numerical  differences
between the  results for the  masses of  the two lighter  Higgs states
from  \texttt{NMSSMCalc}  and our  calculation  will  be discussed  at
different      orders     for      the     scenarios      given     in
tab.~\ref{tab:SampleScenario}  and  tab.~\ref{tab:P1P9}.  Both  codes,
\texttt{NMSSMCalc}       and      our       calculation,      labelled
\texttt{NMSSM-FeynHiggs} in the following,  have been adapted for this
comparison.   The two  codes  interpret the  input  parameters in  the
stop-sector  as  defined  for  on-shell  renormalised  masses  of  the
stops\footnote{We thank Kathrin Walz  for providing a modified version
  of \texttt{NMSSMCalc} for  this feature.}.  Since \texttt{NMSSMCalc}
uses  a different  charge  renormalisation associated  with the  value
$\alpha{(M_Z)}$  for the  electromagnetic coupling  constant, we  have
reparametrised       our       result      as       described       in
sec.~\ref{sec:reparametrisation}.     The    numerical   values    for
$\alpha{(M_Z)}$  and $\Delta{\alpha}$  have been  taken directly  from
\texttt{NMSSMCalc} for this comparison,
\begin{align}
  \Delta{\alpha} = 
  \Delta{\alpha^{(5)}_{\text{had}}}  +  \Delta{\alpha_{\text{lep}}}  =
  5.89188 \cdot 10^{-2}, \quad  \alpha{\left(M_Z\right)} = 1/128.962\ .
\end{align}
\noindent
After the reparametrisation is applied the only difference between the
one-loop  Higgs-mass predictions  of  \NFH\ and  \NC\  stems from  the
finite contribution of $\delta{v}$ used in \texttt{NMSSMCalc}.  Beyond
the   one-loop    level   only   MSSM    two-loop   contributions   of
$\mathcal{O}{(\alpha_t    \alpha_s)}$    (calculated   for    on-shell
renormalised    top-    and    stop-masses)    are    considered    in
\texttt{NMSSM-FeynHiggs}   for   this   comparison,  as   only   their
NMSSM-counterparts  are implemented  in  \texttt{NMSSMCalc}.  Two-loop
corrections beyond  the ones of  $\mathcal{O}{(\alpha_t \alpha_s)}$ as
well as the  resummation of logarithms, which are  incorporated in the
default version of \texttt{NMSSM-FeynHiggs},  are not included for the
analysis  in  this  section  (for  a  discussion  of  their  size  see
sec.~\ref{sec:2L}). For simplicity, we  will refer to this reduced set
of  two-loop  contributions  as  ``two-loop  order''  throughout  this
section.    The   remaining   differences   between   the   Higgs-mass
calculations  of \NC\  and  \NFH\  in this  set-up  are summarised  in
tab.~\ref{tab:DiffNCNFH}.  The  applied modifications ensure  that the
comparison between the codes will quantify the numerical impact of the
genuine  NMSSM two-loop  corrections  of $\mathcal{O}{\left(Y_t\lambda
  \alpha_s, \lambda^2 \alpha_s\right)}$.
\begin{table}[htb]
  \centering
  \begin{tabular}{lccc}
    & \texttt{NMSSMCalc} & & \texttt{NMSSM-FeynHiggs}\\\toprule
    one-loop & $\alpha_{\text{em}}{\left( M_Z \right)}$ renormalised
    & $\leftrightarrow$ &
    $\alpha_{\text{em}}{\left( M_Z \right)}$ reparametrised
    \\\midrule
    two-loop & NMSSM $\mathcal{O}{\left(\alpha_t \alpha_s\right)}$
    & $\leftrightarrow$ &
    MSSM $\mathcal{O}{\left(\alpha_t \alpha_s\right)}$
    \\\bottomrule\\
  \end{tabular}
  \caption{Main calculational differences between  \NC\ and our result
    (labelled  \NFH)  in  the  set-up   used  for  the  comparison  in
    sec.~\ref{subsec:comparison}.  The difference at one-loop order is
    caused  only  by the  different  renormalisation  of the  electric
    charge,   described   in   sec.~\ref{sec:reparametrisation}.    At
    two-loop order the codes in this set-up only differ by the genuine
    NMSSM  contributions  of  $\mathcal{O}{\left(Y_t\lambda  \alpha_s,
      \lambda^2  \alpha_s\right)}$.   The  two-loop  MSSM  corrections
    beyond   $\mathcal{O}{\left(\alpha_t  \alpha_s\right)}$   and  the
    resummation  of  logarithms are  switched  off  in \NFH\  for  the
    comparison in sec.~\ref{subsec:comparison}.}
  \label{tab:DiffNCNFH}
\end{table}

We used the SM parameters as  specified in the built-in standard input
files of \texttt{NMSSMCalc}  for this comparison.  We  passed over the
input values in  the quark- and squark-sectors  as on-shell parameters
from   \NFH\  to   \NC.   The   pole  mass   for  the   top,  $m_t   =
173.2\ \giga\electronvolt$, has been used in the loop contributions in
this section, and the renormalisation  scale has been chosen as $m_t$.
For       the        comparison       the        identical       value
$\alpha_s^{\MSbar}{(m_t^{(\text{OS})})} = 0.1069729$ has been used for
both  codes  (using  the  evaluation in  \texttt{NMSSMCalc}  with  the
routines of~\cite{Chetyrkin:2000yt}).

In a first step  the one- and two-loop results of  \NC\ and \NFH\ have
been compared in  the MSSM-limit, where $\lambda$  and $\kappa$ vanish
simultaneously.   Both the  effects of  the different  renormalisation
schemes and  the reparametrisation  have to vanish  in this  limit and
thus the results have to be  identical.  The one- and two-loop results
for the  mass of the  lightest \cp-even  field obtained in  this limit
with both  codes, given in tab.~\ref{tab:MSSMLimit},  are in agreement
with each other with a precision of better than $1~\mega\electronvolt$
for  each  scenario  (the  same  holds in  this  limit  also  for  the
predictions for the other neutral Higgs bosons).
\begin{table}[htbp]
  \centering
  \begin{tabular}{cccccc}
    & & sample scenario & scenario P1 & scenario P9 & scenario A1
    \\\toprule
    \multirow{2}[3]{*}{$\frac{m_{h_1}^{\rm}}{\giga\electronvolt}$ MSSM-limit} & two-loop &
    $116.902$ & $109.579$ & $115.155$ & $109.685$
    \\\cmidrule{2-6}
    & one-loop & 
    $140.742$ & $115.154$ & $152.526$ & $151.293$
    \\\bottomrule
  \end{tabular}
  \caption{Mass of the lightest \cp-even  Higgs fields obtained in the
    MSSM-limit  with  \NC\ and  \NFH\  with  the reparametrisation  to
    $\alpha{(M_Z)}$. Both  codes yield  the identical results  in this
    limit.}
  \label{tab:MSSMLimit}
\end{table}

This confirms  that the MSSM-contributions are  treated identically in
both calculations.  Thus all  observed differences between the results
for non-vanishing values  of $\lambda$ and $\kappa$  can be associated
to  the  treatment of  the  genuine  NMSSM-contributions and  residual
higher-order effects of the different renormalisation of $v$ after the
reparametrisation.

\subsubsection{Sample Scenario}

\begin{figure}[htbp]
  \centering
  \includegraphics[width=.49\textwidth]{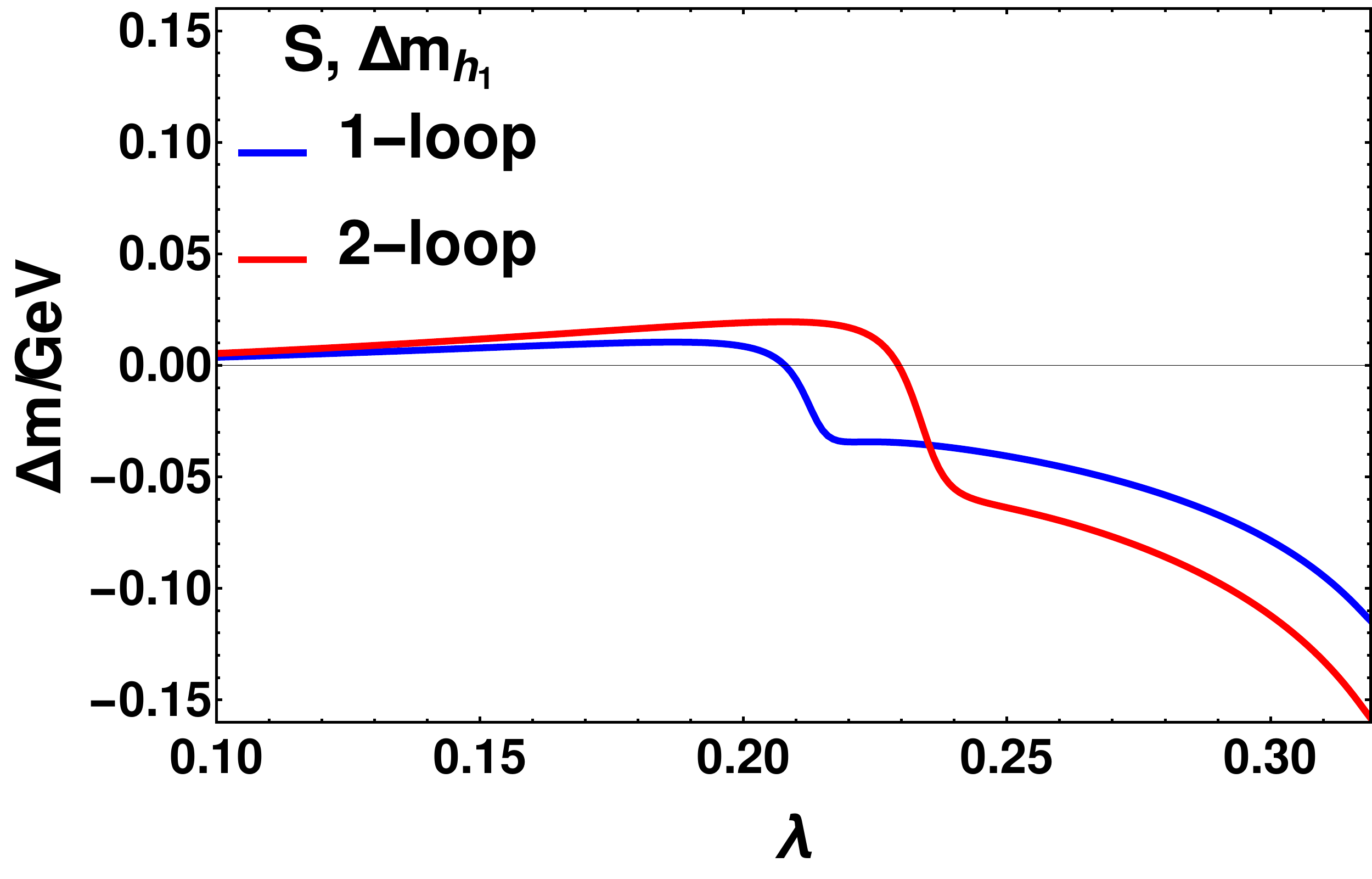}
  \includegraphics[width=.49\textwidth]{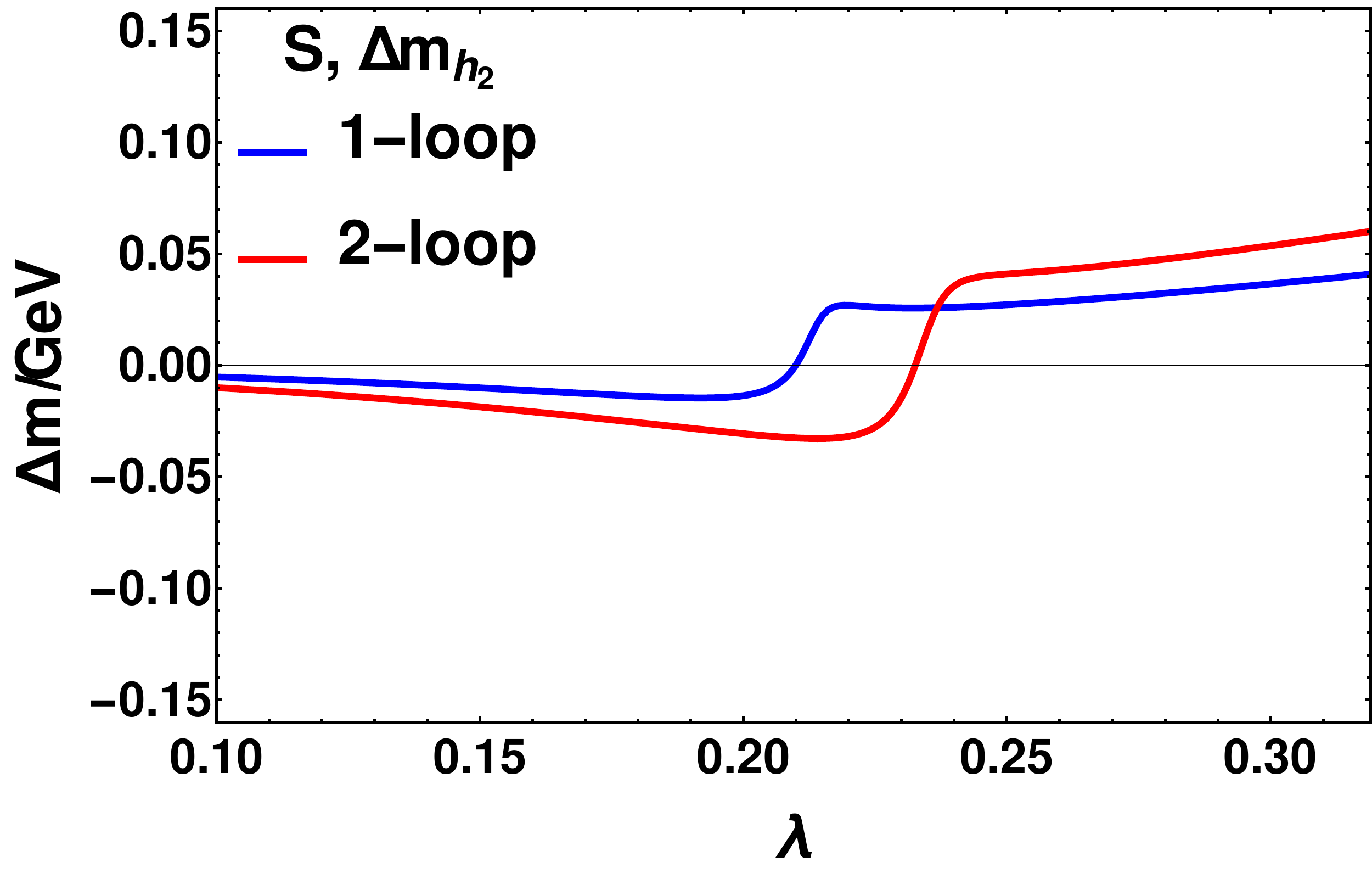}
  \caption{Difference between the mass predictions for the two lighter
    \cp-even  fields  $h_1$  and  $h_2$  from  \texttt{NMSSMCalc}  and
    \texttt{NMSSM-FeynHiggs}    at    one-   and    two-loop    order,
    $\Delta{m_{h_i}}       =       m_{h_i}^{\texttt{NMSSM-FH}}       -
    m_{h_i}^{\texttt{NMSSMCalc}}$,  for  the   sample  scenario.   The
    result  of  \texttt{NMSSM-FeynHiggs}  has been  reparametrised  to
    $\alpha{\left(M_Z\right)}$.   The  points   where  the  cross-over
    behaviour  of  the fields  $h_1$  and  $h_2$  occurs at  one-  and
    two-loop  order  are  $\lambda_{\rm  c}^{(1)}  \approx  0.21$  and
    $\lambda_{\rm c}^{(2)} \approx 0.24$.}
  \label{fig:h12vsNMSSMCalcNoAbs}
\end{figure}

For the  sample scenario defined in  tab.~\ref{tab:SampleScenario} the
differences  between   the  two   mass  predictions  are   plotted  in
fig.~\ref{fig:h12vsNMSSMCalcNoAbs} as  functions of $\lambda$  for the
two   lighter   \cp-even   states   at  one-   and   two-loop   order,
$\Delta{m_{h_i}}        =         m_{h_i}^{\texttt{NMSSM-FH}}        -
m_{h_i}^{\texttt{NMSSMCalc}}$.        The      left       plot      in
fig.~\ref{fig:h12vsNMSSMCalcNoAbs}  shows  the  mass for  the  lighter
state $h_1$, and the mass for the  heavier state $h_2$ is shown in the
right plot.  The state $h_1$  behaves doublet-like for values $\lambda
\lesssim \lambda_{\rm  c}^{(n)}$ and singlet-like for  values $\lambda
\gtrsim  \lambda_{\rm  c}^{(n)}$  (the   behaviour  of  $h_2$  is  the
opposite),   where   $\lambda_{\rm    c}^{(1)}   \approx   0.21$   and
$\lambda_{\rm c}^{(2)} \approx 0.24$.
\footnote{These  values for  $\lambda_{\rm  c}^{(n)}$ slightly  differ
  from the  ones quoted  in \reffi{fig:h12Total}  since our  result in
  \reffi{fig:h12Total} has been parametrised  in terms of $G_F$, while
  for the comparison with  \texttt{NMSSMCalc} we have parametrised the
  result   in    terms   of    $\alpha{(M_Z)}$.}    The    values   of
$\Delta{m_{h_i}}$ are seen  to be negative for  the doublet-like field
and positive for the singlet-like  field.  We find that the difference
between the  two results is small  for both mass predictions  over the
whole  range  of $\lambda$.   As  expected,  the largest  differences,
reaching  about $90~\mega\electronvolt$  for $\Delta{m_{h_1}}$  at the
one-loop level, occur  for the mass of the singlet-like  state for the
largest values of $\lambda$ in the plot.  The mass of the doublet-like
state  is   affected  to   a  lesser   extent  by   approximating  the
$\mathcal{O}{(\alpha_t   \alpha_s)}$  correction   by  the   MSSM-type
contributions.  The  general shape of the  one-loop difference, caused
by the different  treatment of the charge renormalisation,  is seen to
be maintained at the two-loop level.  The main feature at the two-loop
level is  the shift in  the cross-over points  by $\Delta{\lambda_{\rm
    c}} = \lambda_{\rm c}^{(2)}  - \lambda_{\rm c}^{(1)} \approx 0.03$
between two- and one-loop order, while otherwise the difference in the
$\mathcal{O}{(\alpha_t \alpha_s)}$  contributions is  found to  have a
very small  effect. This  can be  seen for  instance by  comparing the
local  and  global extrema  at  $\lambda_{\rm  c}^{(n)}$ and  for  the
largest values around $\lambda_{\rm max}$.  Specifically, for $\lambda
= 0.32$ we find that the  impact of the genuine NMSSM contributions of
$\mathcal{O}{(\alpha_t  \alpha_s)}$  that   are  implemented  only  in
\NC\ amounts  to less  than $50~\mega\electronvolt$  for $h_1$  (for a
Higgs mass of $m_{h_1} \approx 40~\giga\electronvolt$).  For $h_2$ the
comparison of the height of the local extrema in the cross-over region
yields a difference below $20~\mega\electronvolt$ (for a Higgs mass of
$m_{h_2} \approx 125~\giga\electronvolt$).

\subsubsection{Scenarios P1, P9 and A1}

\begin{figure}[htbp]
  \centering
  \includegraphics[width=.49\textwidth]{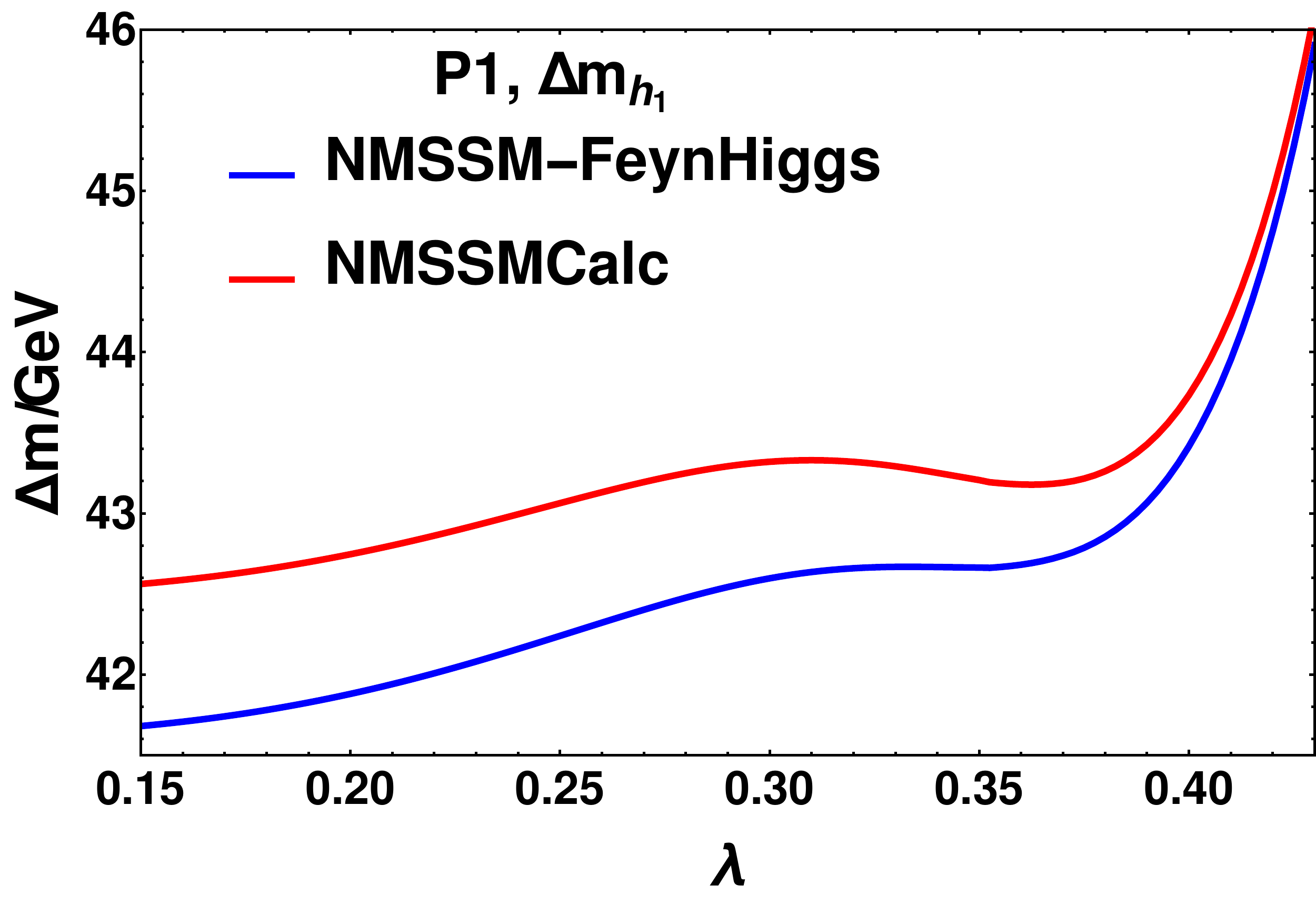}
  \includegraphics[width=.49\textwidth]{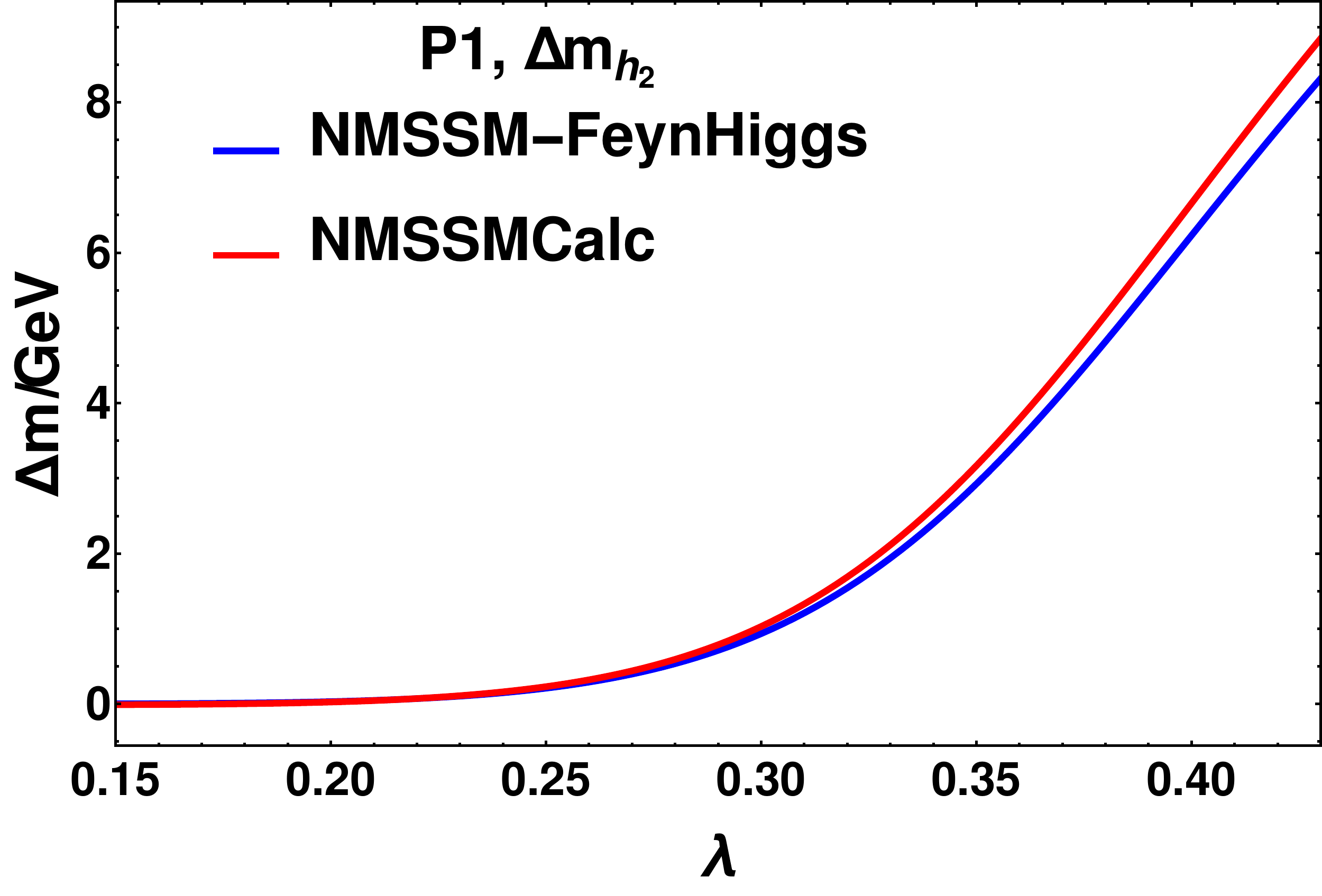}
  \\
  \includegraphics[width=.49\textwidth]{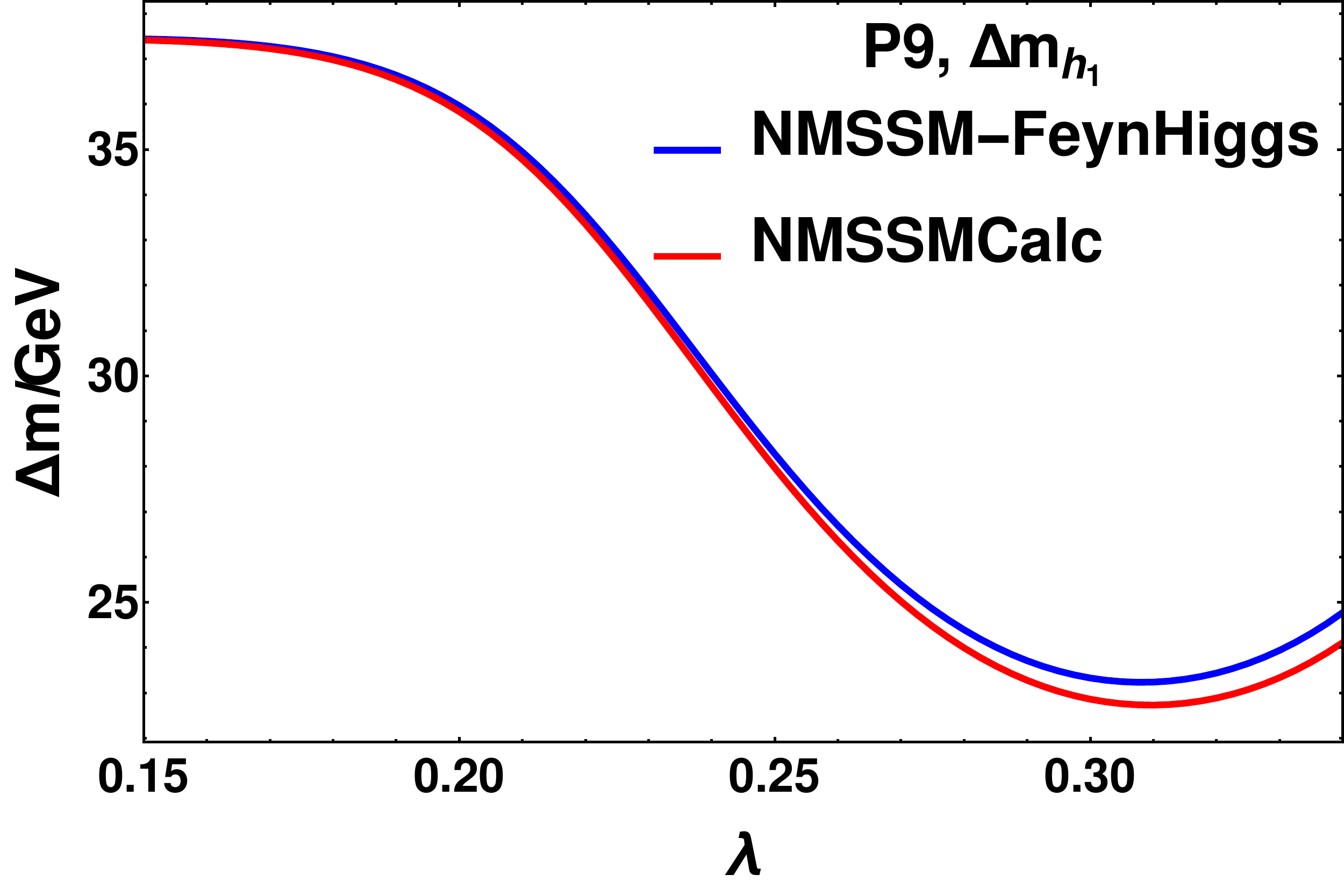}
  \includegraphics[width=.49\textwidth]{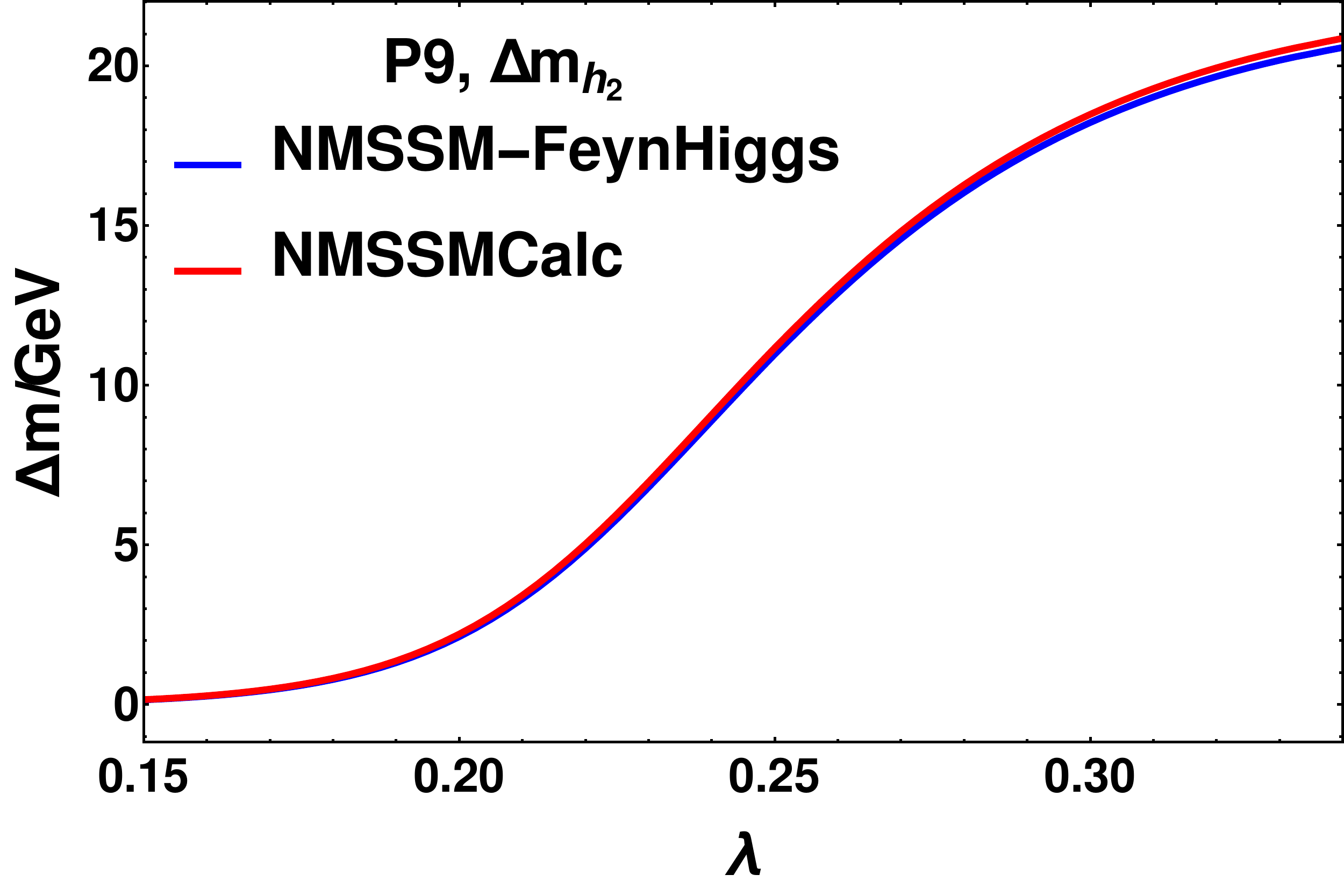}
  \\
  \includegraphics[width=.49\textwidth]{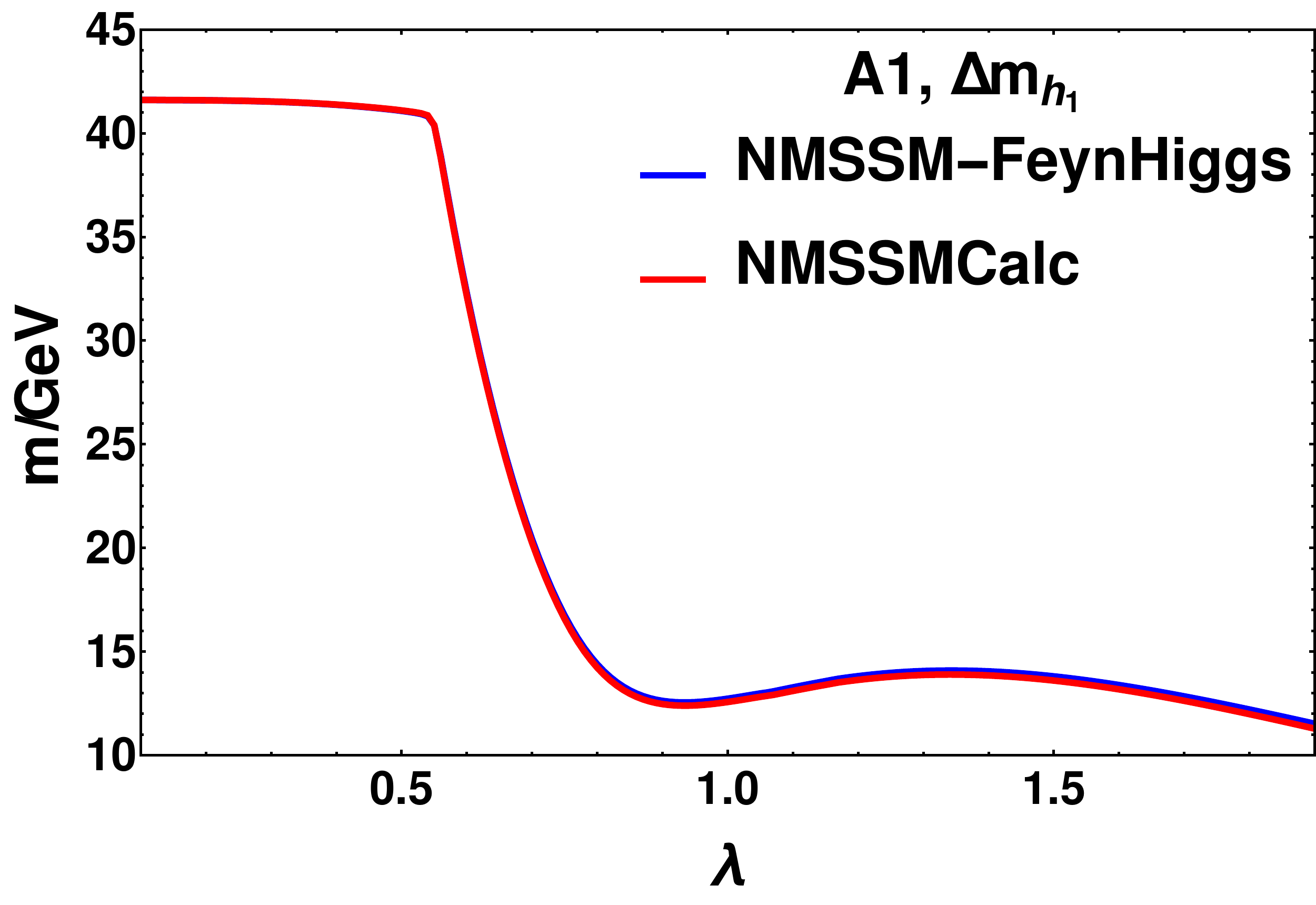}
  \includegraphics[width=.49\textwidth]{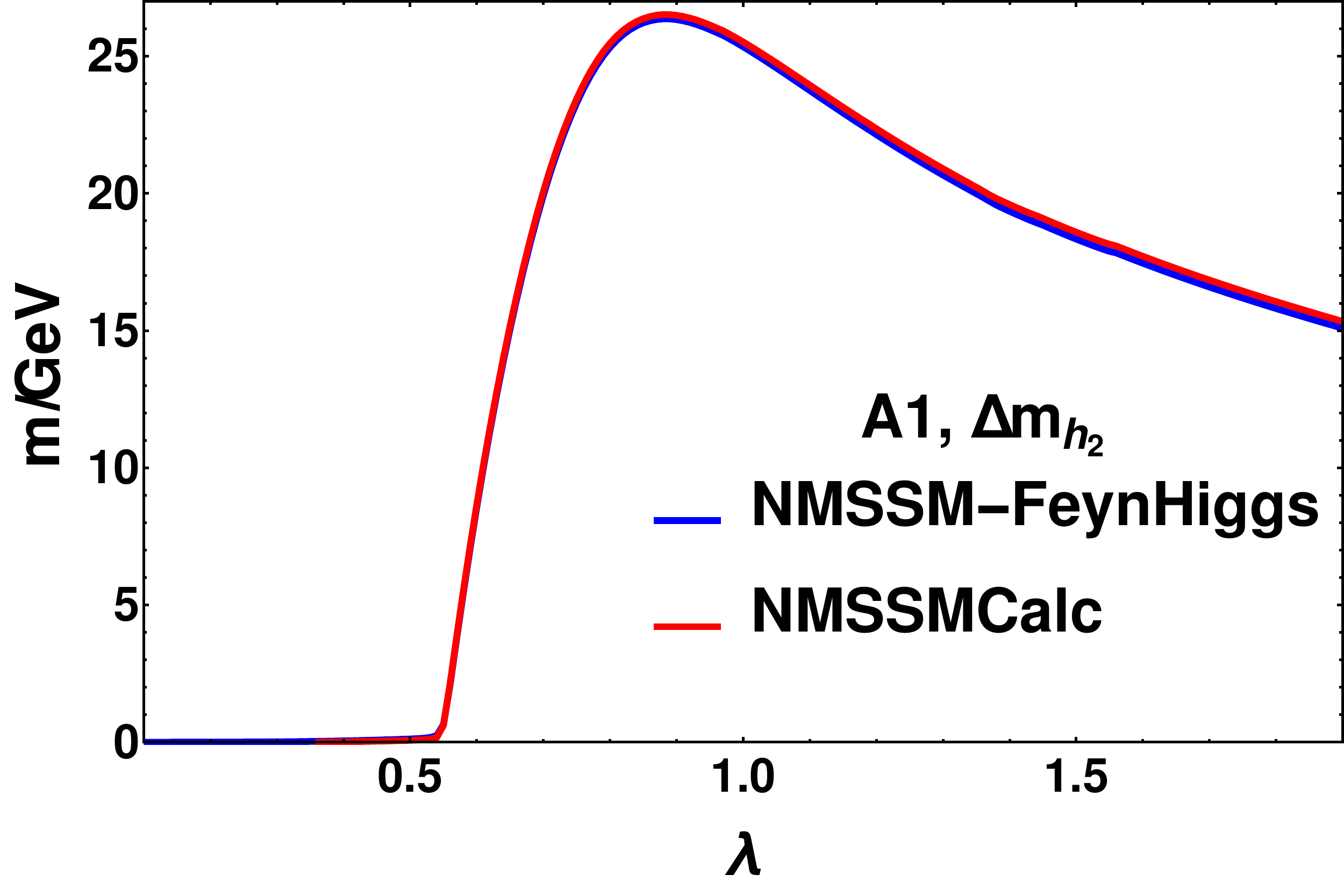}
  \caption{Size  of  the  two-loop contributions,  $\Delta{m_{h_i}}  =
    m_{h_i}^{\rm 1L}  - m_{h_i}^{\rm 2L}$ evaluated  with \NFH\ (blue)
    and with  \NC\ (red) for  the masses  of the two  lighter \cp-even
    fields $h_1$ (left)  and $h_2$ (right) in the  scenarios P1 (upper
    row), P9 (second row) and A1 (third row).}
  \label{fig:2Lminus1L}
\end{figure}

The comparisons between \NFH\ and \NC\ for the scenarios P1 and P9 are
shown in \reffi{fig:2Lminus1L} in the  first and second row, while the
scenario A1  is shown in  the lower  row.  For better  illustration we
plot here the  size of the two-loop  contributions, $\Delta{m_{h_i}} =
m_{h_i}^{\rm 1L} -  m_{h_i}^{\rm 2L}$, as obtained with  the two codes
as a  function of  $\lambda$.  As  for the  sample scenario,  the main
effect  in the  comparison  arises  from a  slight  relative shift  in
$\lambda$ between the  predictions of the two codes.   At the one-loop
level this shift amounts typically to $\Delta{\lambda_{\rm c}} \approx
10^{-4}$ (the corresponding plots are  not shown here since the curves
for  the   predictions  of   the  two   codes  would   be  essentially
indistinguishable).   For  the  two-loop  contributions  displayed  in
\reffi{fig:2Lminus1L} one can see that the genuine NMSSM-type two-loop
corrections  that are  implemented in  \NC\  give rise  to a  slightly
different  dependence on  $\lambda$, which  becomes visible  for large
values of $\lambda$.

As  discussed  above,  in  the  P1 scenario  the  displayed  range  of
$\lambda$ corresponds  to the region  below the cross-over  point. For
the sample scenario  we found in this region a  slight increase in the
absolute      difference      between       the      results,      see
\reffi{fig:h12vsNMSSMCalcNoAbs}.  The difference  between the two-loop
contributions  shown  in \reffi{fig:2Lminus1L}  is  seen  to follow  a
similar pattern.  For  $h_1$ (upper left plot)  the difference between
the two contributions exceeds the  level of $0.5~\gev$ for the highest
values of $\lambda$ that are possible  in this scenario because of the
steep slope  of the  curves (which are  slightly shifted  in $\lambda$
with  respect  to   each  other)  in  this   region.   The  dominantly
doublet-like state  $h_1$ has a significant  singlet-admixture in this
region, which increases up to more than 30\% for the highest $\lambda$
values.   It should  be  noted  that such  a  large singlet  admixture
severely  worsens  the  compatibility  of the  state  $h_1$  with  the
observed Higgs signal at about  $125~\gev$ (independently of its mass,
which is  incompatible with the signal  in this part of  the plot, see
\reffi{fig:h12Total}).    The   differences   are  smaller   for   the
(dominantly singlet-like) state  $h_2$ (upper right plot)  and reach a
significant  level only  for $\lambda$  values that  are close  to the
boundary of the allowed range.

For the  scenario P9, where  above the cross-over region  a relatively
large admixture  of more  than 40\% between  the doublet-like  and the
singlet-like state occurs, the  differences stay relatively small over
the whole  displayed range  of $\lambda$ both  for $h_1$  (middle left
plot) and $h_2$ (middle right plot).  The largest deviations occur for
the  dominantly  singlet-like state  $h_1$  (with  a sizeable  doublet
admixture) for  the highest values  of $\lambda$ above  the cross-over
region, where the two-loop contributions  differ from each other by up
to $0.8~\gev$.

For the  scenario A1, where  above the cross-over region  a relatively
large admixture  of more  than 30\% between  the doublet-like  and the
singlet-like  state occurs,  the differences  nevertheless stay  small
over the whole displayed range of $\lambda$ both for $h_1$ (lower left
plot) and  $h_2$ (lower right  plot). Even  for the largest  values of
$\lambda$ the  difference between the two  contributions remains below
$0.26~\giga\electronvolt$.   Our analysis  shows  that  even for  this
extreme  scenario  with very  high  values  of $\lambda$  the  genuine
NMSSM-type two-loop corrections that are  only implemented in \NC\ are
of minor  numerical significance.  From our  analysis at  the one-loop
level,  on  the   other  hand,  it  is  expected   that  the  two-loop
contributions beyond the fermion/sfermion sector are very important in
this parameter region,  so that the theoretical  uncertainties of both
codes are expected to be rather large in this region.

\subsubsection{Conclusion}

The    results    shown   in    \reffi{fig:h12vsNMSSMCalcNoAbs}    and
\reffi{fig:2Lminus1L}  confirm  that  the approximation  in  terms  of
MSSM-type contributions  at the two-loop level  induces an uncertainty
that is  numerically small, if  $\lambda <  Y_t$, as discussed  in the
previous sections.  As expected, the  approximation works best for the
MSSM-like (doublet-like)  fields, and we  have found for  the analysed
scenarios that  the deviations stay  below the level of  $1~\gev$ even
for the  highest possible values  of $\lambda$  and in regions  with a
large admixture between double-  and singlet-like states.  An improved
prediction for singlet-like states for large values of $\lambda$ would
require the incorporation  of two-loop contributions from  the Higgs /
higgsino and gauge-boson / gaugino  sectors, which is beyond the scope
of our present analysis.

\subsection{Impact of additional Corrections beyond \boldmath{$\mathcal{O}{\left(\alpha_t \alpha_s\right)}$}}
\label{sec:2L}

\begin{figure}[htbp]
  \centering
  \includegraphics[width=.49\textwidth]{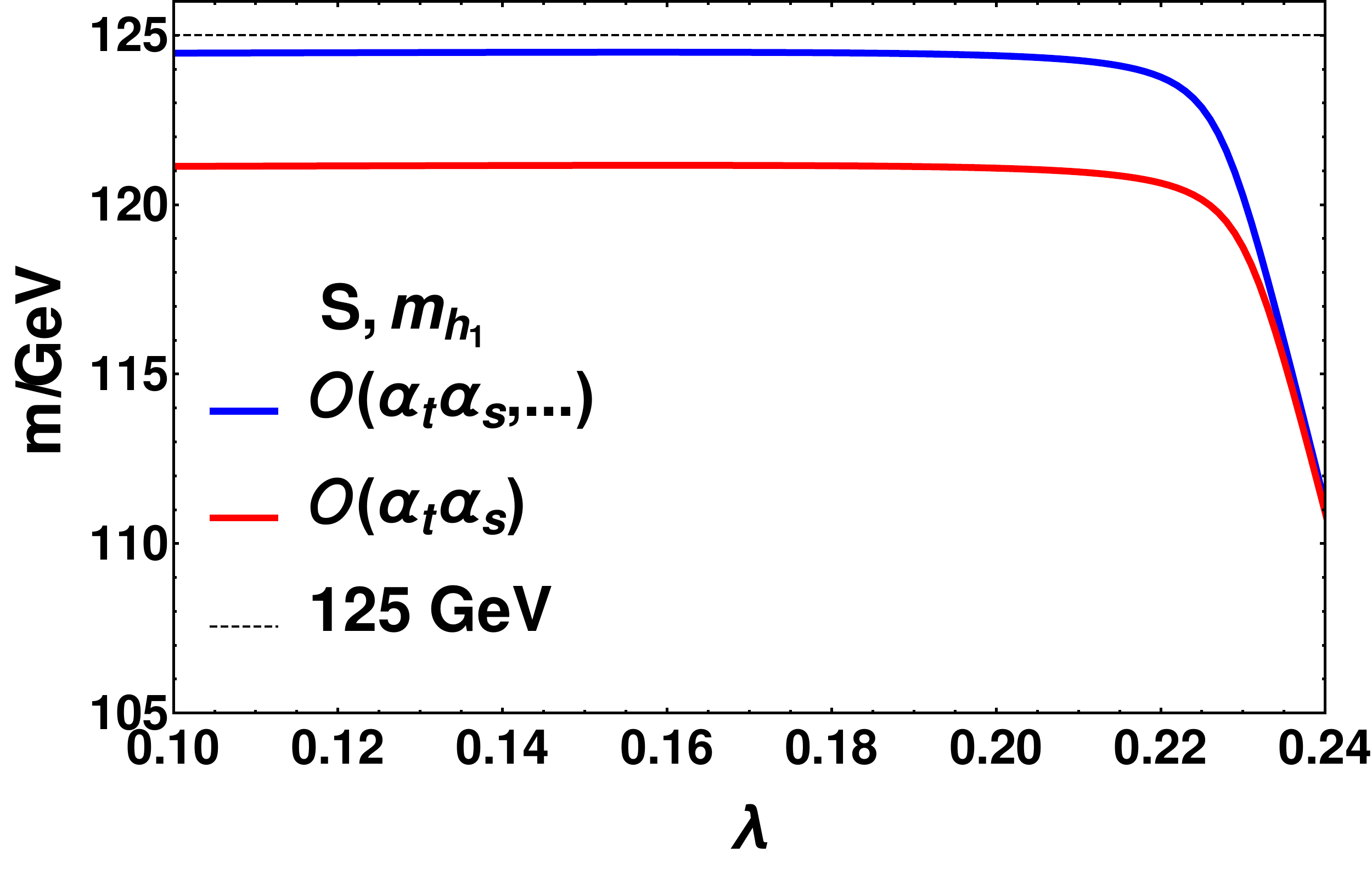}
  \includegraphics[width=.49\textwidth]{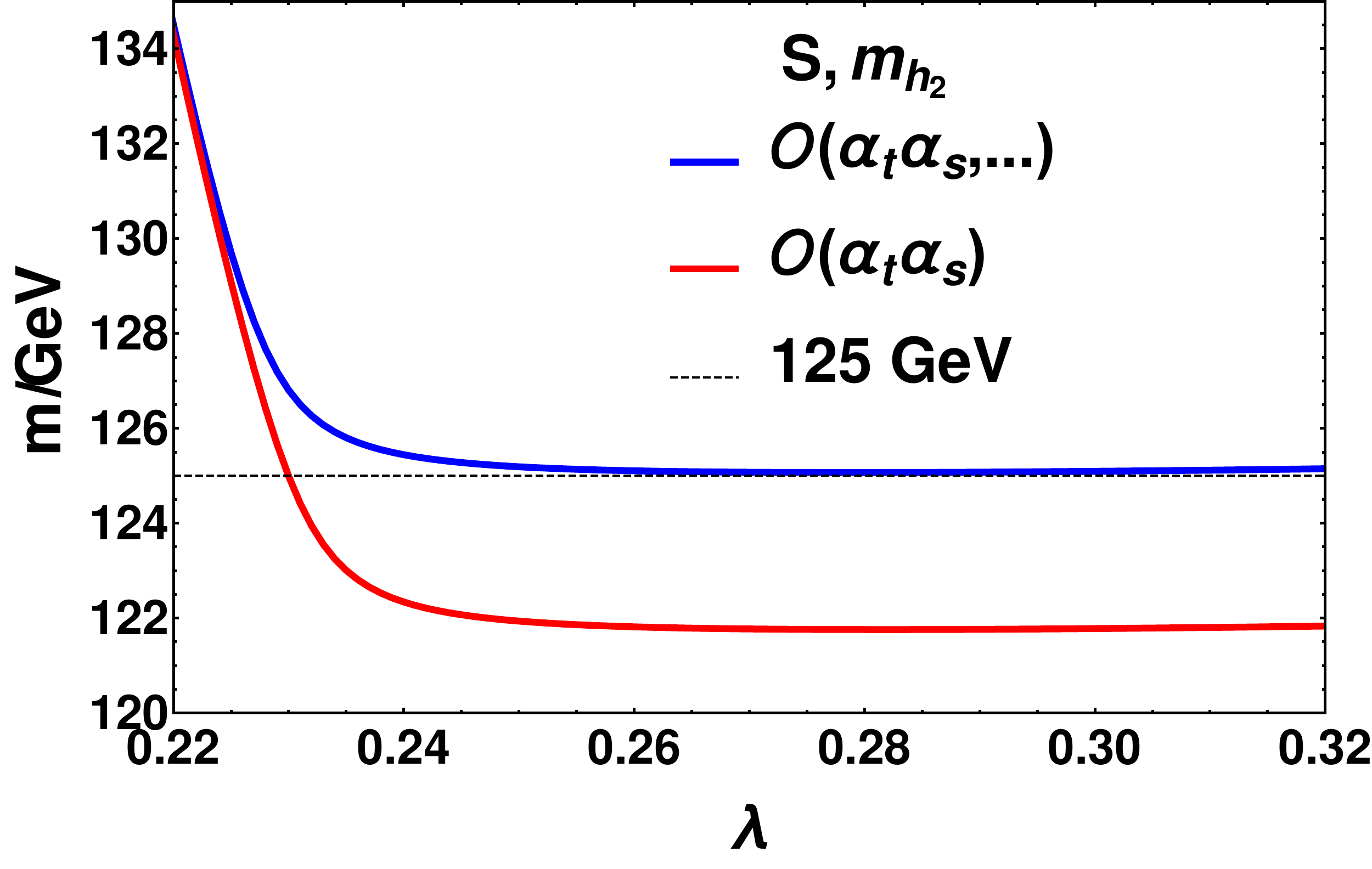}
  \includegraphics[width=.49\textwidth]{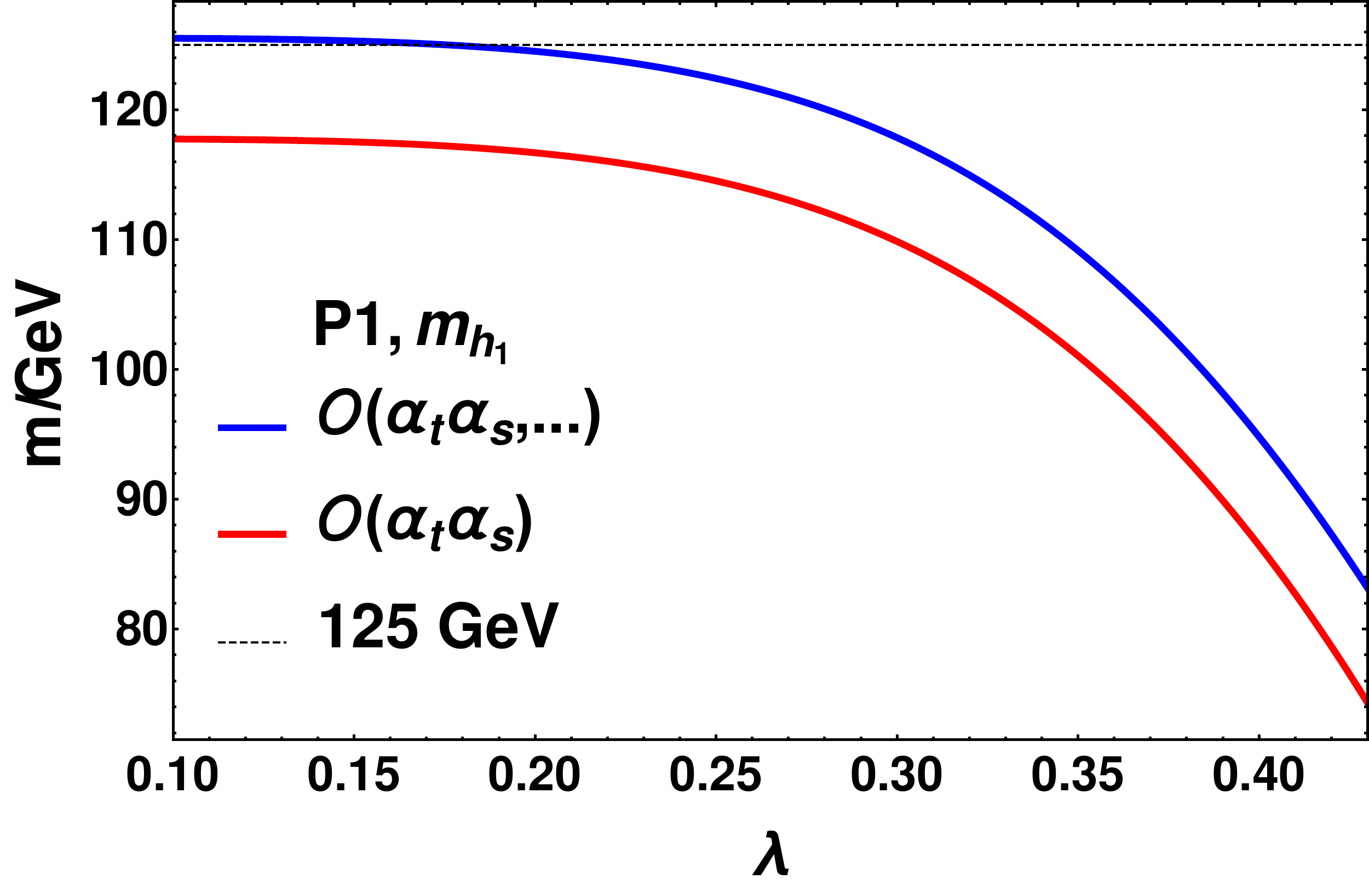}
  \includegraphics[width=.49\textwidth]{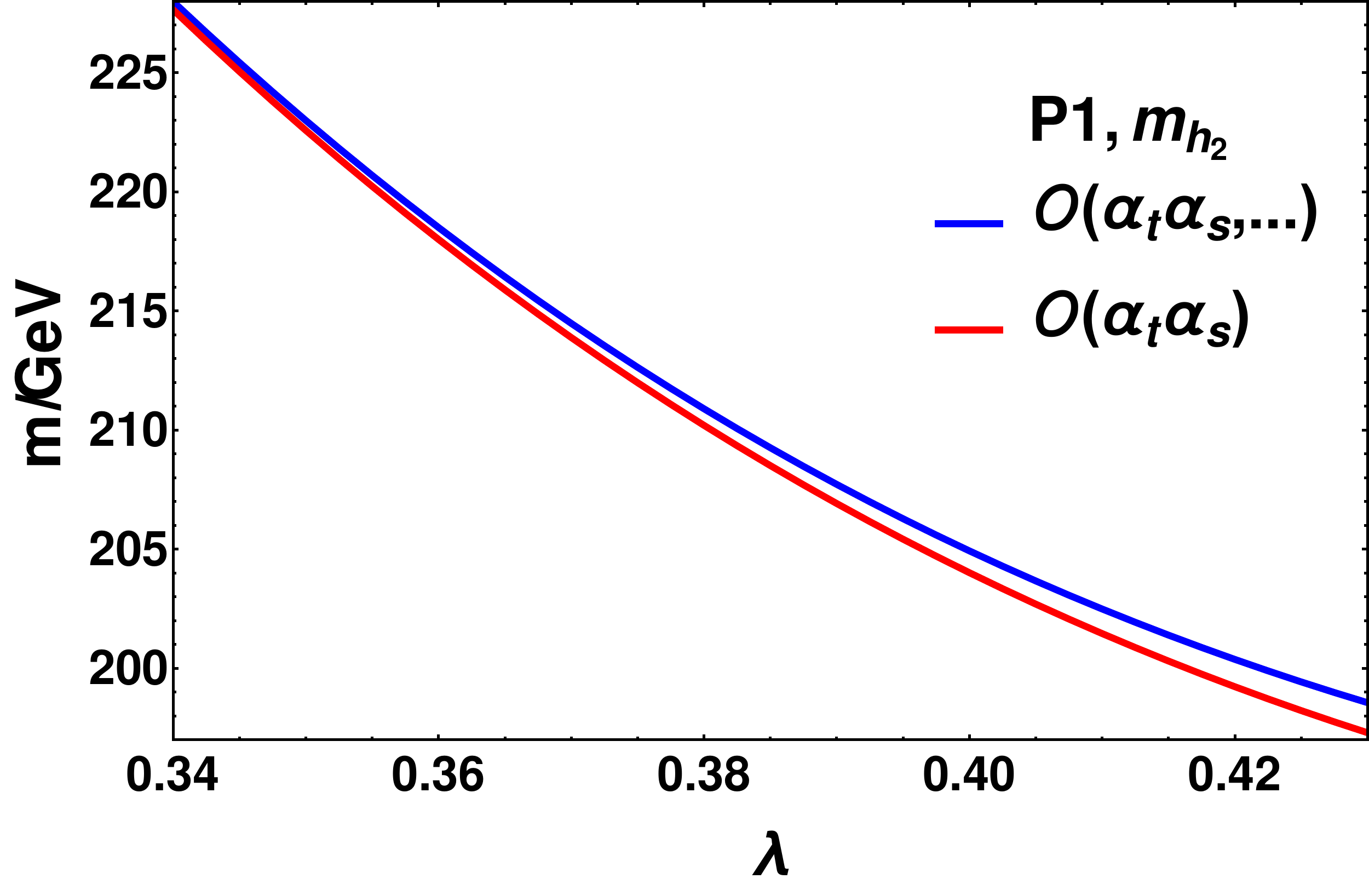}
  \includegraphics[width=.49\textwidth]{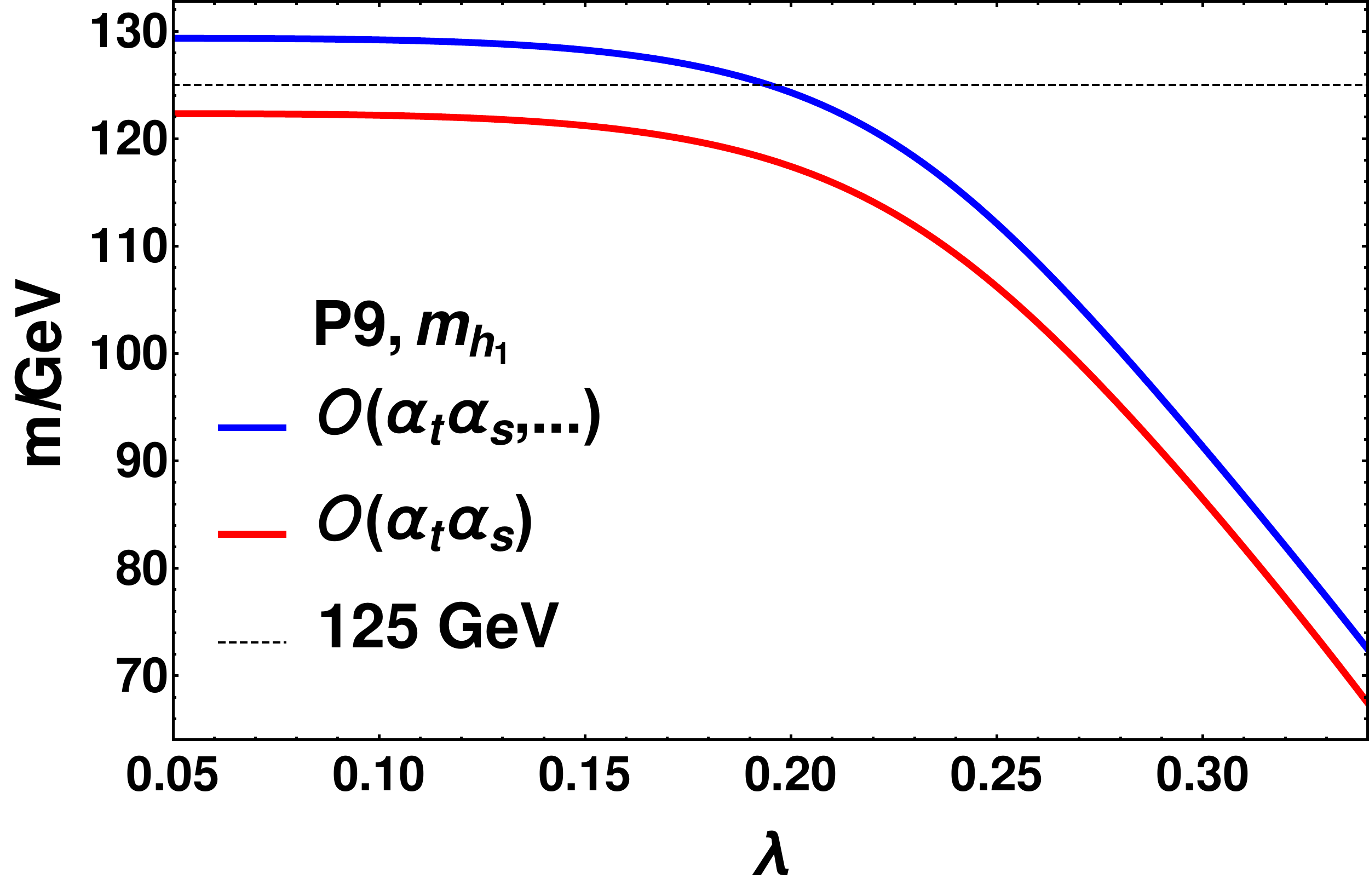}
  \includegraphics[width=.49\textwidth]{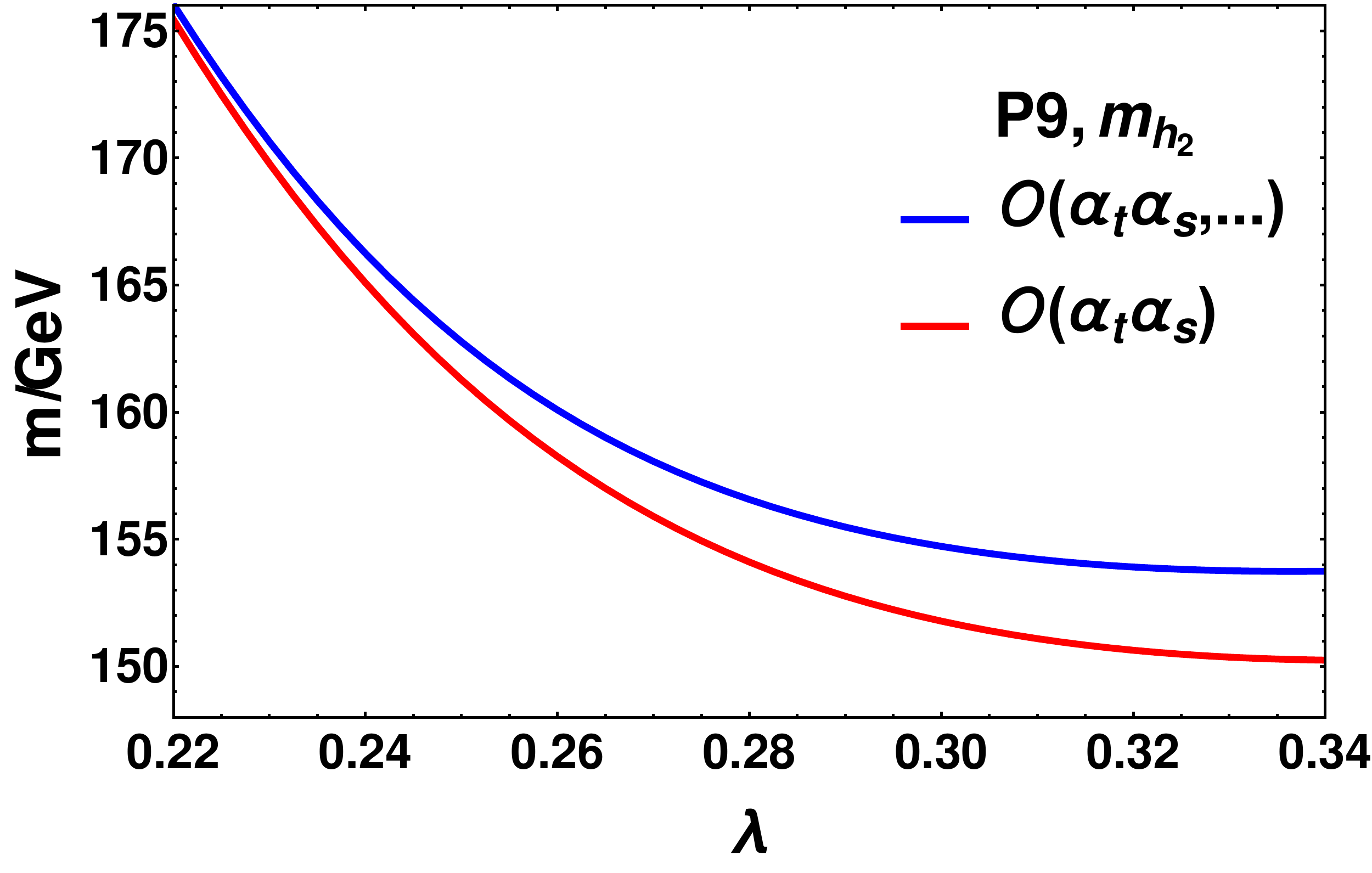}
  \caption{Mass predictions for the  two lighter \cp-even fields $h_1$
    and $h_2$  for different  contributions at  two-loop order  in the
    sample scenario (first row) and  the scenarios P1 (second row) and
    P9 (third row).  The blue  lines include all MSSM-type corrections
    of           $\mathcal{O}{\left(\alpha_t\alpha_s,\alpha_b\alpha_s,
      \alpha_t^2,\alpha_t\alpha_b\right)}$  and   the  resummation  of
    large logarithms  as included in \texttt{FeynHiggs  2.10.2}, while
    for   the   red  curves   only   the   MSSM-type  corrections   of
    $\mathcal{O}{(\alpha_t\alpha_s)}$  are  included beyond  the  full
    one-loop   contributions.    The   thin  horizontal   line   marks
    $125~\giga\electronvolt$.}
  \label{fig:TotalL2}
\end{figure}

\begin{figure}[htbp]
  \centering
  \includegraphics[width=.49\textwidth]{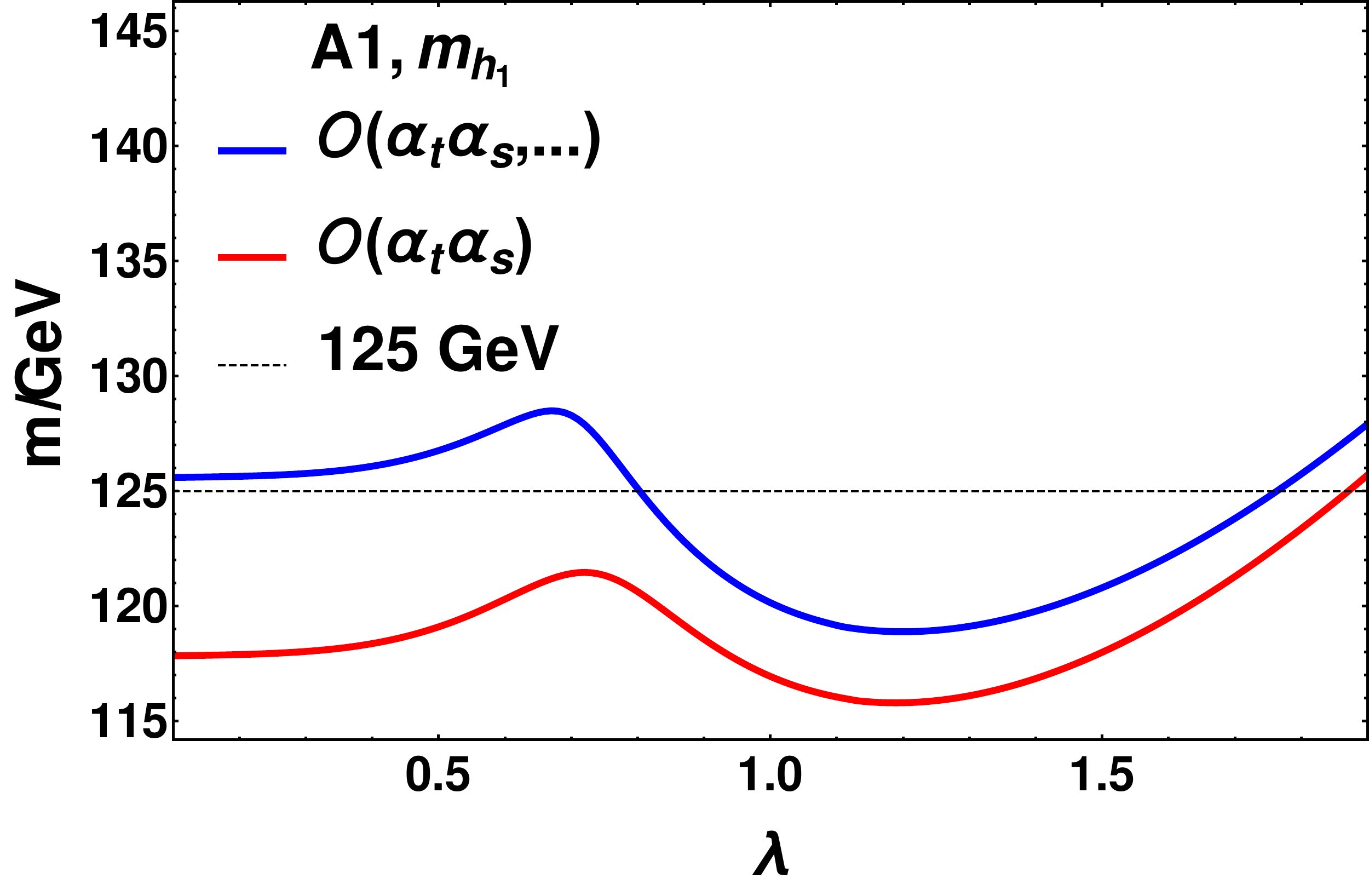}
  \includegraphics[width=.49\textwidth]{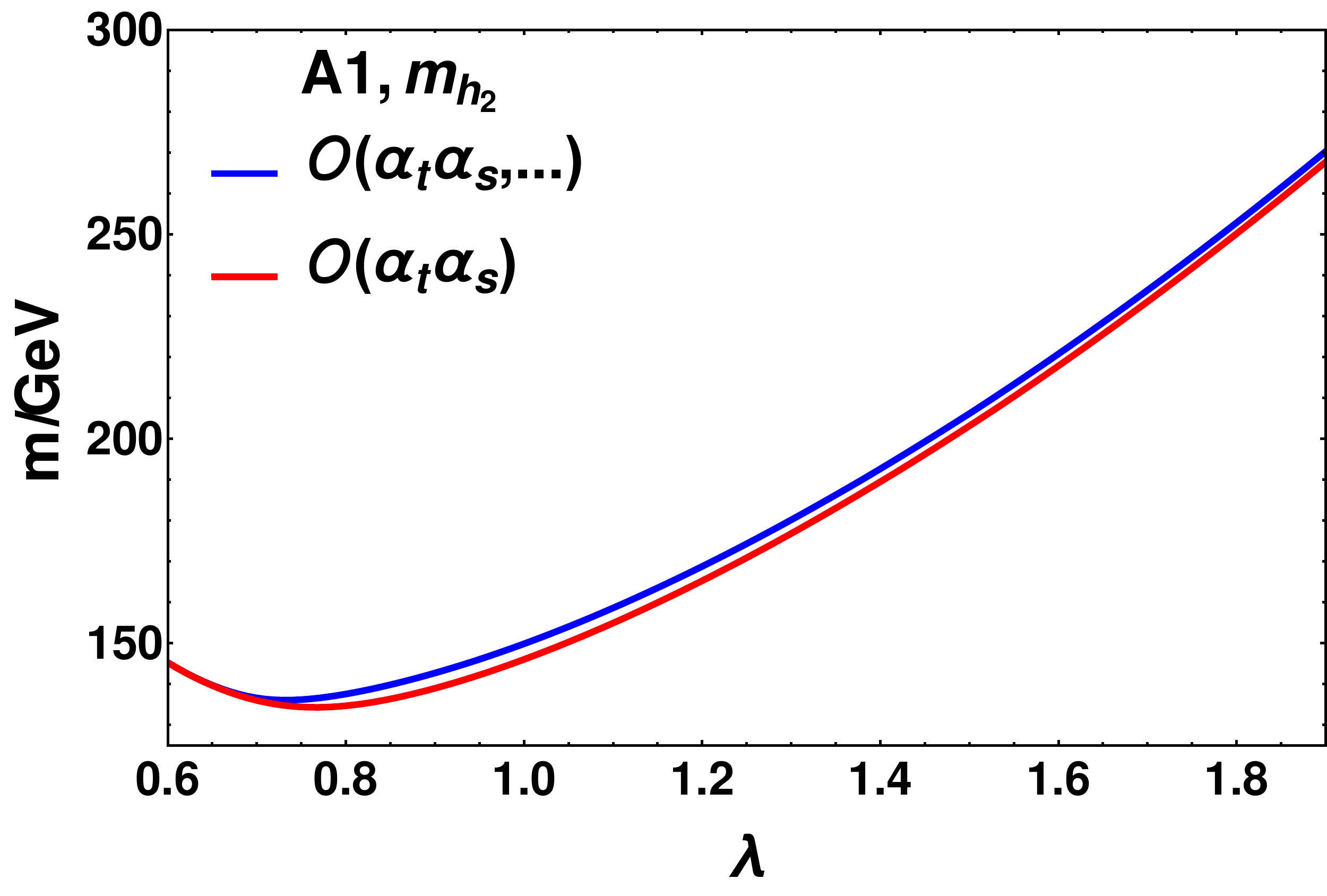}
  \caption{Mass predictions for the  two lighter \cp-even fields $h_1$
    and $h_2$  for different  contributions at  two-loop order  in the
    scenario A1. The meaning of the displayed curves is the same as in
    fig.~\ref{fig:TotalL2}.}
  \label{fig:TotalL2PD}
\end{figure}

While  the genuine  NMSSM  two-loop  corrections of  $\mathcal{O}{(Y_t
  \lambda  \alpha_s, \lambda^2\alpha_s  )}$ induce  small effects,  as
discussed  in  the previous  section,  the  MSSM two-loop  corrections
beyond $\mathcal{O}{(\alpha_t \alpha_s)}$ and the resummation of large
logarithms  can  result  in  a  shift   for  the  mass  of  the  light
doublet-like  field  of  several $\giga\electronvolt$.   In  order  to
quantify   the  impact   of  the   additional  MSSM-contributions   of
$\mathcal{O}{(\alpha_t^2, \alpha_b \alpha_s,  \alpha_t \alpha_b)}$ and
the resummation  of logarithms,  which are  incorporated in  \NFH, the
results with and without these corrections are plotted as functions of
$\lambda$ in  figs.~\ref{fig:TotalL2} and~\ref{fig:TotalL2PD}  for the
discussed scenarios.
Here    the     one-loop    $\MSbar$-value    of     the    top-quark,
$m_t^{\MSbar}{(m_t)}$, is used in  the loop contributions.  A sizeable
shift of about 3--8~$\giga\electronvolt$ can  be observed for the mass
of the doublet-like  field.  As expected, the impact  of the MSSM-type
two-loop  contributions on  the mass  prediction for  the singlet-like
field remains  small.  In comparison with  the contributions discussed
in the  previous section  we find  that the  effect of  the additional
corrections  beyond $\mathcal{O}{(\alpha_t\alpha_s)}$  can exceed  the
numerical impact of the genuine NMSSM-corrections of $\mathcal{O}{(Y_t
  \lambda \alpha_s,  \lambda^2 \alpha_s)}$ by  more than one  order of
magnitude.

\section{Conclusions}
\label{sec:concl}

We have presented predictions for  the Higgs-boson masses in the NMSSM
obtained within the Feynman-diagrammatic  approach.  They are based on
the full NMSSM one-loop corrections supplemented with the dominant and
sub-dominant    two-loop   corrections    of   MSSM-type,    including
contributions   at   $\mathcal{O}{\left(\alpha_t  \alpha_s,   \alpha_b
  \alpha_s,   \alpha_t^2,\alpha_t\alpha_b\right)}$,  as   well  as   a
resummation of  leading and subleading logarithms  from the top/scalar
top  sector.   In  order  to  enable  a  direct  comparison  with  the
corresponding results in the MSSM,  the renormalisation scheme and all
parameters and conventions  have been chosen such  that the well-known
MSSM result  of the code  \fh\ is recovered in  the MSSM limit  of the
NMSSM.

In our phenomenological analysis we have first investigated a scenario
where depending on  the value of $\lambda$ either the  lightest or the
next-to-lightest neutral Higgs state can  be identified with a SM-like
Higgs boson  at about  $125~\giga\electronvolt$.  Furthermore  we have
investigated  two  scenarios  (originally   proposed  in  a  different
context,  see  the  discussion  in  \refse{subsec:numScenario})  where
larger  values  of  $\lambda$  than  in the  sample  scenario  can  be
realised, and  sizeable admixtures  between singlet-  and doublet-like
states  can occur  also  outside of  the  ``cross-over'' region.   The
lightest neutral  Higgs-state can be  identified with a  SM-like Higgs
boson at about $125~\giga\electronvolt$ in  both scenarios for low and
moderate  values of  $\lambda$.  As  expected, the  state that  can be
identified    with    the    observed   Higgs    signal    at    about
$125~\giga\electronvolt$  is  doublet-like  in  all  cases,  i.e.\  it
receives only relatively small contributions from the singlet state of
the  NMSSM.   In  order  to  investigate the  impact  of  the  various
contributions for even higher values of $\lambda$, we have furthermore
analysed  another variation  of  these scenarios  in  which values  of
$\lambda$ up  to $\lsim 1.5$  can be  realised.  The inclusion  of the
higher-order contributions which are known for the MSSM is crucial for
all scenarios in  order to obtain an accurate prediction  for the mass
spectrum.

We have investigated different approximations at the one-loop level in
comparison with our full one-loop  result for the NMSSM. We have found
that the approximation of the  result for the top/stop sector in terms
of  the leading MSSM-type  contributions works  well in  the parameter
regions  where   the  top/stop  sector  itself   yields  a  reasonable
approximation of  the full  result.  It therefore  appears to  be well
motivated to  make use  of this approximation  at the  two-loop level.
The genuine NMSSM  top/stop sector contributions of \order{\yt\lambda,
  \lambda^2} can  be significant for singlet-like  fields if $\lambda$
is  large.    For  such  large  values  of   $\lambda$,  however,  the
improvement  achieved by including  those genuine  NMSSM contributions
from  the top/stop  sector is  by far  overshadowed by  the  fact that
contributions from the Higgs- and higgsino-sector become more and more
important for a singlet-like Higgs field.

We   have   compared   our    predictions   with   the   public   code
\texttt{NMSSMCalc}  for on-shell  parameters in  the top/stop  sector.
For  the  purpose of  this  comparison  we  have done  an  appropriate
reparametrisation  of the  electromagnetic coupling  constant, and  we
have  switched  off  the  two-loop  corrections  beyond  the  ones  of
$\mathcal{O}{(\alpha_t  \alpha_s)}$  as  well as  the  resummation  of
leading and subleading logarithms in our code. After those adaptations
the  predictions  of   the  two  codes  only  differ   in  the  charge
renormalisation  at  the  one-loop  level and  in  the  genuine  NMSSM
top/stop    sector    contributions   of    \order{\yt\lambda\alpha_s,
  \lambda^2\alpha_s} at  the two-loop level.  Since  these differences
arise only from  contributions beyond the MSSM,  agreement between the
predictions of  the two  codes is  expected in the  MSSM limit  of the
NMSSM. We  have indeed found  that the  results obtained with  the two
codes perfectly  agree with each other  in this case. For  the case of
the  NMSSM we  have compared  the predictions  of the  two codes  as a
function of~$\lambda$.  We have found  that the differences stay small
over the whole range of  $\lambda$, with a maximum absolute difference
in the mass  of the singlet- or the doublet-like  state below $1~\gev$
in the considered  scenarios.  The difference is mainly  caused by the
different  treatment of  the  charge renormalisation  at the  one-loop
level,  while  the  effect  of   the  genuine  NMSSM  top/stop  sector
contributions  of   \order{\yt\lambda\alpha_s,  \lambda^2\alpha_s}  is
found  to  be generally  smaller  except  for  the highest  values  of
$\lambda$ that  can be realised in  the scenarios.  The impact  of the
genuine      NMSSM      top/stop     sector      contributions      of
\order{\yt\lambda\alpha_s, \lambda^2\alpha_s}  turned out to  be small
even in parameter regions where  the dominantly doublet-like state has
a singlet-admixture  of more  than 30\%.   A more  detailed comparison
between the two codes will be presented in a forthcoming publication.

As a  final step of  our numerical  analysis we have  investigated the
impact   of   the   MSSM-corrections   beyond   $\mathcal{O}{(\alpha_t
  \alpha_s)}$  and  the  resummation  of  large  logarithms  that  are
incorporated in our  code but not in  \texttt{NMSSMCalc}.  While those
corrections are small for the mass of a dominantly singlet-like state,
they amount to  an effect of 3--8~\giga\electronvolt\ for  the mass of
the doublet-like state in the considered scenarios.  This is typically
more than an  order of magnitude larger than  the corresponding effect
of   the  genuine   NMSSM-corrections  of   $\mathcal{O}{(Y_t  \lambda
  \alpha_s, \lambda^2 \alpha_s)}$.

The results presented  in this paper will  be used as a  basis for the
extension of  the code \fh\  to the  NMSSM. Our analysis  has revealed
that  for singlet-like  states in  the parameter  region of  very high
values of  $\lambda$ two-loop corrections beyond  the fermion/sfermion
sector are expected to be sizeable. In order to reduce the theoretical
uncertainties in  this parameter region the  incorporation of two-loop
contributions  from the  Higgs /  higgsino and  gauge-boson /  gaugino
sectors will be desirable. Partial results of this kind have only been
obtained in a pure \DRbar\ scheme up  to now. We leave a more detailed
discussion of this issue to future work.

\section*{Acknowledgements}

We want to thank Ramona Gr\"ober, Margarete Mühlleitner, Heidi Rzehak,
Oscar St\r{a}l  and Kathrin Waltz for  interesting discussions, useful
input and  interfaces with their  codes.  We thank Melina  Gomez Bock,
Rachid Benbrik and Pietro Slavich for helpful communication.  The work
of S.H.\ is supported in part by CICYT (grant FPA 2013-40715-P) and by
the  Spanish  MICINN's  Consolider-Ingenio 2010  Program  under  grant
MultiDark CSD2009-00064.   The authors acknowledge support  by the DFG
through  the SFB~676  ``Particles, Strings  and the  Early Universe''.
This research was supported in part by the European Commission through
the ``HiggsTools'' Initial Training Network PITN-GA-2012-316704.



\pagestyle{plain}             

\bibliographystyle{h-elsevier}     
\bibliography{literature}

\end{document}